\documentclass[11pt, letterpaper]{article}
\pdfoutput=1

\usepackage{amsfonts}
\usepackage{amsmath}
\usepackage{amsthm}
\usepackage{bibentry}
\usepackage{color}
\usepackage{epsfig}
\usepackage{graphics}
\usepackage{hyperref}
\hypersetup{ colorlinks = true, citecolor = blue, urlcolor = blue}
\usepackage[utf8]{inputenc}
\usepackage[left, modulo]{lineno}
\usepackage{lmodern}
\usepackage{multirow}
\usepackage{relsize}
\usepackage{rotating}
\usepackage{subfigure}
\usepackage{subfiles}
\usepackage{url}
\usepackage{verbatim}
\usepackage{wrapfig}
\usepackage{floatrow}
\newfloatcommand{capbtabbox}{table}[][\FBwidth]

\usepackage[round]{natbib}
\usepackage[noend]{algpseudocode}
\usepackage{algorithm}

\usepackage{graphicx,psfrag,epsf}
\usepackage{enumerate}

\makeatletter
\@addtoreset{equation}{section}   

\makeatother

\makeatletter
\@addtoreset{figure}{section}
\renewcommand{\thefigure}{\thesection.\@arabic\c@figure}
\makeatother

\makeatletter
\@addtoreset{table}{section}
\renewcommand\thetable{\thesection.\@arabic\c@table}
\makeatother



\usepackage{mycommands2}
\theoremstyle{definition}
\newtheorem{Algorithm}{Algorithm}

\newtheorem{Remark}{Remark}[section]

\begin{document}

\def\spacingset#1{\renewcommand{\baselinestretch}%
{#1}\small\normalsize} \spacingset{1}


\title{Ultrahigh-dimensional Robust and Efficient Sparse Regression using Non-Concave Penalized Density Power Divergence}
\date{}
\author{
	Abhik Ghosh\thanks{Corresponding author. E-mail: {\tt  abhik.ghosh@isical.ac.in}}, Indian Statistical Institute, Kolkata, India\\
	and\\
	Subhabrata Majumdar\thanks{Currently in AT\&T Labs Research. Email: {\tt subho@research.att.com}},
	University of Florida, Gainesville, FL, USA
}
\maketitle

\begin{abstract}
We propose a sparse regression method based on the non-concave penalized density power divergence loss function 
which is robust against infinitesimal contamination in very high dimensionality. 
Present methods of sparse and robust regression are based on $\ell_1$-penalization, and their theoretical properties are not well-investigated. 
In contrast, we use a general class of folded concave penalties that ensure sparse recovery and consistent estimation of regression coefficients. 
We propose an alternating algorithm based on the Concave-Convex procedure to obtain our estimate, 
and demonstrate its robustness properties using influence function analysis. 
Under some conditions on the fixed design matrix and penalty function, 
we prove that this estimator possesses large-sample oracle properties in an ultrahigh-dimensional regime. 
The performance and effectiveness of our proposed method for parameter estimation and 
prediction compared to state-of-the-art are demonstrated through simulation studies.
\end{abstract}

\section{Introduction}
Consider the standard linear regression model (LRM):
\begin{eqnarray}
\boldsymbol{y} &=& \bfX \boldsymbol{\beta} + \boldsymbol{\epsilon},
\label{EQ:LRM}
\end{eqnarray}
where $\boldsymbol{y}=(y_1, \ldots, y_n)^T$ are observations from a response variable $Y$, $\bfX = (\boldsymbol{x}_1 \cdots \boldsymbol{x}_n)^T$  is the design matrix containing associated observed values of the explanatory variable $\BX \in \mathbb{R}^p$, and $\bfepsilon = (\epsilon_1, \ldots, \epsilon_n)^T \sim \cN_n({\bf 0}, \sigma^2 \bfI_n)$ are the random error components. Under this setup, sparse estimators of $\boldsymbol{\beta} = (\beta_1, \ldots, \beta_p)^T$, i.e. the estimators which set some entries in the estimated coefficient vector to exactly 0, can be defined as the minimizer of the following objective function:
\begin{align}\label{eqn:PenLoss}
\sum_{i=1}^n \rho (y_i - \bfx_i^T \bfbeta) + \lambda_n \sum_{j=1}^p p (| \beta_j|),
\end{align}
where $\rho(.)$ is a loss function, $p(.)$ is the sparsity inducing penalty function, and $\lambda_n \equiv \lambda$ is the regularization parameter depending on the sample size $n$. Starting from the Least Absolute Shrinkage and Selection Operator (LASSO) method of \cite{Tibshirani96}, the area of sparse penalized regression has seen a flurry of research in the past two decades, owing to their ability of simultaneously perform variable selection and estimation in high-dimensional datasets from diverse areas such as genome biology, chemometrics and economics. A number of further studies led to improvement in estimation accuracy of the penalized estimates \citep{FanLi01,Zou06,CHZhang10}, as well as algorithmic refinement \citep{KimChoiOh08, ZouLi08, WangKimLi13}.

All the above methods are, however, based on penalizing the least square (or normal log-likelihood) loss function, and as a consequence are not robust against data contamination and model misspecifications. This is a serious concern, since real-world high-dimensional datasets, for example gene expression data, often contain variables with heavy-tailed distributions, or observations contaminated due to technical reasons \citep{OsborneOverbay04,ShiehHung09,ZangEtal17}.

Among robust methods for sparse regression, \cite{KhanEtal07} proposed a robust version of the LARS algorithm for calculating the Lasso solution path (RLARS), while \cite{WangLiJiang07} and \cite{AlfonsEtal13} proposed sparse versions of the regression with Least Absolute Deviation loss (LAD-Lasso) and the Least Trimmed Square method (sLTS), respectively. More recently, \cite{ZangEtal17} sparsified the density power divergece (DPD) loss-based regression, and \cite{Kawashima17} did the same for the log-DPD loss function. DPD-based methods are a robust generalization of maximum likelihood-based inference, and are known to produce highly robust parameter estimates with nominal loss in efficiency in several non-sparse regression problems \citep{BasuEtal98,GhoshBasu13,GhoshBasu16}. The above three papers numerically illustrated that such advantages of DPD-based inference continue to hold in high-dimensional settings.

A major issue with the robust high-dimensional regression techniques proposed till now is that all of them are based on (adaptive) $\ell_1$-penalization. The bias of Lasso estimators is well known for penalized least square regression \citep{ZhangHuang08,JavanmardMontanari18}. Nonconcave penalties, e.g. the Smoothly Clipped Absolute Deviation (SCAD, \cite{FanLi01}) remedy this, giving estimators that are variable selection consistent {\it and} provide asymptotically unbiased penalized (least square) estimates for non-zero elements of the coefficient vector $\bfbeta$. Advantage of nonconcave penalized methods have been demonstrated for the LRM in \eqref{EQ:LRM} (see \cite{WangKimLi13} and the references therein), and multiple-response regression \citep{MajumdarChatterjeeStat}.

In this paper, we combine the strengths of non-concave penalties and the DPD loss function to simultaneously perform variable selection and obtain robust estimates of $\bfbeta$ under the model \eqref{EQ:LRM}. We ensure robustness of our procedure against contaminations of infinitesimal magnitude using influence function analysis, and establish theoretical consistency properties of the proposed estimator in a high-dimensional context with non-polynomial (NP) dimensionality. To this end, we use the DPD loss function with suitable non-concave penalties \citep{FanLi01,CHZhang10} in the setup of \eqref{eqn:PenLoss} to obtain estimators with reduced false positives, estimation and prediction error as compared to $\ell_1$-penalization, along with robustness advantages against data contamination. As an added advantage of the generalized formulation of DPD, we obtain new results on the forms of influence functions and asymptotic properties of penalized maximum likelihood LRM estimates of both the regression coefficient and the error variance.

Properties of the closely related class of penalized high-dimensional M-estimators have previously been analyzed by \cite{NeghabanEtal12,BeanEtal13, DonohoMontanari16, LozanoEtal16, LohWainwright17}, but without detailed attention to the robustness aspects. The assumptions on the design matrix they impose are largely same as those imposed for high-dimensional analysis assuming the least-square loss function (e.g. the restricted eigenvalue condition \citep{BickelRitovTsybakov09}). However, in presence of arbitrary contamination in $\bfX$ one does not expect it to adhere to the nice properties that enable a {\it non-robust} analysis to go through. In contrast, our theoretical conditions in Section~\ref{SEC:Theory} are perhaps the first attempt to recognize the need for modified conditions for theoretical analysis in a robust M-estimation framework in high dimensions.

In recent work, \cite{AveMeRonchetti18} discussed the robustness and asymptotic properties of a particular M-estimator based on a quasi-likelihood approach. However, their high-dimensional asymptotics and illustrations are restricted only to the estimation of $\boldsymbol{\beta}$ using adaptive lasso penalties. In comparison, we derive the general theory for {\it simultaneous} robust estimation of $\boldsymbol{\beta}$ and $\sigma$ along with consistent variable selection using general non-concave penalties for a class of location-scale error model of linear regression, and also develop an efficient computational algorithm for our proposal. As seen later, the simplicity and rigorousness of our proposal makes it feasible to extend for more general parametric regression models. This generalizability is a major strength of our theoretical developments.

\paragraph*{Structure of the paper}
We start with the general framework of DPD-based robust methods, then introduce our estimator for a general location-scale class of error distribution in the LRM \eqref{EQ:LRM} in Section~\ref{SEC:penalized_estimation}, followed by a Concave-Convex Procedure (CCCP)\citep{KimChoiOh08,WangKimLi13}-based computational algorithm. We elaborate on the robustness properties of our estimator using influence function analysis in Section~\ref{SEC:IF}. Section~\ref{SEC:Theory} presents a detailed analysis of the theoretical properties of our estimator that underlies its oracle properties. We compare the finite sample performance of our methods with existing sparse robust regression techniques in Section~\ref{sec:SimSection} using numerical experiments. Finally, we finish our paper with a discussion in Section~\ref{sec:conclusion}. Proofs of all theoretical results and additional simulations are given in the online supplementary material.

\paragraph*{Notation}
We denote vectors $\bfa \in \BR^b$ and matrices $\bfA \in \BR^{b \times b_1}$ by bold small and bold capital letters, respectively, and their elements by non-bold small letters. For a vector $\bfa$, we denote its non-zero support by $\supp( \bfa) = \{ c: a_c \neq 0 \}$, its $\ell_1$ norm by $\| \bfa \|_1 = \sum_{c=1}^b | a_c| $, $\ell_2$ norm by $\| \bfa \| = \sqrt{ \sum_{c=1}^b a_c^2}$
and $\ell_\infty$ norm by $\| \bfa \|_\infty = \max_{c=1}^b | a_c |$. For a matrix $\bfA$, the $\ell_\infty$ norm is defined as the maximum of the $\ell_1$ norm of each row, i.e., $\| \bfA \|_\infty = \max_{c=1}^b {\sum}_{c_1=1}^{b_1} | a_{cc_1} |$, and the mixed matrix norm is defined as $\|\boldsymbol{A} \|_{2,\infty} = \max_{\| \boldsymbol{v} \|=1} \| \boldsymbol{A} \boldsymbol{v} \|_\infty$. For a positive definite matrix $\bfP$, we denote its smallest and largest eigenvalue by $\Lambda_{\min} (\bfP)$ and $\Lambda_{\max} (\bfP)$, respectively.
 
\section{Model formulation}
\label{SEC:penalized_estimation}

The density power divergence (DPD) \citep{BasuEtal98} between two densities $g$ and $f$, with respect to some common dominating measure, is defined as 
\begin{eqnarray}
d_\alpha(g,f) &=& \int\left\{f^{1+\alpha} -
\left(1 + \frac{1}{\alpha}\right) f^\alpha g + \frac{1}{\alpha}
g^{1+\alpha}\right\},\alpha> 0, \notag\\
\displaystyle d_0(g,f) & = & \int g \log\left(\frac{g}{f}\right), 
\label{EQ:alpha=0}
\end{eqnarray}
where $\alpha$ is a tuning parameter controlling the trade-off between robustness and efficiency of the resulting estimator. \citet{BasuEtal98} initially proposed a robust estimator under the  parametric setup of independent and identically distributed (IID) data by minimizing the DPD measure between 
model density and observed data density. In particular, if $X_1, \ldots, X_n$ are IID observations from a population having true density $g$ which is being modeled by a parametric family of densities $\left\{f_{\boldsymbol{\theta}} : \boldsymbol{\theta} \in \Theta  \right\}$, the minimum DPD estimator (MDPDE) of $\boldsymbol{\theta}$ is obtained by minimizing the DPD measure $d_{\alpha}(\widehat{g}, f_{\boldsymbol{\theta}})$ with respect to $\boldsymbol{\theta}\in \Theta$, where  $\widehat{g}$ is an empirical estimate of $g$ based on data.

The advantage of DPD that makes it popular among divergences generating robust inference is the following. The third term in \eqref{EQ:alpha=0} above can be neglected when minimizing $d_\alpha(g,f)$ with respect to $\boldsymbol{\theta}$, and the second term can be rewritten as $ \int  f^\alpha g = \int  f^\alpha dG$ with $G$ being the true distribution function of $g$. It thus suffices to estimate $G$ using the empirical distribution function, avoiding any nonparametric estimation of $g$ and associated complications. This leads to the simplified objective function (or the loss function)
\begin{eqnarray}
H_n(\boldsymbol{\theta}) = \int f_{\boldsymbol{\theta}}^{1+\alpha} - \left(1+\frac{1}{\alpha}\right) 
\frac{1}{n}\sum_{i=1}^n f_{\boldsymbol{\theta}}^\alpha(X_i) + \frac{1}{\alpha}.
\label{EQ:DPD_obj}
\end{eqnarray}
%
Consequently, the MDPDE is a robust generalization of the maximum likelihood estimate (MLE) for $\alpha >0$, with the two coinciding as $\alpha \downarrow 0$ (this requires the additional $1/\alpha$ term in (\ref{EQ:DPD_obj})). See \cite{Basu/etc:2011} for more details and examples.

\citet{Durio/Isaia:2011} extended the concept of the minimum DPD estimation to the problem of robust estimation in the LRM (\ref{EQ:LRM}) with normal errors. 
To obtain the MDPDE of $\boldsymbol{\theta}=(\boldsymbol{\beta}^T, \sigma)^T$, they proposed to minimize the loss function
\begin{align}
L_n^\alpha (\boldsymbol{\theta}) =
L_n^\alpha (\boldsymbol{\beta}, \sigma)
= \frac{1}{(2\pi)^{\alpha/2} \sigma^\alpha\sqrt{1+\alpha} }
- \frac{1+\alpha}{\alpha} \frac{1}{n (2\pi)^{\alpha/2} \sigma^\alpha
} \sum_{i=1}^n e^{-\alpha \frac{(y_i - x_i^T\beta)^2}{2\sigma^2}}
+ \frac{1}{\alpha}.\label{EQ:DPD_loss}
\end{align}
When both $Y$ and $\boldsymbol{X}$ are random but we only assume the parametric model for the conditional distribution of $Y$ given $\boldsymbol{X}$, the loss function $L_n^\alpha (\boldsymbol{\beta}, \sigma)$ in (\ref{EQ:DPD_loss}) can be seen as an empirical estimate of the expectation (with respect to the unknown covariate distribution) of the DPD objective function (\ref{EQ:DPD_obj}) between the conditional data and model densities of $Y$ given $\boldsymbol{X}$. However, under the fixed design setup with non-stochastic covariates $\boldsymbol{x}_i$, $i=1, \ldots, n$, the only random observations $y_1, \ldots, y_n$ are independent but non-homogeneous (INH). \citet{GhoshBasu13} recently studied this problem, where they suggested minimizing the average DPD measure between the data and the model densities for each given covariate value. Interestingly, this also leads to the same loss function, i.e. (\ref{EQ:DPD_loss}). Therefore the loss function $L_n^\alpha (\bftheta) = L_n^\alpha (\boldsymbol{\beta}, \sigma)$ in (\ref{EQ:DPD_loss}), referred from here as the DPD loss function (with tuning parameter $\alpha$), can be used to obtain robust MDPDE under the LRM (\ref{EQ:LRM}) with normal errors for both stochastic and fixed design matrices.

\subsection{Non-concave penalized DPD}
\label{SEC:penalty}

In the present paper we propose a penalized version of the DPD loss function, with an appropriate class of non-concave penalties \citep{FanLi01} to simultaneously perform robust parameter estimation
and variable selection for the LRM (\ref{EQ:LRM}) in a high-dimensional setup under data contamination.
The choice of penalty functions plays an important role in characterizing the properties of the resulting estimator.

A good penalty function  $p_\lambda(|s|)$, with $s\in \mathbb{R}$ and $\lambda>0$, should have three basic properties \citep{FanLi01}:
(i) Unbiasedness to remove modeling biases (holds when $p_\lambda'(|s|)=0$ for large $s$),
(ii) Sparsity for variable selection (holds when $\min(|s|+p_\lambda'(|s|))>0$), and 
(iii) Continuity of the resulting estimator for greater stability 
(holds when $\min(|s|+p_\lambda'(|s|))$ is attained at $s=0$). 
The SCAD penalty \citep{FanLi01} satisfies all the above properties. It is defined based on two fixed parameters $a>2$ and $\lambda>0$ as:
\begin{equation}
p_\lambda(|s|) = \left\{
\begin{array}{ll}
\lambda|s| & \mbox{if }~|s|\leq \lambda,\\
\frac{2a\lambda|s|-|s|^2 - \lambda^2}{2(a-1)} & \mbox{if }~\lambda< |s|\leq a\lambda,\\
\frac{(a+1)\lambda^2}{2} & \mbox{if }~|s|> a\lambda.
\end{array}
\right.
\label{EQ:SCAD}
\end{equation}
\citet{CHZhang10} developed another important penalty function satisfying properties (i)--(ii) but not (iii), known as the minimax concave penalty (MCP):
\begin{equation}\label{EQ:MCP}
p_\lambda(|s|) = \left\{
\begin{array}{ll}
\lambda|s| - \frac{|s|^2}{2a} & \mbox{if }~|s|\leq a\lambda,\\
\frac{a\lambda^2}{2} & \mbox{if }~|s|> a\lambda, ~~~~ a>1, \lambda>0.
\end{array}
\right.
\end{equation}

Combining an appropriate penalty function $P_{n,\lambda} (\bfbeta) = \sum_j p_\lambda (|\beta_j|)$ with the DPD loss function (\ref{EQ:DPD_loss}), we now propose to minimize a general penalized objective function
\begin{align}
Q_{n,\lambda}^{\alpha}(\boldsymbol{\theta}) &=
L_n^\alpha (\boldsymbol{\beta}, \sigma) + P_{n,\lambda} (\bfbeta) \notag\\
&= \frac{1}{(2\pi)^{\alpha/2} \sigma^\alpha\sqrt{1+\alpha} }
- \frac{1+\alpha}{\alpha} \frac{1}{n (2\pi)^{\alpha/2} \sigma^\alpha
} \sum_{i=1}^n e^{-\alpha \frac{(y_i - x_i^T\beta)^2}{2\sigma^2}} + \frac{1}{\alpha}
&+ \sum_{j=1}^{p} p_\lambda (|\beta_j|).
\label{EQ:penalDPD_loss}
\end{align}
Note that as $\alpha\downarrow 0$, $L_{n}^{\alpha}(\boldsymbol{\beta}, \sigma)$ coincides (in a limiting sense) with the negative log-likelihood and hence $Q_{n,\lambda}^{\alpha}(\boldsymbol{\beta}, \sigma)$ becomes the (non-robust) nonconcave penalized negative log-likelihood, which was studied in \cite{FanLi01,KimChoiOh08,WangKimLi13} among others. Thus the proposed objective function $Q_{n,\lambda}^{\alpha}(\boldsymbol{\theta})$ generalizes the penalized likelihood objective function, with the extra advantage of robustness against data contamination at $\alpha>0$ (as shown in Section \ref{SEC:IF}). We refer to the estimator obtained by minimizing the objective function $Q_{n,\lambda}^{\alpha}(\boldsymbol{\beta}, \sigma)$ in (\ref{EQ:penalDPD_loss}) as the \textit{Minimum Non-concave Penalized DPD estimator} (MNPDPDE) with tuning parameter $\alpha$.

Based on the objective function (\ref{EQ:DPD_obj}), it is straightforward to generalize our proposal for any suitable error distribution. Under the LRM (\ref{EQ:LRM}), assume that the errors $\epsilon_1, \ldots, \epsilon_n$ are IID from a distribution with mean 0 and variance $\sigma^2>0$ having density $(1/\sigma) f(\epsilon/\sigma)$, where $f$ is any univariate density with zero mean and unit variance. 
Then, given covariate value $\boldsymbol{x}_i$, $y_i$ has the density $f_{\boldsymbol{\theta}}(y|\boldsymbol{x}_i) = (1/\sigma) f ((y - \boldsymbol{x}_i^T\boldsymbol{\beta}) / \sigma )$. Based on (\ref{EQ:DPD_obj}) the DPD loss function for this general LRM can now be defined as
\begin{align}
L_n^\alpha(\boldsymbol{\theta}) = L_n^\alpha (\boldsymbol{\beta}, \sigma)
= \frac{1}{\sigma^\alpha}M_f^{(\alpha)}
- \frac{1+\alpha}{\alpha} \frac{1}{n\sigma^\alpha} 
\sum_{i=1}^n f^\alpha\left(\frac{y_i - \boldsymbol{x}_i^T\boldsymbol{\beta}}{\sigma}\right)
+ \frac{1}{\alpha},
\label{EQ:DPD_lossGen}
\end{align}
where $M_f^{(\alpha)} = \int f(\epsilon)^{1+\alpha}d\epsilon$ is assumed to exist finitely. 
Thus, the MNPDPDE of $\boldsymbol{\theta}$ for the LRM (\ref{EQ:LRM}) with general error density $f$ can be defined 
as the minimizer of the penalized DPD loss function given by
\begin{align}
Q_{n,\lambda}^{\alpha}(\boldsymbol{\theta}) = 
\frac{1}{\sigma^\alpha}M_f^{(\alpha)}
- \frac{1+\alpha}{\alpha} \frac{1}{n\sigma^\alpha} 
\sum_{i=1}^n f^\alpha\left(\frac{y_i - \boldsymbol{x}_i^T\boldsymbol{\beta}}{\sigma}\right)
+ \frac{1}{\alpha} + \sum_{j=1}^{p} p_{\lambda}(|\beta_j|).
\label{EQ:penalDPD_lossGen}
\end{align} 
In this paper, we derive the theoretical properties of the proposed MNPDPDE for this general set-up with appropriate error density $f$. 
Although the MNPDPDE can be defined via (\ref{EQ:penalDPD_lossGen}) ideally for all $f$ with $M_f^{(\alpha)}<\infty$, throughout the present paper,
we will assume a few condition of $f$ as follows.

\vspace{1em}
\noindent {\bf (A0)} The error density $f$ is thrice continuously differentiable almost everywhere,  
	such that $M_{f, i, j}^{(\alpha)} = \int s^i u(s)^j f(s)^{1+\alpha}ds$ 	and $M_{f, i}^{(\alpha)\ast} = \int s^i u'(s) f(s)^{1+\alpha}ds$ 
	exist finitely for $i,j = 0, 1, 2$, where $u=f'/f$ and the superscript $'$ in $f$ or $u$ denote their respective first order derivatives.

\vspace{1em}
\noindent Assumption (A0) is quite general covering most common error distributions. This is a major advantage in the context of robust regression for high-dimensional data, and covers all generalized linear models with identity link function. It can also be extended to more general parametric regression problems with some additional notations and technicalities. However, for brevity, we restrict the computational algorithm and empirical illustrations to normal errors only.

\subsection{Computational Algorithm}
\label{SEC:computation}

We iteratively compute robust solutions $\hat \bftheta = (\hat \bfbeta, \hat \sigma^2)$ to the optimization problem in \eqref{EQ:penalDPD_loss} 
using an alternating iterative algorithm that minimizes the following quantities, namely
\begin{align}
&L_n^\alpha (\bfbeta, \hat \sigma) + \sum_{j=1}^p p_\lambda (| \beta_j|),~~ \mbox{ with respect to } \boldsymbol{\beta},
\label{eqn:subProblemBeta}\\
\mbox{and } ~~&L_n^\alpha (\hat \bfbeta, \sigma), ~~~~~~~~~~~~~~~~~~~~\mbox{ with respect to } \sigma. 
\label{eqn:subProblemsigma}
\end{align}

In the $k^{\Th}$ step of the iterative procedure ($k \geq 1$), given $\hat \sigma^{(k)}$ we use the Concave-Convex Procedure (CCCP) to solve \eqref{eqn:subProblemBeta}. The CCCP algorithm \citep{AnTao97,YuilleRangarajan03} can be applied when an objective function can be decomposed into a sum of convex and concave functions. \citet{KimChoiOh08} used it to calculate solutions to nonconvex penalized linear regression problems, while \cite{WangKimLi13} showed in the same context that a calibrated CCCP algorithm provides theoretical guarantees of convergence to the oracle solution, and has better finite sample performance than local linear approximation \citep{ZouLi08} or local quadratic approximation \citep{FanLi01}.

The CCCP algorithm hinges on the following decomposition and subsequent first-order approximation of the penalty function:
\begin{align*}
p_\lambda ( | \beta_j| ) = \tilde J_\lambda (| \beta_j |) + \lambda | \beta_j |
\simeq  \tilde J_\lambda' (| \beta_j^c|) \beta_j + \lambda | \beta_j |;
\end{align*}
for $j = 1, \ldots, p$, where $\tilde J_\lambda (\cdot)$ is a differentiable concave function with derivative $\tilde J_\lambda' (\cdot)$, 
and $\bfbeta^c = (\beta_1^c, \ldots, \beta_p^c)^T$ is a current solution. Taking the present iterate as current solution: $\hat \bfbeta^c = \hat \bfbeta^{(k)}$, we update $\hat \bfbeta$ by solving a convex relaxation of the original  \eqref{eqn:subProblemBeta}:
\begin{align}
\hat \bfbeta^{(k+1)} = \argmin_\bfbeta \left\{
L_n^\alpha \left(\bfbeta, \hat \sigma^{(k)} \right)
+ \sum_{j=1}^p \left[
\tilde J_\lambda' (| \beta_j^{(k)}|) \beta_j + \lambda | \beta_j | \right] \right\}.\label{eqn:subProblemBetaStep}
\end{align}

To obtain $\hat \sigma^{(k+1)}$ we consider the following derivative:
\begin{align*}
\frac{d }{d\sigma}  L_n^\alpha (\hat \bfbeta, \sigma) = 0
\Rightarrow \frac{1}{n} \sum_{i=1}^n \left[ 1 - \frac{(y_i - \bfx_i^T \hat \bfbeta)^2}{\sigma^2} \right] e^{ -\frac{ \alpha (y_i - \bfx_i^T \hat \bfbeta)^2}{\sigma^2}} =
\frac{\alpha}{ ( 1 + \alpha)^{3/2}},
\end{align*}
and update the estimate as
\begin{align}
\hat \sigma^{2 (k+1)} =
\left[ \sum_{i=1}^n w_i^{(k)} - \frac{\alpha}{ ( 1 + \alpha)^{3/2}} \right]^{-1}
\left[ \frac{1}{n} \sum_{i=1}^n w_i^{(k)} \left(y_i - \bfx_i^T \bfbeta^{(k+1)} \right)^2 \right], 
\label{eqn:subProblemSigmaStep}
\end{align}
with $w_i^{(k)} := \exp \{ -\alpha (y_i - \bfx_i^T \bfbeta^{(k)})^2/ \sigma^{2 (k)} \}$. Putting everything together, we summarize the steps of our alternating procedure in Algorithm \ref{algo:dpdncv-algo}.

\begin{Algorithm}\label{algo:dpdncv-algo}
	(Robust non-concave penalized regression using density power divergence)
	
	\noindent 1. Set $k=0$. Fix initial values $(\hat \bfbeta^{(0)}, \hat \sigma^{(0)})$, tuning parameter $\lambda$ and tolerance $\epsilon>0$.
	
	\noindent 2. Update $\hat \bfbeta$ using \eqref{eqn:subProblemBetaStep}.
	
	\noindent 3. Update $\hat \sigma$ using \eqref{eqn:subProblemSigmaStep}.
	
	\noindent 4. If $\left| L_n^\alpha \left(\hat \bfbeta^{(k+1)}, \hat \sigma^{(k+1)} \right) - L_n^\alpha \left(\hat \bfbeta^{(k)}, \hat \sigma^{(k)} \right) \right| < \epsilon$
	
	\hspace{2em} Stop.
	
	\noindent \hspace{1em} Else set $k \leftarrow k+1$, go to step 2.
\end{Algorithm}

\paragraph*{Convergence}
Note that Algorithm~\ref{algo:dpdncv-algo} above is a Majorization-Minimization (MM) algorithm. 
This is because the CCCP algorithm solves a tight convex upper bound of the objective function in (\ref{eqn:subProblemBetaStep}) \citep{KimChoiOh08}, 
and minimizing $L_n^\alpha (\hat \bfbeta, \sigma)$ obtains the global minimum with respect to $\sigma$. 
Consequently, Algorithm~\ref{algo:dpdncv-algo} possesses the descent property, i.e. each full iteration decreases the objective function:
\begin{align*}
Q_{n,\lambda}^\alpha \left( \hat \bfbeta^{(k)}, \hat \sigma^{(k)} \right)
\geq Q_{n,\lambda}^\alpha \left( \hat \bfbeta^{(k+1)}, \hat \sigma^{(k)} \right)
\geq Q_{n,\lambda}^\alpha \left( \hat \bfbeta^{(k+1)}, \hat \sigma^{(k+1)} \right).
\end{align*}
The properties of MM-algorithms ensure that Algorithm~\ref{algo:dpdncv-algo} eventually converges to a stationary point \citep{Wu83}. However the overall objective function being solved is biconvex (with $\bfbeta$ and $\sigma$ as blocks) and not convex, so convergence to a global minimum is not guaranteed. At this point, accurate enough starting values for the block parameters are important in ensuring convergence to true global parameters \citep{Kawashima17,Lange16Book}. To this end, we take solutions obtained by one of the following methods of high-dimensional robust regression- RLARS, sLTS or random sample consensus (RANSAC), as starting points. For a detailed analysis of the theoretical details of the algorithm additional technicalities will be involved, possibly in the lines of \citep{LohWainwright17,LohWainwright15}, which we shall consider in future work.

\paragraph*{Choice of $\lambda$}
To choose the regularization tuning parameter $\lambda$, we use the High-dimensional Bayesian Information Criterion (HBIC) \citep{WangKimLi13,KimKwonChoi12} 
which has demonstrably better performance compared to standard BIC under NP dimensionality with 
$\log p/n \rightarrow 0$ as $p \rightarrow \infty$ \citep{FanTang13}. We define a robust version of the HBIC:
\begin{equation}\label{eqn:hbic}
\text{HBIC} (\lambda) = \log (\hat \sigma^2) + \frac{\log \log (n) \log p}{n} \| \hat \bfbeta \|_0,
\end{equation}
and select the optimal tuning parameter $\lambda^*$ that minimizes the HBIC over a pre-determined set of values $\Lambda_n$: $\lambda^* = \argmin_{\lambda \in \Lambda_n} \text{HBIC} (\lambda)$.
\section{Influence Function analysis}
\label{SEC:IF}

In this section, we study the robustness properties of the MNPDPDE through influence function analysis. The influence function (IF: \cite{Hampel:1968,Hampel:1974}) is a classical tool of measuring the asymptotic local robustness of any estimator, and has been applied to several non-penalized regression estimators, for example \cite{Huber:1983, Hampel/etc:1986}. However, 
a fully rigorous definition of the influence function for the class of penalized estimators has been proposed only recently \citep{Avella-Medina:2017}.

To study the IF of the proposed MNPDPDE with tuning parameter $\alpha$, we need to define its corresponding statistical functional $\boldsymbol{T}_{\alpha}(G)$, at the true joint distribution $G$ of $(Y, \boldsymbol{X})$,
as the in-probability limit of the MNPDPDE. In other words, the MNPDPDE should equal $\boldsymbol{T}_\alpha(G_n)$,
where $G_n$ is the empirical estimate of $G$ having mass $1/n$ at the $n$ data-points $(y_i, \boldsymbol{x}_i); i=1, \ldots,n$. Based on the objective function (\ref{EQ:penalDPD_lossGen}) with a general error distribution satisfying assumption (A0), we define this functional $\boldsymbol{T}_{\alpha}(G)$ as the minimizer of
\begin{eqnarray}
Q_{\lambda}^{\alpha}(\boldsymbol{\theta}) 
= L_{\alpha}(\boldsymbol{\theta}) + \sum_{j=1}^{p} p_{\lambda}(|\beta_j|),
\label{EQ:penalDPD_lossFunc}
\end{eqnarray}
with respect to $\boldsymbol{\theta} = (\boldsymbol{\beta}^T, \sigma)^T$, where the penalty $p_\lambda(\cdot)$ is now assumed to be independent of $n$ 
(or, can be taken as the limiting form of $n$-dependent penalties) and
%
$L_{\alpha}(\boldsymbol{\theta}) = \int L_{\alpha}^\ast((y, \boldsymbol{x});\boldsymbol{\theta}) dG(y,\boldsymbol{x})$,
where 
\begin{align*}
L_{\alpha}^\ast((y, \boldsymbol{x});\boldsymbol{\theta}) = 
\frac{1}{\sigma^\alpha}M_f^{(\alpha)} - \frac{1+\alpha}{\alpha} \frac{1}{\sigma^\alpha} 
f^\alpha\left(\frac{y - \boldsymbol{x}^T\boldsymbol{\beta}}{\sigma}\right) + \frac{1}{\alpha}.
\end{align*} 
Note that substituting $G=G_n$, $L_{\alpha}(\boldsymbol{\theta})$ coincides with the DPD loss $L_{n}^\alpha (\boldsymbol{\theta})$ in (\ref{EQ:DPD_lossGen}).
Also if $g^{\ast}(y|\boldsymbol{x})$ denotes the true conditional density of $Y$ given $\boldsymbol{X}=\boldsymbol{x}$, the MNPDPDE functional is equivalently the minimizer of 
$$
E_{\boldsymbol{X}}\left[d_{\alpha}\left(f_{\boldsymbol{\theta}}(\cdot|\boldsymbol{x}), g^{\ast}(\cdot|\boldsymbol{x})\right)\right]
+ \sum_{j=1}^{p} p_{\lambda}(|\beta_j|)
$$
with respect to $\boldsymbol{\theta}$. This justify the use of our specific loss function in defining the MDPDPDE. Further, $\boldsymbol{T}_{\alpha}(G)$ belongs to the class of M-estimator considered in \cite{Avella-Medina:2017}, with their $L(Z, \theta)$ function coinciding with our $L_{\alpha}^\ast((y, \boldsymbol{x});\boldsymbol{\theta})$.

\subsection{Twice differentiable penalties}
\label{SEC:IF1}

If the penalty function $\widetilde{p}_{\lambda}(s) = p_{\lambda}(|s|)$ is twice differentiable in $s$ (e.g., $L_2$ penalty), we can indeed apply the classical definition of the IF \citep{Hampel/etc:1986}.
Consider a contaminated version of  the true distribution $G$ given by $G_{\epsilon} = (1-\epsilon)G + \epsilon\wedge_{(y_t, \boldsymbol{x}_t)}$ where $\epsilon$ is the contamination proportion and $\wedge_{(y_t, \boldsymbol{x}_t)}$ is the degenerate distribution at the contamination point ${(y_t, \boldsymbol{x}_t)}$.
Then, the IF of the MNPDPDE functional $\boldsymbol{T}_{\alpha}$ at $G$ is defined as the limiting standardized bias due to infinitesimal contamination:
\begin{align}
\mathcal{IF}({(y_t, \boldsymbol{x}_t)}, \boldsymbol{T}_\alpha, G)
= \lim_{\epsilon\rightarrow 0}\frac{\boldsymbol{T}_\alpha(G_\epsilon) - \boldsymbol{T}_\alpha(G)}{\epsilon}
= \left.\frac{\partial}{\partial\epsilon}\boldsymbol{T}_\alpha(G_\epsilon)\right|_{\epsilon=0},
\label{EQ:IF_Def}
\end{align}
mathematically characterized by a particular Gateaux derivative of the functional $\boldsymbol{T}_{\alpha}$ \citep{Hampel:1968, Hampel:1974}. To derive this IF with twice differentiable penalty $\widetilde{p}_{\lambda}$ and differentiable error density $f$, we start with the estimating equations of $\boldsymbol{T}_{\alpha}(G)$
as given by 
$\nabla Q_{\lambda}^{\alpha}(\boldsymbol{\theta}) =\boldsymbol{0}_{p+1}$, where $\nabla$ denotes the derivative with respect to $\boldsymbol{\theta}$. 
Let us also define 
\begin{align}
\widetilde{\boldsymbol{P}}_{\lambda}^{\ast}(\boldsymbol{\beta}) 
&= \left(\widetilde{p}_{\lambda}'({\beta}_1), \ldots, \widetilde{p}_{\lambda}'({\beta}_p)\right)^T,\notag\\
\widetilde{\boldsymbol{P}}_{\lambda}^{\ast\ast}(\boldsymbol{\beta}) 
&= \diag \left\{\widetilde{p}_{\lambda}''({\beta}_1), \ldots, \widetilde{p}_{\lambda}''({\beta}_p)\right\}, \notag\\
\psi_{\alpha}((y, \boldsymbol{x});\boldsymbol{\theta}) &=
\nabla L_{\alpha}^\ast((y, \boldsymbol{x});\boldsymbol{\theta})
= \frac{(1+\alpha)}{\sigma^{\alpha+1}}
\begin{bmatrix}
\begin{array}{c}
\psi_{1, \alpha}\left(\frac{y - \boldsymbol{x}^T\boldsymbol{\beta}}{\sigma}\right)\boldsymbol{x}
\\
\psi_{2,\alpha}\left(\frac{y - \boldsymbol{x}^T\boldsymbol{\beta}}{\sigma}\right)
\end{array}
\end{bmatrix},
\end{align}
where, with the notation from assumption (A0), 
\begin{align*}
\psi_{1, \alpha}(s) &= u(s)f^{\alpha}(s),\\
\psi_{2, \alpha}(s) &= \{s u(s)+1\} f^{\alpha}(s) -\frac{\alpha}{{\alpha+1}} M_f^{(\alpha)}.
\end{align*}
Then, assuming the existence of relevant integrals, 
the estimating equations of $\boldsymbol{T}_{\alpha}(G)$ are given by 
\begin{eqnarray}
\begin{array}{lll}
\frac{(1+\alpha)}{\sigma^{\alpha+1}}
\int \psi_{1, \alpha}\left(\frac{y - \boldsymbol{x}^T\boldsymbol{\beta}}{\sigma}\right)\boldsymbol{x}dG(y, \boldsymbol{x})
& + \widetilde{\boldsymbol{P}}_{\lambda}^{\ast}(\boldsymbol{\beta}) & = \boldsymbol{0}_{p},\\
\frac{(1+\alpha)}{\sigma^{\alpha+1}}
\int \psi_{2, \alpha}\left(\frac{y - \boldsymbol{x}^T\boldsymbol{\beta}}{\sigma}\right)dG(y, \boldsymbol{x})
&  & = 0.
\end{array} 
\end{eqnarray}
By standard calculations (\cite{Hampel/etc:1986} or Lemma 1 in \cite{Avella-Medina:2017}), 
we can obtain the influence function of the MNPDPDE functional $\boldsymbol{T}_{\alpha}$ with a twice differentiable penalty function.

\begin{Theorem}\label{THM:IF_MNPDPDE}
Suppose that the penalty function  $\widetilde{p}_{\lambda}(s)$ is twice differentiable in $s$.
Consider the LRM (\ref{EQ:LRM}) with a general error density $f$ satisfying assumption (A0) for which 
$\int \psi_{\alpha}((y, \boldsymbol{x});\boldsymbol{\theta})dG(y, \boldsymbol{x})$ and 
$\boldsymbol{J}_{\alpha}(G;\boldsymbol{\theta}) = \int \nabla \psi_{\alpha}((y, \boldsymbol{x});\boldsymbol{\theta}) dG(y, \boldsymbol{x})$ exist finitely, 
and the matrix
$$
\boldsymbol{J}_{\alpha}^{\ast}(G;\boldsymbol{\theta}) :=
\left[\boldsymbol{J}_{\alpha}(G;\boldsymbol{\theta}) 
+ \diag\{\widetilde{\boldsymbol{P}}_{\lambda}^{\ast\ast}(\boldsymbol{\beta}), 0\}\right]
$$
is invertible at $\boldsymbol{\theta}=\boldsymbol{\theta}^g=\boldsymbol{T}_{\alpha}(G)=(\boldsymbol{\beta}^g, \sigma^g)^T$. 
Then, whenever it exists, the influence function of the MNPDPDE functional $\boldsymbol{T}_{\alpha}$ at $G$ is given by
\begin{align}
\mathcal{IF}({(y_t, \boldsymbol{x}_t)}, \boldsymbol{T}_\alpha, G) &=
- \boldsymbol{J}_{\alpha}^{\ast}(G;\boldsymbol{\theta}^g)^{-1}
\begin{bmatrix}
\begin{array}{c}
\frac{(1+\alpha)}{(\sigma^g)^{\alpha+1}}\psi_{1, \alpha}\left(\frac{y_t - \boldsymbol{x}_t^T\boldsymbol{\beta}^g}{\sigma^g}\right)\boldsymbol{x}_t 
+ \widetilde{\boldsymbol{P}}_{\lambda}^{\ast}(\boldsymbol{\beta}^g)
\\
\frac{(1+\alpha)}{(\sigma^g)^{\alpha+1}}\psi_{2,\alpha}\left(\frac{y_t - \boldsymbol{x}_t^T\boldsymbol{\beta}^g}{\sigma^g}\right)
\end{array}
\end{bmatrix}.
\label{EQ:IF_MNPDPDE}
\end{align}
If the parameter space $\Theta$ is compact, the above IF exists for all contamination points $(y_t, \boldsymbol{x}_t)$.
\end{Theorem}

If $\boldsymbol{T}_{\alpha}^\beta$ and $T_{\alpha}^\sigma$ denote the MNPDPDE functional 
for the parameters $\boldsymbol{\beta}$ and $\sigma$, respectively, so that 
$\boldsymbol{T}_{\alpha}= (\boldsymbol{T}_{\alpha}^\beta, T_{\alpha}^\sigma)^T$,
their individual IF can be obtained separately from (\ref{EQ:IF_MNPDPDE})
based on the forms of $\boldsymbol{J}_{\alpha}^{\ast}(G;\boldsymbol{\theta}^g)$
or equivalently of
\begin{align*}
\boldsymbol{J}_{\alpha}(G;\boldsymbol{\theta}^g) &=
- \frac{(1+\alpha)}{\sigma^{\alpha+2}} E_G\begin{bmatrix}
\begin{array}{cc}
J_{11, \alpha}\left(\frac{y - \boldsymbol{x}^T\boldsymbol{\beta}}{\sigma}\right)\boldsymbol{x}\boldsymbol{x}^T &
J_{12, \alpha}\left(\frac{y - \boldsymbol{x}^T\boldsymbol{\beta}}{\sigma}\right)\boldsymbol{x}\\
J_{12, \alpha}\left(\frac{y - \boldsymbol{x}^T\boldsymbol{\beta}}{\sigma}\right)\boldsymbol{x}^T &
J_{22, \alpha}\left(\frac{y - \boldsymbol{x}^T\boldsymbol{\beta}}{\sigma}\right)
\end{array}
\end{bmatrix},
\end{align*}
where
\begin{align*}
J_{11, \alpha}(s) &=  \{\alpha u^2(s) + u'(s)\}f^{\alpha}(s),\\
J_{12, \alpha}(s) &=  \{(1+\alpha) u(s) + \alpha s u^2(s) +su'(s)\}f^{\alpha}(s), \\
J_{22, \alpha}(s) &= -\alpha M_f^{(\alpha)} + \{
(1+\alpha) (1 + 2s u(s)) + \alpha s^2 u^2(s) + s^2 u'(s) \} f^{\alpha}(s).
\end{align*}
Whenever the conditional density of $Y$ given $\boldsymbol{X}$ belongs to the LRM (\ref{EQ:LRM}) with general error density $f$ satisfying assumption (A0), i.e., $g^{\ast} = f_{\boldsymbol{\theta}_0}$ for some $\boldsymbol{\theta}_0 = (\boldsymbol{\beta}_0, \sigma_0)^T$, 
then $\boldsymbol{\theta}^g = \boldsymbol{\theta}_0$ and  
we can further simplify the above matrix to have the form 
\begin{align*}
\boldsymbol{J}_{\alpha}^{(0)}(\boldsymbol{\theta}_0)
:= \boldsymbol{J}_{\alpha}(F_{\boldsymbol{\theta}_0};\boldsymbol{\theta}_0)
= - \frac{(1+\alpha)}{\sigma_0^{\alpha+2}}
\begin{bmatrix}
\begin{array}{cc}
J_{11, \alpha}^{(0)}E\left(\boldsymbol{x}\boldsymbol{x}^T\right) &
J_{12, \alpha}^{(0)}E\left(\boldsymbol{x}\right)\\
&\\
J_{12, \alpha}^{(0)}E\left(\boldsymbol{x}\right)^{T} &
J_{22, \alpha}^{(0)}
\end{array}
\end{bmatrix},
\end{align*}
where, under the notation of assumption (A0), we define
\begin{align*}
J_{11, \alpha}^{(0)} &=  \alpha M_{f,0,2}^{(\alpha)} + M_{f,0}^{(\alpha)\ast},\\
J_{12, \alpha}^{(0)} &= (1+\alpha) M_{f,0,1}^{(\alpha)}  + \alpha M_{f,1,2}^{(\alpha)} + M_{f,1}^{(\alpha)\ast},\\
J_{22, \alpha}^{(0)} &= M_f^{(\alpha)} +
2(1+\alpha) M_{f,1,1}^{(\alpha)} + \alpha M_{f,2,2}^{(\alpha)} +
+ M_{f,2}^{(\alpha)\ast}.
\end{align*}
Additionally, if the error density $f$  satisfies $J_{12, \alpha}^{(0)}=0$ or $E(\boldsymbol{X}) =\boldsymbol{0}_p$, it is straightforward to separate out the IFs of $\boldsymbol{T}_{\alpha}^\beta$ and $T_{\alpha}^\sigma$ at the model (\ref{EQ:LRM}) as given by
\begin{align*}
\mathcal{IF}({(y_t, \boldsymbol{x}_t)}, \boldsymbol{T}_\alpha^\beta,  F_{\boldsymbol{\theta}_0})
&= \left[\frac{(1+\alpha)}{\sigma_0^{\alpha+2}}
J_{11, \alpha}^{(0)}E\left(\boldsymbol{x}\boldsymbol{x}^T\right) - 
\widetilde{\boldsymbol{P}}_{\lambda}^{\ast\ast}(\boldsymbol{\beta}_0)\right]^{-1} \\
& \quad \left[\frac{(1+\alpha)}{\sigma_0^{\alpha+1}}\psi_{1, \alpha}\left(\frac{y_t - \boldsymbol{x}_t^T\boldsymbol{\beta}_0}{\sigma_0}\right)\boldsymbol{x}_t 
+ \widetilde{\boldsymbol{P}}_{\lambda}^{\ast}(\boldsymbol{\beta}_0)\right],\\
\mathcal{IF}({(y_t, \boldsymbol{x}_t)}, T_\alpha^\sigma, F_{\boldsymbol{\theta}_0})
&= \frac{\sigma_0}{J_{22, \alpha}^{(0)}}\psi_{2,\alpha}\left(\frac{y_t - \boldsymbol{x}_t^T\boldsymbol{\beta}_0}{\sigma_0}\right).
\end{align*}
An example where $J_{12, \alpha}^{(0)}=0$ is $f=\phi$, the standard normal error density. 
In other cases, we need to use the formula for the inverse of a block matrix to get 
the individual IFs although the form will be more complicated.

\begin{Remark}\label{REM:1}
Note that Theorem \ref{THM:IF_MNPDPDE} can only be applied to twice differentiable penalties,
which do not often satisfy all three basic optimality properties (i)--(iii) described earlier in Section \ref{SEC:penalty}. For example, $\ell_q$ penalty is twice differentiable for any $q\geq 2$ (and hence Theorem \ref{THM:IF_MNPDPDE} holds) but it does not satisfy the sparsity condition (ii). 
However, given twice differentiability, the invertiblity the matrix $\boldsymbol{J}_{\alpha}^{\ast}(G;\boldsymbol{\theta})$ no longer depends on the penalty function; rather it depends mostly on the error density $f$ and holds for most common choices of error distributions. 
\end{Remark}

\subsection{Influence Functions for non-concave penalties}

As noted in Remark \ref{REM:1}, a major problem with Theorem \ref{THM:IF_MNPDPDE} is that it 
cannot be used to study the robustness of the MNPDPDEs with sparsity-inducing penalties like $\ell_1$, SCAD or MCP as they are not twice differentiable. 
To tackle this problem, we define their IFs by the limiting form of IFs with differentiable penalty functions, as proposed by \cite{Avella-Medina:2017}. 
Consider a sequence of continuous and infinitely differentiable penalty functions $p_{m, \lambda}(s)$ converging to $p_{\lambda}(|s|)$ 
in the Sobolev space (as $m\rightarrow\infty$), and denote the MNPDPDE functional corresponding to the penalty $p_{m, \lambda}(s)$ by 
$\boldsymbol{T}_{m, \alpha}(G)$ (which can be obtained from Theorem \ref{THM:IF_MNPDPDE}). 
Then, we define the IF of the MNPDPDE $\boldsymbol{T}_{\alpha}$ corresponding to the 
actual (possibly non-differentiable) penalty function  $p_{\lambda}(|s|)$ as given by
\begin{eqnarray}
\mathcal{IF}({(y_t, \boldsymbol{x}_t)}, \boldsymbol{T}_\alpha, G)
= \lim_{m\rightarrow \infty}\mathcal{IF}({(y_t, \boldsymbol{x}_t)}, \boldsymbol{T}_{m, \alpha}, G).
\label{EQ:IF_DefLim}
\end{eqnarray}
Whenever $\boldsymbol{T}_{m, \alpha}$ satisfies the assumptions of Theorem \ref{THM:IF_MNPDPDE}, 
$\Theta$ is compact and the quantities $\int \psi_{\alpha}((y, \boldsymbol{x});\boldsymbol{\theta})dG(y, \boldsymbol{x})$ and 
$\boldsymbol{J}_{\alpha}(G;\boldsymbol{\theta})$ are continuous in $\boldsymbol{\theta}$, 
then the limiting IF in \eqref{EQ:IF_DefLim} exists and is independent of the choice of the penalty sequence $p_{m, \lambda}(s)$ 
\cite[][Proposition 1]{Avella-Medina:2017}. 
Moreover, this limiting influence function can be rigorously seen as the distributional derivative of $\boldsymbol{T}_{\alpha}(G_{\epsilon})$ 
with respect to $\epsilon$ at $\epsilon=0$  \cite[][Proposition 6]{Avella-Medina:2017}.

Based on the above definition we now derive the IF of our MNPDPDE for general (possibly non-differentiable) penalty functions $\tilde p_\lambda$ that have differentiable $p_{\lambda}$, e.g. most common penalties including SCAD, $\ell_1$ and MCP. To account for the possible non-differentiability at $s=0$, we separately consider the cases where true regression coefficient is zero.

\begin{Theorem}\label{THM:IFlim_MNPDPDE}
Consider the above set-up with the general penalty function of the form $p_{\lambda}(|s|)$, where $p_{\lambda}(s)$ is twice differentiable in $s$,
and a general error density $f$ satisfying assumption (A0) for which 
$\int \psi_{\alpha}((y, \boldsymbol{x});\boldsymbol{\theta})dG(y, \boldsymbol{x})$ and $\boldsymbol{J}_{\alpha}(G;\boldsymbol{\theta})$ exist finitely. 
Further, for any $\boldsymbol{v}=(v_1, \ldots, v_q)^T$ with all non-zero elements, 
we define the quantities
$\boldsymbol{P}_{\lambda}^{\ast}(\boldsymbol{v}) = \left( p_{\lambda}'(|v_1|) \sign(v_1), \ldots, p_{\lambda}'(|v_q|)\sign (v_q)\right)^T$
 and $\boldsymbol{P}_{\lambda}^{\ast\ast}(\boldsymbol{v}) = \diag \{p_{\lambda}''(|v_1|), \ldots, p_{\lambda}''(|v_q|)\}$.
	
\vspace{1em}\noindent
1. Suppose that the true value of the regression coefficient $\boldsymbol{\beta}^g=\boldsymbol{T}_{\alpha}^\beta(G)$
		has all components non-zero (which forces $p\leq n$). Then, whenever the associated quantities exists, 
		the influence function of the MNPDPDE functional $\boldsymbol{T}_{\alpha}$ at $G$ 
		is given by (\ref{EQ:IF_MNPDPDE}), with $\boldsymbol{P}_{\lambda}^{\ast}(\boldsymbol{\beta})$ and $\boldsymbol{P}_{\lambda}^{\ast\ast}(\boldsymbol{\beta})$ in place of $\widetilde{\boldsymbol{P}}_{\lambda}^{\ast}(\boldsymbol{\beta})$ and $\widetilde{\boldsymbol{P}}_{\lambda}^{\ast \ast}(\boldsymbol{\beta})$, respectively.		
		
\vspace{1em}\noindent
2. Suppose that the true value of $\boldsymbol{\beta}^g$ is {\it sparse}  with only $s (<n)$ non-zero components (which allows $p>>n$). 	Without loss of generality, assume $\boldsymbol{\beta}^g=(\boldsymbol{\beta}_1^{gT}, \boldsymbol{0}_{p-s}^T)^T$, where $\boldsymbol{\beta}_1^{g}$ contains all and only $s$-non-zero elements of $\boldsymbol{\beta}^{g}$. Denote $\boldsymbol{\theta}^g= \boldsymbol{T}_{\alpha}(G) = (\boldsymbol{\beta}_1^{gT}, \boldsymbol{0}_{p-s}^T, \sigma^g)^T$, and the corresponding partition of the MNPDPDE functional $\boldsymbol{T}_{\alpha}(G)$ by $(\boldsymbol{T}_{1,\alpha}^\beta(G)^T, \boldsymbol{T}_{2,\alpha}^\beta(G)^T, {T}_{\alpha}^\sigma(G))^T$. Then, whenever the associated quantities exists, the influence function of $\boldsymbol{T}_{2,\alpha}^\beta$ is identically zero at $G$ and that of  $(\boldsymbol{T}_{1,\alpha}^\beta, {T}_{\alpha}^\sigma)$ at $G$ is given by
\begin{align}
\mathcal{IF}({(y_t, \boldsymbol{x}_t)}, (\boldsymbol{T}_{1,\alpha}^\beta, {T}_{\alpha}^\sigma), G)	=
- \bfS_{\alpha}(G;\boldsymbol{\theta}^g)^{-1}
\begin{bmatrix}
\begin{array}{c}
\frac{(1+\alpha)}{(\sigma^g)^{\alpha+1}}\psi_{1, \alpha}\left(\frac{y_t - \boldsymbol{x}_t^T\boldsymbol{\beta}^g}{\sigma^g}\right)\boldsymbol{x}_{1,t} 
+ {\boldsymbol{P}}_{\lambda}^{\ast}(\boldsymbol{\beta}_1^g) \\
\frac{(1+\alpha)}{(\sigma^g)^{\alpha+1}}\psi_{2,\alpha}\left(\frac{y_t - \boldsymbol{x}_t^T\boldsymbol{\beta}^g}{\sigma^g}\right)
\end{array}
\end{bmatrix},
\label{EQ:IF_MNPDPDE_Sparse}
\end{align}
where $\bfS_{\alpha}(G;\boldsymbol{\theta}) \in \BR^{(s+1)\times (s+1)}$ is defined as
\begin{align*}
\bfS_{\alpha}(G;\boldsymbol{\theta}) =
- \frac{(1+\alpha)}{\sigma^{\alpha+2}}
E_G\begin{bmatrix}
\begin{array}{cc}
J_{11, \alpha}\left(\frac{y - \boldsymbol{x}^T\boldsymbol{\beta}}{\sigma}\right)\boldsymbol{x}_1\boldsymbol{x}_1^T &
J_{12, \alpha}\left(\frac{y - \boldsymbol{x}^T\boldsymbol{\beta}}{\sigma}\right)\boldsymbol{x}_1\\
J_{12, \alpha}\left(\frac{y - \boldsymbol{x}^T\boldsymbol{\beta}}{\sigma}\right)\boldsymbol{x}_1^T &
J_{22, \alpha}\left(\frac{y - \boldsymbol{x}^T\boldsymbol{\beta}}{\sigma}\right)
\end{array}
\end{bmatrix}
+ \begin{bmatrix}
\begin{array}{cc}
\boldsymbol{P}_{\lambda}^{\ast\ast}(\boldsymbol{\beta}_1) &
\boldsymbol{0}_s\\
\boldsymbol{0}_s^T & 0
\end{array}
\end{bmatrix},
\end{align*}
with $\boldsymbol{x}_1$, $\boldsymbol{x}_{1,t}$ and $\boldsymbol{\beta}_1$ being the $s$-vectors of the first $s$ elements of the $p$-vectors $\boldsymbol{x}$, $\boldsymbol{x}_t$ and $\boldsymbol{\beta}$, respectively.

\vspace{1em}\noindent
3. If the parameter space $\Theta$ is compact, then the IF exists in both cases for all $(y_t, \boldsymbol{x}_t)$.
\end{Theorem}

One can further simplify the IFs in Theorem \ref{THM:IFlim_MNPDPDE} at the conditional model distribution given by LRM (\ref{EQ:LRM}) and separately write down the IFs of $\boldsymbol{T}_{1,\alpha}^\beta$ and ${T}_{\alpha}^\sigma$ following the discussions after Theorem \ref{THM:IF_MNPDPDE}. These IFs depend on the non-differentiable penalty functions $p_{\lambda}(|s|)$ through the quantities $\boldsymbol{P}_{\lambda}^{\ast}(\boldsymbol{\beta})$ and $\boldsymbol{P}_{\lambda}^{\ast\ast}(\boldsymbol{\beta})$. Table \ref{TAB:Penalties} gives their explicit forms for the three common and useful penalties.

\begin{table*}[t]
\centering
\caption{$\boldsymbol{P}_{\lambda}^{\ast}(\boldsymbol{\beta})$ and $\boldsymbol{P}_{\lambda}^{\ast\ast}(\boldsymbol{\beta})$ for common non-differentiable penalties of the form $p_{\lambda}(|s|)$}
\resizebox{\textwidth}{!}{
\begin{tabular}{l|l|l|l} \hline
Penalty	& $p_{\lambda}(u)$	&	$j^{\Th}$ element of $\boldsymbol{P}_{\lambda}^{\ast}(\boldsymbol{\beta})$ 
& $j^{\Th}$ diagonal of $\boldsymbol{P}_{\lambda}^{\ast\ast}(\boldsymbol{\beta})$	\\	 \hline \hline
$\ell_1$ & $\lambda u$  & $\lambda \sign(\beta_j)$ & 0
\\\hline
SCAD, \eqref{EQ:SCAD} & 
$\left\{
\begin{array}{ll}
\lambda u & \mbox{if }~ u\leq \lambda,\\
\frac{2a\lambda u - u^2 - \lambda^2}{2(a-1)} & \mbox{if }~\lambda < u \leq a\lambda,\\
\frac{(a+1)\lambda^2}{2} & \mbox{if }~ u > a\lambda.
\end{array}
\right.$ 
&
$\left\{
\begin{array}{ll}
\lambda~sign(\beta_j) & \mbox{if }~|\beta_j|\leq \lambda,\\
\frac{a\lambda - |\beta_j|}{(a-1)}~ \sign(\beta_j) & \mbox{if }~\lambda< |\beta_j|\leq a\lambda,\\
0 & \mbox{if }~|\beta_j|> a\lambda.
\end{array}
\right.$ 
&
$-\frac{1}{(a-1)} I\left(\lambda< |\beta_j|\leq a\lambda\right)$ 
\\\hline
MCP, \eqref{EQ:MCP} & 
$\left\{
\begin{array}{ll}
	\lambda u - \frac{u^2}{2a} & \mbox{if }~u\leq a\lambda,\\
	\frac{a\lambda^2}{2} & \mbox{if }~ u> a\lambda, ~~~~ a>1, \lambda>0.
\end{array}
\right.$ 
&
$\left\{
\begin{array}{ll}
\frac{a\lambda - |\beta_j|}{a}~ \sign(\beta_j) & \mbox{if }~|\beta_j|\leq a \lambda,\\
0 & \mbox{if }~|\beta_j|> a\lambda.
\end{array}
\right.$ 
&
$-\frac{1}{a} I\left( |\beta_j|\leq a\lambda\right)$ 
\\\hline
\end{tabular}}
\label{TAB:Penalties}
\end{table*}


The IFs of different MNDPDEs obtained in Theorem \ref{THM:IF_MNPDPDE} or \ref{THM:IFlim_MNPDPDE} depend on the contamination points $(y_t, \boldsymbol{x}_t)$ only through the functions $\psi_{1, \alpha}$ and $\psi_{2, \alpha}$.
For most common differentiable error densities $f$, it can easily be verified that they are bounded for any $\alpha>0$ but unbounded at $\alpha=0$. Thus, the proposed MNPDPDE is robust for any $\alpha>0$. This also justifies the well-known non-robust nature of the penalized MLE (at $\alpha=0$). This boundedness of the IF is clearly independent of the penalty functions, although in the sparse cases described in part 2 of Theorem \ref{THM:IFlim_MNPDPDE} it is enough only to examine the boundedness of $\psi_{1, \alpha} ((y_t - \boldsymbol{x}_t^T \boldsymbol{\beta})/\sigma ) \boldsymbol{x}_{1,t}$ to study the robustness of the MNPDPDE of the non-zero regression coefficient, which can only be achieved using a suitable sparse penalty function.

\begin{Remark}
At the special case $\alpha\downarrow 0$, the MNPDPDE coincides with the non-concave penalized maximum likelihood estimate \citep{FanLi01,FanLv11}. Thus, at $\alpha=0$, Theorems \ref{THM:IF_MNPDPDE} and \ref{THM:IFlim_MNPDPDE} additionally yield their influence function as well, which was not studied by \cite{FanLv11}.
\end{Remark}

\subsection{Influence Functions for normal errors}
As an illustration, we explicitly compute the IFs of the MNPDPDE for the case of normal errors under the sparse assumption in part 2 of Theorem \ref{THM:IFlim_MNPDPDE}. In particular, at the model distribution with $f=\phi$, we have $J_{12, \alpha}^{(0)}=0$. Hence we can separately write down the IF of the MNDPDE functionals $\boldsymbol{T}_{1,\alpha}^\beta$ and ${T}_{\alpha}^\sigma$, corresponding to the non-zero regression coefficients $\boldsymbol{\beta}_1$ and $\sigma$, respectively, from Equation (\ref{EQ:IF_MNPDPDE_Sparse}).
\begin{align*}
\mathcal{IF}({(y_t, \boldsymbol{x}_t)}, \boldsymbol{T}_{1, \alpha}^\beta,  F_{\boldsymbol{\theta}})
&= \left[\frac{1}{\sigma^{2\alpha+4} (2\pi)^{\alpha/2} (1+\alpha)^{1/2}}
E\left(\boldsymbol{x}_1\boldsymbol{x}_1^T\right) + 
{\boldsymbol{P}}_{11,\lambda}^{\ast\ast} (\boldsymbol{\beta})\right]^{-1} \\
& \left[ -\frac{1+\alpha}{\sigma^{2\alpha+3} (2\pi)^{\alpha/ 2}} \left(
{y_t - \boldsymbol{x}_t^T\boldsymbol{\beta}}\right)
e^{-\frac{\alpha\left(y_t - \boldsymbol{x}_t^T\boldsymbol{\beta}\right)^2}{2\sigma^2}} \boldsymbol{x}_{1,t} + {\boldsymbol{P}}_{1,\lambda}^{\ast}(\boldsymbol{\beta})\right], \\
\mathcal{IF}({(y_t, \boldsymbol{x}_t)}, \boldsymbol{T}_\alpha^\sigma, F_{\boldsymbol{\theta}}) 
&= \frac{\sigma(1+\alpha)^{5/2}}{(2+\alpha^2)}
\left[\left\{1 - \left(\frac{y_t - \boldsymbol{x}_t^T\boldsymbol{\beta}}{\sigma}\right)^2\right\}
e^{-\frac{\alpha\left(y_t - \boldsymbol{x}_t^T\boldsymbol{\beta}\right)^2}{2\sigma^2}}
 - \frac{\alpha}{(1+\alpha)^{1/2}}\right].
\end{align*}

\noindent
The IFs depend on the contamination in $Y$ only through the residual $r_t := y_t - \boldsymbol{x}_t^T\boldsymbol{\beta} $, 
and their boundedness over contamination points $(y_t, \boldsymbol{x}_t)$ depends on the terms 
$r_t \exp[ - \alpha r_t^2/(2 \sigma^2)] \bfx_{i,t}$ and $(1 - r_t^2/\sigma^2 ) \exp[ - \alpha r_t^2/(2 \sigma^2)]$. 
Both are bounded in the contaminated residual $r_t$ or contamination in covariate space at $\boldsymbol{x}_t$ only at $\alpha>0$, 
illustrating the claimed robustness of the proposed MNPDPDE with $\alpha>0$.

\begin{figure*}[t]
\centering
\includegraphics[width=.3\textwidth]{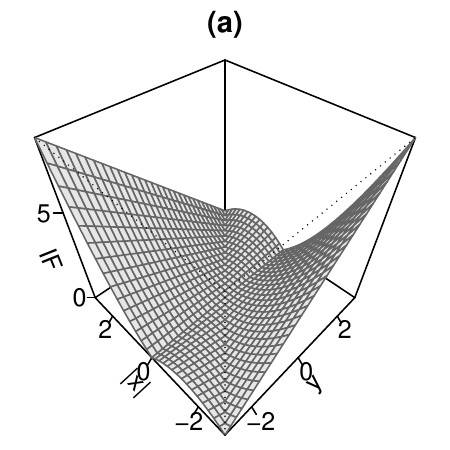}
\includegraphics[width=.3\textwidth]{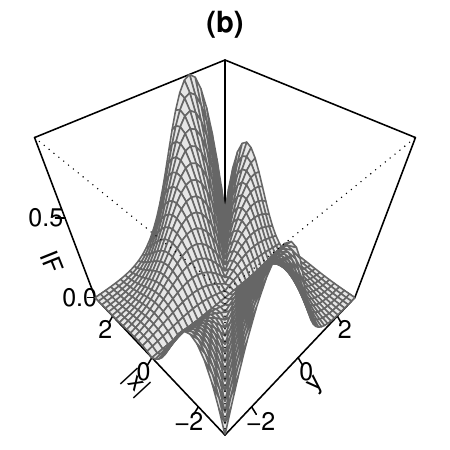}\\
\includegraphics[width=.3\textwidth]{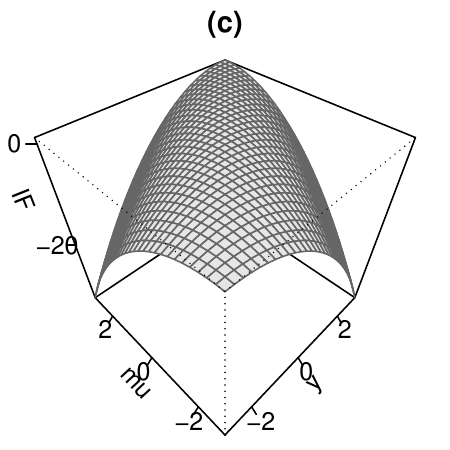}
\includegraphics[width=.3\textwidth]{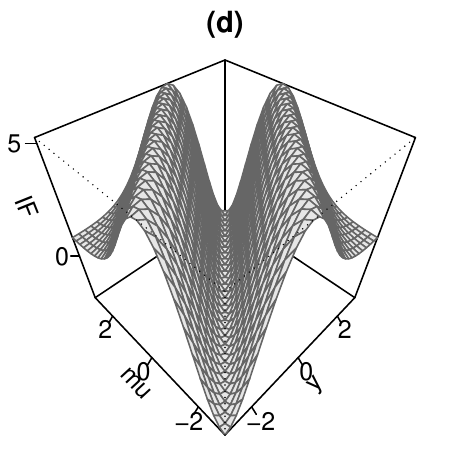}
\caption{Influence function plots for $\bfbeta $ (panels a and b, $(y_t,\| \bfx_{1t} \|_1)$ on the $(x,y)$ axes, and $\ell_2$ norms of IFs are plotted) and $\sigma$ (panels c and d, $(y_t, \bfx_t^T \bfbeta)$ on the axes). We assume $\bfx_{1t}$ is drawn from $\cN_5({\bf 0}, \bfI)$, and $\bfbeta_1 = (1,1,1,1,1)^T, \sigma=1$. Panels a and c are for $\alpha=0$, while b and d are for $\alpha=0.5$.
}
\label{fig:IFplot}
\end{figure*}

The supremum of the norm of its IFs over the contamination points measures the local robustness of the MNPDPDE functional $\boldsymbol{T}_{\alpha}$ (known as `sensitivity'; see \cite{Hampel/etc:1986}). One can easily verify that this measure is infinity at $\alpha=0$, translating to an unbounded IF, and decreases with increasing $\alpha>0$. This further implies the increasing robustness of the proposed MNPDPDE as $\alpha>0$ increases, and is clear from the explicit form of the normal error case above. In the example of Figure~\ref{fig:IFplot}, influence functions of the MLE (corresponding to $\alpha=0$) increase and decrease unboundedly for $\bfbeta$ and $\sigma$, respectively, while for $\alpha=0.5$ IF values of the MNPDPDE return to 0 for higher $(y_t,\bfx_{1t})$.

\section{Consistency and Oracle Properties with NP-High dimensionality}
\label{SEC:Theory}

We now study the estimation and variable selection consistency of the proposed MNPDPDE. We use a similar line of argument as \cite{FanLv11} to generalize optimality results for the classical likelihood loss to the DPD loss function. However, in contrast to \cite{FanLv11} who focused on the asymptotics of $\boldsymbol{\beta}$ only, we develop joint asymptotic results for the simultaneous estimation of $\boldsymbol{\beta}$ and $\sigma$, which will be seen not to be independent for general error distributions. Thus, our results also provide an extended asymptotic theory for the non-concave penalized likelihood estimator (at the choice $\alpha=0$).

Here is a brief roadmap to the theoretical results in this section. In Section~\ref{sec:theory-intro}, we introduce some notations, and state a result (Proposition~\ref{prop:prop-KKT}) that ensures the existence of a solution to our optimization problem \eqref{EQ:penalDPD_lossGen}. Section~\ref{subsec:results} states the assumptions and results concerning oracle properties of the MNPDPDE. The MNPDPDE satisfies weak oracle properties (Theorem~\ref{THM:AsympNP_weakOracle}) under assumptions (A1)-(A4), while strengthening the assumptions (A2) and (A3) leads to strong oracle properties being satisfied (Theorem~\ref{THM:AsympNP_strongOracle}). Finally, adding assumption (A5) helps esablish an asymptotic distribution of our robust estimator (Theorem~\ref{THM:AsympNP_Normality}).

\subsection{Background and notations}
\label{sec:theory-intro}
Our optimality results hold for a general class of folded concave penalties, as considered in \cite{FanLv11,LvFan09}. This covers all three important penalty functions, i.e. $\ell_1$, SCAD and MCP. Note that the three desired properties of \cite{FanLi01} as stated in Section \ref{SEC:penalized_estimation} are only satisfied by the SCAD penalty, but not the other two. More precisely, assume that the penalty function $p_{\lambda}(s)$ is as considered in Section \ref{SEC:IF} (i.e., does not depend on $n$ except through $\lambda \equiv \lambda_n$), and satisfies the following assumption.

\vspace{1em}
\noindent {\bf (P)} $p_{\lambda}(s)$ is increasing, continuously differentiable and concave in $s \in [0, \infty)$. 
Also $p_{\lambda}'(s)/ \lambda$ is an increasing function of $\lambda$ with $\rho(p_\lambda) := p_{\lambda}'(0+)/\lambda$ positive and independent of $\lambda$.

\vspace{1em}
\noindent The $\ell_1$ penalty is the only convex function satisfying Assumption (P), along with the non-concave SCAD and MCP penalties.
Thus, assumption (P) is satisfied by a larger class of useful penalty function compared to the three conditions of \cite{FanLi01} 
as described in Section \ref{SEC:penalty}.

Next, consider the quantity $\boldsymbol{P}_{\lambda}^{\ast}(\boldsymbol{\beta})$ as in Theorem \ref{THM:IFlim_MNPDPDE},
and define the following quantities, along the lines of \cite{FanLv11, CHZhang10, LvFan09}.

\begin{Definition}
The local concavity of the penalty $p_\lambda$ at $\boldsymbol{b} \in \BR^p$ is defined as
\begin{align*}
\zeta(p_\lambda; \boldsymbol{b}) &= \lim_{\epsilon\downarrow 0} \max_{1\leq j\leq p}\sup\limits_{t_1<t_2 \in (|b_j|-\epsilon, |b_j|+\epsilon)}
- \frac{p_\lambda'(t_2) - p_\lambda'(t_1)}{t_2 - t_1}.
\end{align*}
The maximum concavity is defined as 
\begin{eqnarray}
\zeta(p_\lambda) =
\sup\limits_{t_1<t_2 \in (0, \infty)} - \frac{p_\lambda'(t_2) - p_\lambda'(t_1)}{t_2 - t_1}.
\end{eqnarray}
\end{Definition}
Assumption (P) ensures that $\zeta(p_\lambda; \boldsymbol{b}) \geq 0$ and $\zeta(p_\lambda)\geq 0$. 
Additionally if $p_\lambda$ has continuous second-order derivative over $t\in(0, \infty)$, 
we have $\zeta(p_\lambda; \boldsymbol{b}) = \max_j \{ - p_\lambda''(|b_j|) \}$. Hence $\zeta(p_\lambda; \boldsymbol{b}) $ 
can be obtained from Table \ref{TAB:Penalties} for the three common penalties.

As a first step towards obtaining oracle properties of the MNPDPDE, we derive necessary and sufficient conditions for the existence of a solution to the estimation problem \eqref{EQ:penalDPD_lossGen}. Denote $r_i(\boldsymbol{\theta}) = (y_i - \boldsymbol{x}_i^T\boldsymbol{\beta}) / \sigma$, and $\boldsymbol{r}(\boldsymbol{\theta}) =(r_1(\boldsymbol{\theta}), \ldots, r_n(\boldsymbol{\theta}))^T$. 

\begin{Proposition}\label{prop:prop-KKT}
Consider the general penalized DPD loss function $Q_{n,\lambda}^{\alpha}(\boldsymbol{\theta})$
as defined in \eqref{EQ:penalDPD_lossGen} for a fixed $\alpha \geq 0$, a general error density $f$ satisfying assumption (A0)
and a penalty function $p_\lambda$ satisfying assumption (P). 
Then, $\widehat{\boldsymbol{\theta}} = (\widehat{\boldsymbol{\beta}}, \widehat{\sigma}) $ is a strict local minimizer of  
$Q_{n,\lambda}^{\alpha}(\boldsymbol{\theta})$ if and only if
\begin{align}
&\frac{1+\alpha}{n\widehat{\sigma}^{\alpha+1}}
\sum_{i=1}^n \psi_{1,\alpha}({r}_i(\widehat{\boldsymbol{\theta}}))\boldsymbol{x}_{1i}
+ \boldsymbol{P}_{\lambda}^{\ast}(\widehat{\boldsymbol{\beta}}_1) 
= \boldsymbol{0}, \label{EQ:Est_EQ_MNPDPDE1}\\
&\left\|\frac{1+\alpha}{n\widehat{\sigma}^{\alpha+1}}
\sum_{i=1}^n \psi_{1,\alpha}({r}_i(\widehat{\boldsymbol{\theta}}))\boldsymbol{x}_{2i}\right\|_{\infty}
< p_{\lambda}'(0+), \label{EQ:Est_EQ_MNPDPDE2}\\
&\frac{1+\alpha}{n\widehat{\sigma}^{\alpha+1}}\sum_{i=1}^n \psi_{2,\alpha}({r}_i(\widehat{\boldsymbol{\theta}}))
= 0, \label{EQ:Est_EQ_MNPDPDE3}\\
&\Lambda_{\min} \left( -\frac{1+\alpha}{n\widehat{\sigma}^{\alpha+2}} \sum_{i=1}^n
\begin{bmatrix}
\begin{array}{cc}
{J}_{11,\alpha}({r}_i(\widehat{\boldsymbol{\theta}}))\boldsymbol{x}_{1i}\boldsymbol{x}_{1i}^T
& {J}_{12,\alpha}({r}_i(\widehat{\boldsymbol{\theta}}))\boldsymbol{x}_{1i}\\
{J}_{12,\alpha}({r}_i(\widehat{\boldsymbol{\theta}}))\boldsymbol{x}_{1i}^T
& {J}_{22,\alpha}({r}_i(\widehat{\boldsymbol{\theta}}))
\end{array}
\end{bmatrix} \right)
> \zeta(p_{\lambda}; \widehat{\boldsymbol{\beta}}_1),
\label{EQ:Est_EQ_MNPDPDE4}
\end{align}
where $\widehat{\boldsymbol{\beta}}_1$ contains non-zero components of $\widehat{\boldsymbol{\beta}}$,
and $\boldsymbol{x}_i = (\boldsymbol{x}_{1i}^T, \boldsymbol{x}_{2i}^T)^T$ is the corresponding partition of 
$\boldsymbol{x}_i$ for each $i$ with $\boldsymbol{x}_{1i}$ having the same dimension as $\widehat{\boldsymbol{\beta}}_1$.
\end{Proposition}

Equations \eqref{EQ:Est_EQ_MNPDPDE1}-\eqref{EQ:Est_EQ_MNPDPDE3}, with strict inequality replaced by non-strict inequality in \eqref{EQ:Est_EQ_MNPDPDE2}, are obtained from the necessary Karush-Kuhn-Tucker (KKT) conditions for existence of a global minimizer of $Q_{n,\lambda}^{\alpha} (\boldsymbol{\theta})$. On the other hand, \eqref{EQ:Est_EQ_MNPDPDE4} is the second-order condition ensuring that the solution is indeed a minimizer. 
Under the classical setting of $p\leq n$, they can be shown to lead to a global minimizer when $\text{rank}(\bfX) = p$.

We define a few quantities before starting on deriving the oracle consistency of the proposed MNPDPDE. 
We assume that the true density belongs to the model family with $g^{\ast} = f_{\boldsymbol{\theta}_0}$ for some $\boldsymbol{\theta}_0=(\boldsymbol{\beta}_0^T, \sigma_0)^T$, 
and consider the sparsity in the true coefficient vector $\boldsymbol{\beta}_0 = (\beta_{1,0}, \ldots, \beta_{p,0})^T$ under NP-dimensional settings. 
Denote the true active set by $\cS = \left\{j : \beta_{j,0} \neq 0 \right\}$ and assume, without loss of generality, that 
$\cS = \left\{1, \ldots, s\right\}$ with $s<<n$ and $\cN := \cS^c = \left\{s+1, \ldots, p \right\}$. 
Denote $\boldsymbol{\beta}_0 = (\boldsymbol{\beta}_{S0}^T, \boldsymbol{0}^T)^T$, 
where $\boldsymbol{\beta}_{S0}=\left(\beta_{1,0}, \ldots, \beta_{s,0}\right)^T \in \BR^s$ 
consists of all and only the $s$ non-zero elements of $\boldsymbol{\beta}_0$. 
Consider the partition $\boldsymbol{\beta} = (\boldsymbol{\beta}_S^T, \boldsymbol{\beta}_N^T)$ and $\bfX=[\bfX_S, \bfX_N]$, 
where $\boldsymbol{\beta}_S \in \BR^s, \bfX_S \in \BR^{n\times s}$, and define the matrices
\begin{align}
\bfX_h^\ast &= \mbox{Block-}\diag \left\{\bfX_h, \boldsymbol{1}_n\right\}; \quad h = S, N,\\
\bfJ_{ij}^{(\alpha)}(\boldsymbol{\theta}) &=
- \frac{1+\alpha}{{\sigma}^{\alpha+2}}
\diag \{{J}_{ij,\alpha}({r}_1(\boldsymbol{\theta})),\ldots, {J}_{ij,\alpha}({r}_n(\boldsymbol{\theta})) \};  i, j = 1, 2,\\
\boldsymbol{\Sigma}_{\alpha}(\boldsymbol{\theta}) &=
\begin{bmatrix}
\begin{array}{cc}
\boldsymbol{J}_{11}^{(\alpha)}(\boldsymbol{\theta}) & \boldsymbol{J}_{12}^{(\alpha)}(\boldsymbol{\theta})\\
\boldsymbol{J}_{12}^{(\alpha)}(\boldsymbol{\theta}) & \boldsymbol{J}_{22}^{(\alpha)}(\boldsymbol{\theta})
\end{array}
\end{bmatrix},
\end{align} 
where $\boldsymbol{1}_n \in \BR^n$ have all components as 1 and ${J}_{ij,\alpha}(\cdot)$ are as defined in Section \ref{SEC:IF1}.

\subsection{Oracle properties}
\label{subsec:results}
We first prove a weak oracle property of the MNPDPDE at any fixed $\alpha\geq 0$ under the following assumptions. 
We assume the dimension of $\bfbeta$, $p \equiv p_n$, and its sparsity level, $s \equiv s_n$, depend on the sample size $n$. 
We denote by $b_s$ a diverging sequence of positive numbers depending on $s$, and $d_n := \min\limits_{j \in \cS}|\beta_{j0}|/2$.

\vspace{1em}
\noindent {\bf (A1)} $||\boldsymbol{x}^{(j)}|| = O(\sqrt{n})$ for $j=1, \ldots, p$, where $\boldsymbol{x}^{(j)}$ is the $j^{\Th}$ column of $\bfX$.

\vspace{1em}
\noindent {\bf (A2)} The design matrix $\bfX$ satisfies 
\begin{align}
&\left\|\left(\bfX_S^{\ast T}\boldsymbol{\Sigma}_{\alpha}(\boldsymbol{\theta}_0)\bfX_S^\ast\right)^{-1}
\right\|_{\infty} = O \left( \frac{b_s}{n} \right),
\label{EQ:A2.1}\\
&\left\|\left(\bfX_N^{\ast T}\boldsymbol{\Sigma}_{\alpha}(\boldsymbol{\theta}_0)\bfX_S^\ast\right)
\left(\bfX_S^{\ast T}\boldsymbol{\Sigma}_{\alpha}(\boldsymbol{\theta}_0)\bfX_S^\ast\right)^{-1}
\right\|_{\infty}
< \min \left\{ \frac{C p_\lambda'(0+)}{p_\lambda'(d_n)}, O(n^{\tau_1}) \right\},
\label{EQ:A2.2}\\
&\max \limits_{(\boldsymbol{\delta}, \sigma)\in \mathcal{N}_0}
\max \limits_{1 \leq j \leq p+1}
\left\{ \Lambda_{\max} \left( \bfX_S^{\ast}
\left[\nabla_{(\boldsymbol{\delta}, \sigma)}^2
{\Gamma}_{j,\alpha}(\boldsymbol{\delta}, \sigma) \right]
\bfX_S^{\ast T}\right) \right\}  = O\left({n}\right),
\label{EQ:A2.3}
\end{align}
for $C\in (0,1)$, $\tau_1 \in [0,0.5]$, $\mathcal{N}_0$ is the set
$$
\left\{(\boldsymbol{\delta}, \sigma)\in \mathbb{R}^s\times\mathbb{R}^{+}
~:~ ||\boldsymbol{\delta}-\boldsymbol{\beta}_{S0}||_\infty \leq d_n, |\sigma - \sigma_0|\leq d_n \right\},$$
$\nabla_{(\boldsymbol{\delta}, \sigma)}^2$ denotes the second-order derivative with respect to 
$(\boldsymbol{\delta}, \sigma)$ and 
\begin{eqnarray}
\boldsymbol{\Gamma}_{\alpha}(\boldsymbol{\delta}, \sigma) &=&\left(\Gamma_{1,\alpha}(\boldsymbol{\delta}, \sigma), 
\ldots, \Gamma_{p,\alpha}(\boldsymbol{\delta},\sigma), \Gamma_{p+1,\alpha}(\boldsymbol{\delta}, \sigma)\right)^T
\nonumber\\
&=&\sum_{i=1}^n \psi_\alpha\left((y_i, \boldsymbol{x}_i), 
(\boldsymbol{\delta}^T, \boldsymbol{0}_{p-s}^T, \sigma)^T\right).
\label{EQ:gamma_j}
\end{eqnarray}

\vspace{1em}
\noindent {\bf (A3)} For some $\tau \in (0, 0.5]$, we have $d_n \geq \log n/ n^\tau,
b_s = o (\min \{n^{1/2-\tau}\sqrt{\log n},n^{\tau}/ s\log n \} )$. Further, with $s=O(n^{\tau_0})$, we define $\tau^\ast = \min\{0.5, 2\tau-\tau_0\} - \tau_1$. Then the regularization parameter $\lambda$ satisfies 
\begin{eqnarray}
p_{\lambda}'(d_n) = o\left(\frac{\log n}{b_s n^\tau}\right),
\lambda \geq \frac{(\log n)^2}{n^{\tau^\ast}}.
\label{EQ:A3.1}
\end{eqnarray}
Also, $\max\limits_{1\leq j \leq p}||\boldsymbol{x}^{(j)}||_\infty = o\left(n^{\tau^\ast}/ \sqrt{\log n} \right)$ and
\begin{align}
\max_{(\boldsymbol{\delta}, \sigma) \in \mathcal{N}_0} \zeta(p_{\lambda}; \boldsymbol{\delta}) &=
o \left(\max_{(\bfdelta, \sigma) \in \mathcal{N}_0} \Lambda_{\min} \left[\frac{1}{n}
\bfX_S^{\ast T}\boldsymbol{\Sigma}_{\alpha}((\boldsymbol{\delta}^T, \boldsymbol{0}_{p-s}^T, \sigma))\bfX_S^\ast\right]\right).
\label{EQ:A3.2}
\end{align}

\vspace{1em}
\noindent {\bf (A4)} For any $\boldsymbol{a} \in \mathbb{R}^n$ and $0<\epsilon < \| \boldsymbol{a}\|/ \|\boldsymbol{a}\|_\infty$, we have the probability bound
\begin{eqnarray}
P\left(\left|\frac{1+\alpha}{{\sigma}_0^{\alpha+1}}
\sum_{i=1}^n a_i \psi_{1,\alpha}({r}_i({\boldsymbol{\theta}}_0))\right| >
\| \boldsymbol{a} \| \epsilon\right)
\leq 2 e^{-c_1\epsilon^2},
\end{eqnarray}
for some $c_1>0$.

The above assumptions generalize those proposed by \cite{FanLv11}. In particular, (A1) and (A3), except Equation (\ref{EQ:A3.2}), are in fact exactly the same as considered in \cite{FanLv11}. Whenever the (fixed) covariates  are standardized, a basic requirement of such high-dimensional analyses, we have $||\boldsymbol{x}^{(j)}|| = \sqrt{n}$ and hence (A1) holds. This assumption allows us to cover the cases of non-standardized (but bounded) covariates. For random design as well, (A1) holds whenever the covariates have finite second-order moments, a commonly used condition in this context.

When only $\boldsymbol{\beta}$ is of interest (e.g., $\sigma$ known or response is standardized), 
(A2) and Equation (\ref{EQ:A3.2}) in (A3) coincides with conditions 2-3 of \cite{FanLv11}, respectively, at $\alpha=0$ and provide their direct generalizations at $\alpha>0$. 
Even in such cases with any $\alpha>0$ and normal error distribution, one can show that $\boldsymbol{\Sigma}_{\alpha}(\boldsymbol{\theta}_0)$
is indeed a constant times identity matrix and hence  (\ref{EQ:A2.1}) in Assumption (A2) holds true for the choice $b_s = s^{1/2}$ whenever $[\Lambda_{\min}(\bfX_S^{T}\bfX_S)]^{-1}=O(n^{-1})$, since $\|(\bfX_S^T \bfX_S )^{-1} \|_{\infty} \leq {\sqrt{s}}{[ \Lambda_{\min} (\bfX_S^{T}\bfX_S ) ]^{-1}}$. 
Clearly the assumption on the eigenvalue is a very mild (and usual) one which holds whenever the covariates has variances away from zero. Equation (\ref{EQ:A2.3}) in (A2) also holds under such mild eigenvalue restriction for normal error distribution; this is also in similar spirit to the condition required for asymptotic analysis of the MDPDE in low-dimensional regression models \citep{GhoshBasu13}. The remaining Equation (\ref{EQ:A2.2}) in (A2) indeed equivalent to ensure weak correlation between the important covariates (in $\bfX_S$) and the unimportant covariates ($\bfX_N$), which is a very common requirement for sparse recovery in high-dimensional context. It is often satisfied for finite (non-diverging) active set size $s=O(1)$. More generally, assumptions (A2)-(A3) indeed also cover the cases of unknown error variance; even then they can be shown to hold for any $\alpha\geq 0$ under very mild conditions in most common cases including the normal error regression.

Finally, assumption (A4) is related to the residual distribution that directly depend on the choice of $f$; this is a direct (robust) generalization of Equation (22) in \cite{FanLv11} for $\alpha>0$ 
and they coincides at $\alpha=0$. On similar lines, by standard probability inequality results, 
it can be shown that (A4) actually holds for gaussian or sub-gaussian errors.

\paragraph*{}
Given the above assumptions hold, we now present the weak oracle property of MNPDPDE in the following theorem.

\begin{Theorem}
\label{THM:AsympNP_weakOracle}
Consider the general penalized DPD loss function $Q_{n,\lambda}^{\alpha}(\boldsymbol{\theta})$ in \eqref{EQ:penalDPD_lossGen} for some fixed $\alpha \geq 0$, 
a general error density $f$ satisfying assumption (A0) and the penalty function satisfying Assumption (P). Suppose Assumptions (A1)-(A4) hold for the given $\alpha$ with  $s=o(n)$ and $\log p = O(n^{1-2\tau^\ast})$. Then, there exist MNPDPDEs $\widehat{\boldsymbol{\beta}}=(\widehat{\boldsymbol{\beta}}_S^T, \widehat{\boldsymbol{\beta}}_N^T)^T$ of $\boldsymbol{\beta}$, 
with $\widehat{\boldsymbol{\beta}}_S \in \BR^s$, and $\widehat{\sigma}$ of $\sigma$ such that $(\widehat{\boldsymbol{\beta}}, \widehat{\sigma})$ is a (strict) local minimizer of $Q_{n,\lambda}^{\alpha}(\boldsymbol{\theta})$, with
\begin{enumerate}
\item $\widehat{\boldsymbol{\beta}}_N = \boldsymbol{0}_{p-s}$, and
\item $ \left\| \widehat{\boldsymbol{\beta}}_S - {\boldsymbol{\beta}}_{S0} \right\|_\infty 
= O\left( \frac{\log n}{n^{\tau}} \right), | \widehat{\sigma} - \sigma_{0} |
= O\left( \frac{\log n}{n^{\tau}} \right),$
\end{enumerate}
holding with probability $ \geq 1 - (2/n) (1 + s + (p-s) \exp[-n^{1-2\tau^\ast}] )$.
\end{Theorem}

\begin{Remark}
If the assumed parametric model satisfies $\boldsymbol{J}_{12}^{(\alpha)}(\boldsymbol{\theta}_0)=\boldsymbol{O}$, e.g.~normal error regression model, one can simplify Assumption (A2) further
and the same line of proof yields $|\widehat{\sigma} - \sigma_{0}| = O( {n^{-1/2}} )$ in Theorem \ref{THM:AsympNP_weakOracle}.
\end{Remark}

We can improve the above rate of convergence and derive strong oracle properties of the estimators by replacing (A2)-(A3) with the stronger assumptions (A2*)-(A3*).

\vspace{1em}
\noindent {\bf (A2*)} The design matrix $\bfX$ satisfies 
\begin{align}
& \min\limits_{(\boldsymbol{\delta},\sigma)\in \mathcal{N}_0} \Lambda_{\min}
\left[\bfX_S^{\ast T}\boldsymbol{\Sigma}_{\alpha}((\boldsymbol{\delta}^T, \boldsymbol{0}_{p-s}, \sigma)^T) \bfX_S^\ast\right] \geq cn,\\
& \left\| \left( \bfX_N^{\ast T}\boldsymbol{\Sigma}_{\alpha}(\boldsymbol{\theta}_0)\bfX_S^\ast \right)
\right\|_{2,\infty} = O(n),\\
& \max\limits_{(\boldsymbol{\delta}, \sigma)\in \mathcal{N}_0}\max_{1\leq j\leq p+1}
\Lambda_{\max}
\left(\bfX_S^{\ast }\left[ \nabla_{(\boldsymbol{\delta}, \sigma)}^2
{\Gamma}_{j,\alpha}(\boldsymbol{\delta}, \sigma)\right]\bfX_S^{\ast T}\right) 
= O\left({n}\right),
\end{align}
for some $c>0$ and $\mathcal{N}_0, \boldsymbol{\Gamma}_{\alpha}(\boldsymbol{\delta}, \sigma)$ as defined in (A2). Further, 
\begin{align*}
E\left\|\frac{1+\alpha}{{\sigma}_{0}^{\alpha+1}}
\sum_{i=1}^n \psi_{1,\alpha}({r}_i({\boldsymbol{\theta}}_0))\boldsymbol{x}_{Si}\right\|_2^2 = O(sn),
E\left|\frac{1+\alpha}{{\sigma}_{0}^{\alpha+1}}\sum_{i=1}^n \psi_{2,\alpha}({r}_i({\boldsymbol{\theta}}_0))\right|^2 = O(n),
\end{align*}	
where the expectation is taken with respect to the true conditional model distribution 
(with parameter $\boldsymbol{\theta}_0$) of $\boldsymbol{y}$ given $\bfX$.

\vspace{1em}
\noindent {\bf (A3*)} For some $\tau \in (0, 0.5]$, we have
$$ p_{\lambda}'(d_n) = O( n^{-1/2} ),
d_n \gg \lambda \gg \min\left\{s^{1/2}n^{-1/2}, n^{\frac{\tau-1}{2}}\sqrt{\log n}\right\}.$$
Also $ \max\limits_{1\leq j \leq p} \| \boldsymbol{x}^{(j)} \|_\infty = o ( n^{(1-\tau)/2} / \sqrt{\log n} )$, and $\max\limits_{(\boldsymbol{\delta},\sigma)\in \mathcal{N}_0} \zeta(p_{\lambda}; \boldsymbol{\delta})= o \left(1\right)$. 

\paragraph*{}
Our next theorem utilizes these assumptions (A2*) and (A3*) to derive a strong oracle property 
of the MNPDPDE extending the results of Theorem \ref{THM:AsympNP_weakOracle}.

\begin{Theorem}
\label{THM:AsympNP_strongOracle}
Suppose  $s \ll n$ and $\log p = O(n^{\tau^*})$ for some $\tau^* \in (0, 0.5)$ and 
assumptions (P), (A0), (A1), (A2*), (A3*) and (A4) hold at a fixed $\alpha\geq 0$. Then, there exists a strict minimizer (strict MNPDPDE) 
$(\widehat{\boldsymbol{\beta}}^T, \widehat{\sigma})^T$ of $Q_{n,\lambda}^\alpha(\boldsymbol{\theta})$ in (2.9) that satisfies the following results with probability tending to 1 as $n \rightarrow\infty$.
\begin{enumerate}
\item $\widehat{\boldsymbol{\beta}}_N = \boldsymbol{0}$, 
where $\widehat{\boldsymbol{\beta}}=(\widehat{\boldsymbol{\beta}}_S^T, \widehat{\boldsymbol{\beta}}_N^T)^T$
with $\widehat{\boldsymbol{\beta}}_S \in \BR^s$,
\item $\left\| \widehat{\boldsymbol{\beta}} - {\boldsymbol{\beta}}_{0} \right\|
= O (\sqrt{s/n} )$ and $|\widehat{\sigma} - \sigma_0 | = O(n^{-1/2} )$.
\end{enumerate}
\end{Theorem}

To derive the asymptotic normality of the MNPDPDE under NP dimensionality, we need an additional Lyapunov-type condition. To this end, we define
\begin{align*}
\boldsymbol{K}_{ij,\alpha}(\boldsymbol{\theta}) &=
\frac{(1+\alpha)^2}{{\sigma}^{2\alpha+2}}
\diag \left\{ \psi_{i, \alpha}({r}_1(\boldsymbol{\theta}))\psi_{j, \alpha}({r}_1(\boldsymbol{\theta})),
\ldots, \psi_{i, \alpha}({r}_n(\boldsymbol{\theta}))\psi_{j, \alpha}({r}_n(\boldsymbol{\theta})) \right\}; i,j=1,2\\
\boldsymbol{\Sigma}_{\alpha}^{\ast}(\boldsymbol{\theta}) &=
\begin{pmatrix}
\begin{array}{cc}
\boldsymbol{K}_{11}^{(\alpha)}(\boldsymbol{\theta}) & \boldsymbol{K}_{12}^{(\alpha)}(\boldsymbol{\theta})\\
\boldsymbol{K}_{12}^{(\alpha)}(\boldsymbol{\theta}) & \boldsymbol{K}_{22}^{(\alpha)}(\boldsymbol{\theta})
\end{array}
\end{pmatrix}.
\end{align*}
Clearly $n^{-1} [\bfX_S^{\ast T}\boldsymbol{\Sigma}_{\alpha}^{\ast}(\boldsymbol{\theta}_0)\bfX_S^\ast ]$ yields a consistent estimate of $\boldsymbol{V}_{\alpha}(\boldsymbol{\theta}) = Var_G [\boldsymbol{\psi}_{\alpha}((Y, \boldsymbol{X});\boldsymbol{\theta}) ]$.
%
%
We assume the following condition to handle the above variance estimate.


\vspace{1em}
\noindent {\bf (A5)} The penalty and loss functions satisfy the following conditions:
\begin{align*}
& p_{\lambda}'(d_n) = O( (sn)^{-1/2} );\\
& \max\limits_{1\leq i \leq n} E\left|\psi_{k,\alpha}({r}_i({\boldsymbol{\theta}}_0))\right|^3
= O\left(1\right), k=1,2,
\end{align*}
and for the design matrix the following hold
\begin{align*}
\min\limits_{(\boldsymbol{\delta},\sigma)\in \mathcal{N}_0}\Lambda_{\min}
\left[\bfX_S^{\ast T}\boldsymbol{\Sigma}_{\alpha^{\ast}}((\boldsymbol{\delta}^T, \boldsymbol{0}_{p-s}, \sigma)^T)
\bfX_S^\ast\right]	&\geq cn,\\
\sum\limits_{i=1}^n \left[\boldsymbol{x}_{Si}^{\ast T}\left(
\bfX_S^{\ast T}\boldsymbol{\Sigma}^{\ast}(\boldsymbol{\theta}_0)\bfX_S^\ast\right)^{-1}\boldsymbol{x}_{Si}^{\ast}
\right]^{3/2} &= o(1),
\end{align*}
with $\mathcal{N}_0$ as defined in (A2*) and  $\boldsymbol{x}_{Si}^{\ast} := (\boldsymbol{x}_{Si}^T, 1)^T$.

\begin{Theorem}
\label{THM:AsympNP_Normality}
In addition to the assumptions of Theorem \ref{THM:AsympNP_strongOracle}, suppose that $s = o(n^{1/3})$ and (A5) holds. Then, the strict MNPDPDE $(\widehat{\boldsymbol{\beta}}, \widehat{\sigma})$ satisfies the following results with probability tending to 1 as $n \rightarrow\infty$.
\begin{enumerate}
\item $\widehat{\boldsymbol{\beta}}_N = \boldsymbol{0}$, where $\widehat{\boldsymbol{\beta}}=(\widehat{\boldsymbol{\beta}}_S^T, \widehat{\boldsymbol{\beta}}_N^T)^T$
with $\widehat{\boldsymbol{\beta}}_S \in \BR^s$.

\item Let $\boldsymbol{A}_n \in \BR^{q\times (s+1)}$ such that $\boldsymbol{A}_n \boldsymbol{A}_n^T \rightarrow \boldsymbol{G}$
as $n\rightarrow\infty$, where $\boldsymbol{G}$ is symmetric and positive definite. Then,
\begin{align}
& \boldsymbol{A}_n
\left(\bfX_S^{\ast T}\boldsymbol{\Sigma}_{\alpha}^{\ast}(\boldsymbol{\theta}_0)\bfX_S^\ast\right)^{-1/2}
\left(\bfX_S^{\ast T}\boldsymbol{\Sigma}_{\alpha}(\boldsymbol{\theta}_0)\bfX_S^\ast\right)
\left((\widehat{\boldsymbol{\beta}}_S, \widehat{\sigma})^T
- ({\boldsymbol{\beta}}_{S0}, \sigma_0)^T \right)
\notag\\
& \displaystyle\mathop{\rightarrow}^\mathcal{D} N_q (\boldsymbol{0}_q,  \boldsymbol{G}).
\label{EQ:AsympNP_Normality}
\end{align}
\end{enumerate}
\end{Theorem}

\begin{Remark}
Whenever $\boldsymbol{J}_{12}^{(\alpha)}(\boldsymbol{\theta}_0)=\boldsymbol{O}$ 
and $\boldsymbol{K}_{12}^{(\alpha)}(\boldsymbol{\theta}_0)=\boldsymbol{O}$ 
for the assumed error density $f$, e.g.~normal, the MDPDEs $\widehat{\boldsymbol{\beta}}_S$ and $\widehat \sigma$ in Theorem \ref{THM:AsympNP_Normality} become asymptotically independent and their limiting distributions can be separated out in (\ref{EQ:AsympNP_Normality}). This can significantly help in deriving (robust) asymptotic testing procedures for the important regression coefficient $\boldsymbol{\beta}_S$ based on the MDPDE $\widehat{\boldsymbol{\beta}}_S$. We hope to pursue this in future work.
\end{Remark}

\section{Numerical Illustrations}\label{SEC:Simulation}
\label{sec:SimSection}

\subsection{Experiment set-up}
We now compare the performance of our method against several robust and non-robust methods of high-dimensional linear regression using a simulation study. We obtain rows of the covariate matrix $\bfX$ as $n=100$ random draws from $\cN(0, \Sigma_X)$, where $\Sigma_X$ is a positive definite matrix with $(i,j)^{\Th}$ element given by $0.5^{|i-j|}$. Given a parameter dimension $p$, we consider two settings for the coefficient vector $\bfbeta$:
\begin{itemize}
\item Setting A (strong signal): For $j \in \{ 1, 2, 4, 7, 11\}$, we set $\beta_j = j$. The rest of the $p-5$ entries of $\bfbeta$ are set at 0;
	
\item Setting B (weak signal): We set $\beta_1 = \beta_7 = 1.5, \beta_2 = 0.5, \beta_4 = \beta_{11} = 1$, and $\beta_j = 0$ otherwise.
\end{itemize}
Finally, we generate the random errors as $\bfepsilon \sim N(0, 0.5^2)$, and set $\bfy = \bfX \bfbeta + \bfepsilon$.

To evaluate efficiency loss against non-robust methods in absence of any contamination, as well as compare performance in presence of contamination in the data, we generate $(\bfy, \bfX, \bfbeta)$ using the above setup in three parallel sets of samples. In one set we do not add any outlying noise in any sample. For the other two, we consider the following two settings of contamination.
\begin{itemize}
\item Y-outliers: We add 20 to the response variables of a random $ 10 \%$ of samples.
	
\item X-outliers: We add 20 to each of the elements in the first 10 rows of $\bfX$ for a random $10 \%$ of samples.
\end{itemize}
The above settings make adherence to the theoretical conditions in Section~\ref{SEC:Theory} likely. 
Specifically, the distribution for columns of $\bfX$ ensure that exponential tail bounds of $\chi^2$ distributions \citep{LaurentMassart00} apply for assumption (A1), and minimum eigenvalue and weak correlation conditions apply for Assumption (A2) (or A2*). Due to the finite parameter situation assumption (A3) (or A3*) is implicitly assumed \citep{FanPeng04}, while sub-gaussian errors with a small proportion of contamination in the Y-outlier case means assumption (A4) holds as well.

We repeat the above for $p = 100, 200, 500$. Finally, given a value of $p$, a setting for signal strength and an outlier type (or no outlier), we repeat the data generation procedure 100 times. For brevity, we report the results for $p=500$ in the main paper and $p=100,200$ in the supplementary material.

\subsection{Competing methods and performance measures}
We compare our method with a host of robust methods of high-dimensional regression, as well as a few non-robust ones available in the literature by applying all of them to each dataset generated from the above setup. We consider the following robust methods for comparison- Robust LARS (RLARS; \cite{KhanEtal07}), sparse least trimmed squares (sLTS; \citep{AlfonsEtal13}, RANSAC, and $\ell_1$-penalized regression using the following loss functions- least absolute deviation (LAD-Lasso; \cite{WangLiJiang07}), DPD (DPD-lasso, \cite{ZangEtal17}) and log DPD (LDPD-lasso, \cite{Kawashima17}). 
We repeat model fitting by our method (DPD-ncv) and LDPD-lasso for different values of $\alpha$ = 0.2, 0.4, 0.6, 0.8, 1, as well as for different values of the starting point, chosen by RLARS, sLTS and RANSAC. DPD-lasso seems to work well for higher $\alpha$, so we repeat for $\alpha$ = 0.4, 0.8, 1.2, 1.6, 2. For DPD-ncv, starting points chosen by RANSAC tend to result in better estimates across all metrics, followed by RLARS and sLTS. However RLARS is much faster, and the advantage becomes clearer as $p$ becomes large. For this reason, here we report outputs of DPD-ncv, DPD-Lasso and LDPD-Lasso corresponding to RLARS starting points. For the proposed DPD-ncv, we used the two most common penalties: SCAD and MCP. The results are very similar, and hence for brevity we only report the findings for the SCAD penalty. Finally, we also use three non-robust methods- Lasso, SCAD and MCP for comparison purposes and to measure efficiency loss. We use 5-fold cross-validation for tuning parameter selection in all the above competing methods except LAD-Lasso, for which we use BIC.

We use the following performance metrics to estimate obtained using each of the above methods: Mean Square Estimation Error (MSEE), Root Mean Square Prediction Error (RMSPE), Estimation Error (EE) of $\sigma$, True Positive proportion (TP), True Negative proportion (TN), and Model Size (MS).
\begin{align*}
\text{MSEE} (\hat \bfbeta) &= (1/p)\| \hat \bfbeta - \bfbeta_0 \|^2,\\
\text{RMSPE}(\hat \bfbeta) &= \sqrt {\|\bfy_{test} - \bfX_{test} \hat \bfbeta \|^2},\\
\text{EE} (\widehat \sigma) &= | \hat \sigma - \sigma_0|,\\
\text{TP} (\hat \bfbeta) &= \frac{ | \supp (\widehat \bfbeta) \cap \supp (\bfbeta_0)|}{|\supp (\bfbeta_0)|},\\
\text{TN} (\hat \bfbeta) &= \frac{ | \supp^c (\widehat \bfbeta) \cap \supp^c (\bfbeta_0)|}{|\supp^c (\bfbeta_0)|},\\
\text{MS} (\hat \bfbeta) &= | \supp (\widehat \bfbeta)|.
\end{align*}
For each of the above metrics, we calculate their average over 100 replications. A good estimator will have low MSEE, RMSPE and EE, values of TP and TN close to 1, and MS close to 5.

\subsection{Results}

Tables \ref{table:outtable_p500e1y}, \ref{table:outtable_p500e1x} and \ref{table:outtable_p500e1} summarize the simulation results for $p = 500$. In all but one (setting A, Y-outliers) of the scenarios, our method has the best performance across all metrics, followed by LDPD-Lasso and RLARS: which select slightly larger models but have worse prediction performance than our method. This is expected, since both these methods are based on $\ell_1$-penalization, which is known to produce biased estimates. Among different DPD-ncv estimates, lower values of $\alpha>0$ (within 0.6) produce the best estimates.

Among other methods, the non-robust methods perform badly as expected. Surprisingly, in spite of being outlier-robust methods, LAD-Lasso does not perform well in our specific outlier settings. DPD-Lasso and LDPD-Lasso perform quite well in our outlier settings. However they tend to be less sparse, in the sense that they are likely to have higher TN and model size even but comparable MSEE and RMSPE than DPD-ncv. This is more severe for DPD-Lasso. They also do not improve upon the RLARS initializations. Our method is the only one that consistently improves upon the starting values given by RLARS among the three DPD-based methods. When no outliers are present in the data, our outputs are same for all values of $\alpha$ considered.

\begin{table}[!h]
	\begin{footnotesize}
		\centering
		\begin{tabular}{r|rrrrrr}
			\hline
			\multicolumn{7}{c}{{\bf Setting A}}\\
			\hline
			Method          & $\text{MSEE} (\hat \bfbeta)$    & $\text{RMSPE}(\hat \bfbeta)$ & $\text{EE} (\widehat \sigma)$ & $\text{TP} (\hat \bfbeta)$   & $\text{TN} (\hat \bfbeta)$   & $\text{MS} (\hat \bfbeta)$\\ 
			~ & ($\times 10^{-4}$) & ($\times 10^{-2}$) &&&& \\\hline
			RLARS & 22.3 & 6.0 & 0.42 & 0.80 & 1.00 & 6.00 \\ 
			sLTS & 4.8 & 6.3 & 0.32 & 1.00 & 1.00 & 6.67 \\ 
			RANSAC & 9.9 & 6.2 & 0.09 & 1.00 & 0.99 & 12.00 \\ 
			LAD-Lasso & 89.4 & 18.1 & 2.78 & 0.94 & 0.99 & 9.53 \\\hline
			DPD-ncv, $\alpha=$ 0.2 & 3.5 & 5.6 & 0.42 & 1.00 & 1.00 & 6.00 \\ 
			DPD-ncv, $\alpha=$ 0.4 & 6.3 & 5.6 & 0.42 & 1.00 & 1.00 & 6.00 \\ 
			DPD-ncv, $\alpha=$ 0.6 & 20.5 & 5.6 & 0.42 & 0.80 & 1.00 & 6.00 \\ 
			DPD-ncv, $\alpha=$ 0.8 & 20.3 & 5.6 & 0.42 & 0.80 & 1.00 & 6.00 \\ 
			DPD-ncv, $\alpha=$ 1 & 21.1 & 5.8 & 0.42 & 0.80 & 1.00 & 6.00 \\ \hline
			DPD-Lasso, $\alpha=$ 0.4 & 35.7 & 12.5 & 0.43 & 1.00 & 0.79 & 108.00 \\ 
			DPD-Lasso, $\alpha=$ 0.8 & 2.6 & 6.1 & 0.43 & 1.00 & 0.89 & 59.00 \\ 
			DPD-Lasso, $\alpha=$ 1.2 & 2.6 & 4.8 & 0.19 & 1.00 & 0.90 & 55.00 \\ 
			DPD-Lasso, $\alpha=$ 1.6 & 2.6 & 4.8 & 0.00 & 1.00 & 0.90 & 53.50 \\ 
			DPD-Lasso, $\alpha=$ 2 & 2.6 & 4.3 & 0.00 & 1.00 & 0.92 & 46.00 \\\hline
			LDPD-Lasso, $\alpha=$ 0.2 & 2.2 & 5.5 & 0.06 & 1.00 & 0.99 & 10.83 \\ 
			LDPD-Lasso, $\alpha=$ 0.4 & 2.5 & 5.7 & 0.08 & 1.00 & 0.99 & 8.59 \\ 
			LDPD-Lasso, $\alpha=$ 0.6 & 3.1 & 5.9 & 0.11 & 1.00 & 0.99 & 7.93 \\ 
			LDPD-Lasso, $\alpha=$ 0.8 & 3.7 & 6.0 & 0.15 & 1.00 & 0.99 & 7.51 \\ 
			LDPD-Lasso, $\alpha=$ 1 & 4.2 & 6.2 & 0.19 & 1.00 & 0.99 & 7.49 \\ \hline
			Lasso & 361.8 & 37.6 & 6.93 & 0.79 & 1.00 & 5.69 \\ 
			SCAD & 139.7 & 22.3 & 3.20 & 0.87 & 0.98 & 13.73 \\ 
			MCP & 138.9 & 21.7 & 3.24 & 0.78 & 0.99 & 6.72 \\ 
			\hline
			\multicolumn{7}{c}{{\bf Setting B}}\\
			\hline
			Method          & $\text{MSEE} (\hat \bfbeta)$    & $\text{RMSPE}(\hat \bfbeta)$ & $\text{EE} (\widehat \sigma)$ & $\text{TP} (\hat \bfbeta)$   & $\text{TN} (\hat \bfbeta)$   & $\text{MS} (\hat \bfbeta)$\\ 
			~ & ($\times 10^{-4}$) & ($\times 10^{-2}$) &&&& \\\hline
			RLARS & 1.1 & 4.6 & 0.09 & 1.00 & 1.00 & 6.00 \\ 
			sLTS & 6.2 & 6.1 & 0.23 & 1.00 & 0.93 & 40.07 \\ 
			RANSAC & 6.2 & 4.8 & 0.24 & 1.00 & 0.92 & 44.00 \\ 
			LAD-Lasso & 68.6 & 15.7 & 2.77 & 0.65 & 0.99 & 6.28 \\\hline
			DPD-ncv, $\alpha=$ 0.2 & 0.8 & 4.3 & 0.06 & 1.00 & 1.00 & 5.00 \\ 
			DPD-ncv, $\alpha=$ 0.4 & 0.8 & 4.3 & 0.06 & 1.00 & 1.00 & 5.00 \\ 
			DPD-ncv, $\alpha=$ 0.6 & 0.8 & 4.5 & 0.06 & 1.00 & 1.00 & 5.00 \\ 
			DPD-ncv, $\alpha=$ 0.8 & 0.7 & 4.6 & 0.06 & 1.00 & 1.00 & 5.00 \\ 
			DPD-ncv, $\alpha=$ 1 & 0.8 & 4.6 & 0.06 & 1.00 & 1.00 & 5.00 \\\hline
			DPD-Lasso, $\alpha=$ 0.4 & 31.4 & 8.4 & 0.47 & 1.00 & 0.37 & 316.00 \\ 
			DPD-Lasso, $\alpha=$ 0.8 & 2.3 & 4.7 & 0.43 & 1.00 & 0.89 & 59.00 \\ 
			DPD-Lasso, $\alpha=$ 1.2 & 2.0 & 4.8 & 0.00 & 1.00 & 0.90 & 55.00 \\ 
			DPD-Lasso, $\alpha=$ 1.6 & 1.8 & 4.7 & 0.00 & 1.00 & 0.91 & 52.00 \\ 
			DPD-Lasso, $\alpha=$ 2 & 1.8 & 4.5 & 0.00 & 1.00 & 0.91 & 49.00 \\\hline
			LDPD-Lasso, $\alpha=$ 0.2 & 2.1 & 5.1 & 0.07 & 1.00 & 0.99 & 10.19 \\ 
			LDPD-Lasso, $\alpha=$ 0.4 & 2.2 & 5.1 & 0.09 & 1.00 & 0.99 & 7.97 \\ 
			LDPD-Lasso, $\alpha=$ 0.6 & 2.3 & 5.1 & 0.11 & 1.00 & 0.99 & 7.62 \\ 
			LDPD-Lasso, $\alpha=$ 0.8 & 2.3 & 5.1 & 0.13 & 1.00 & 1.00 & 7.38 \\ 
			LDPD-Lasso, $\alpha=$ 1 & 2.3 & 5.2 & 0.14 & 1.00 & 1.00 & 7.38 \\ \hline
			Lasso & 134.1 & 22.4 & 4.54 & 0.02 & 1.00 & 0.24 \\ 
			SCAD & 128.6 & 21.0 & 3.60 & 0.32 & 0.99 & 8.72 \\ 
			MCP & 141.6 & 21.1 & 3.69 & 0.24 & 0.99 & 4.52 \\ 
			\hline
		\end{tabular}
		\caption{Table of outputs for $p=500$ and Y-outliers}
		\label{table:outtable_p500e1y}
	\end{footnotesize}
\end{table}

\begin{table}[!h]
	\begin{footnotesize}
		\centering
		\begin{tabular}{r|rrrrrr}
			\hline
			\multicolumn{7}{c}{{\bf Setting A}}\\
			\hline
			Method          & $\text{MSEE} (\hat \bfbeta)$    & $\text{RMSPE}(\hat \bfbeta)$ & $\text{EE} (\widehat \sigma)$ & $\text{TP} (\hat \bfbeta)$   & $\text{TN} (\hat \bfbeta)$   & $\text{MS} (\hat \bfbeta)$\\ 
			~ & ($\times 10^{-4}$) & ($\times 10^{-2}$) &&&& \\\hline
			RLARS & 0.5 & 3.7 & 0.06 & 1.00 & 1.00 & 6.00 \\ 
			sLTS & 6.8 & 5.7 & 0.25 & 1.00 & 1.00 & 6.00 \\ 
			RANSAC & 6.3 & 5.6 & 0.11 & 1.00 & 0.99 & 10.00 \\ 
			LAD-Lasso & 561.8 & 42.3 & 1.15 & 0.83 & 0.84 & 81.57 \\ \hline 
			DPD-ncv, $\alpha=$ 0.2 & 0.4 & 4.0 & 0.04 & 1.00 & 1.00 & 5.00 \\ 
			DPD-ncv, $\alpha=$ 0.4 & 0.5 & 3.9 & 0.05 & 1.00 & 1.00 & 5.00 \\ 
			DPD-ncv, $\alpha=$ 0.6 & 0.5 & 3.9 & 0.06 & 1.00 & 1.00 & 5.00 \\ 
			DPD-ncv, $\alpha=$ 0.8 & 0.5 & 3.8 & 0.08 & 1.00 & 1.00 & 5.00 \\ 
			DPD-ncv, $\alpha=$ 1 & 0.5 & 3.8 & 0.10 & 1.00 & 1.00 & 5.00 \\ \hline 
			DPD-Lasso, $\alpha=$ 0.4 & 1211.6 & 29.0 & 0.43 & 1.00 & 0.81 & 99.50 \\ 
			DPD-Lasso, $\alpha=$ 0.8 & 3.2 & 7.8 & 0.43 & 1.00 & 0.87 & 67.50 \\ 
			DPD-Lasso, $\alpha=$ 1.2 & 3.0 & 6.8 & 0.30 & 1.00 & 0.88 & 64.00 \\ 
			DPD-Lasso, $\alpha=$ 1.6 & 3.1 & 6.4 & 0.00 & 1.00 & 0.89 & 60.00 \\ 
			DPD-Lasso, $\alpha=$ 2 & 3.0 & 6.5 & 0.00 & 1.00 & 0.89 & 57.00 \\\hline 
			LDPD-Lasso, $\alpha=$ 0.2 & 1.8 & 5.1 & 0.05 & 1.00 & 0.99 & 8.84 \\ 
			LDPD-Lasso, $\alpha=$ 0.4 & 1.7 & 5.1 & 0.06 & 1.00 & 0.99 & 7.81 \\ 
			LDPD-Lasso, $\alpha=$ 0.6 & 1.7 & 5.1 & 0.07 & 1.00 & 1.00 & 7.30 \\ 
			LDPD-Lasso, $\alpha=$ 0.8 & 1.7 & 5.1 & 0.07 & 1.00 & 1.00 & 7.11 \\ 
			LDPD-Lasso, $\alpha=$ 1 & 1.7 & 5.1 & 0.08 & 1.00 & 1.00 & 6.93 \\ \hline 
			LASSO & 444.6 & 41.4 & 1.77 & 0.86 & 0.86 & 75.45 \\ 
			SCAD & 887.4 & 53.9 & 3.72 & 0.66 & 0.97 & 15.87 \\ 
			MCP & 863.7 & 51.2 & 3.59 & 0.66 & 0.97 & 16.26 \\
			\hline
			\multicolumn{7}{c}{{\bf Setting B}}\\
			\hline
			Method          & $\text{MSEE} (\hat \bfbeta)$    & $\text{RMSPE}(\hat \bfbeta)$ & $\text{EE} (\widehat \sigma)$ & $\text{TP} (\hat \bfbeta)$   & $\text{TN} (\hat \bfbeta)$   & $\text{MS} (\hat \bfbeta)$\\ 
			~ & ($\times 10^{-4}$) & ($\times 10^{-2}$) &&&& \\\hline
			RLARS & 2.0 & 4.2 & 0.14 & 1.00 & 0.99 & 12.00 \\ 
			sLTS & 8.7 & 5.3 & 0.24 & 1.00 & 0.92 & 42.50 \\ 
			RANSAC & 5.8 & 5.9 & 0.26 & 1.00 & 0.98 & 15.00 \\ 
			LAD-Lasso & 108.0 & 20.4 & 2.87 & 0.38 & 0.99 & 7.71 \\ \hline 
			DPD-ncv, $\alpha=$ 0.2 & 1.2 & 4.1 & 0.08 & 1.00 & 1.00 & 7.00 \\ 
			DPD-ncv, $\alpha=$ 0.4 & 1.1 & 4.0 & 0.10 & 1.00 & 1.00 & 7.00 \\ 
			DPD-ncv, $\alpha=$ 0.6 & 1.1 & 4.2 & 0.12 & 1.00 & 1.00 & 7.00 \\ 
			DPD-ncv, $\alpha=$ 0.8 & 1.4 & 4.2 & 0.14 & 1.00 & 1.00 & 7.00 \\ 
			DPD-ncv, $\alpha=$ 1 & 1.5 & 4.2 & 0.15 & 1.00 & 1.00 & 7.00 \\ \hline 
			DPD-Lasso, $\alpha=$ 0.4 & 38.4 & 9.7 & 0.43 & 1.00 & 0.83 & 87.50 \\ 
			DPD-Lasso, $\alpha=$ 0.8 & 3.2 & 7.3 & 0.43 & 1.00 & 0.87 & 67.00 \\ 
			DPD-Lasso, $\alpha=$ 1.2 & 3.0 & 6.0 & 0.00 & 1.00 & 0.88 & 63.00 \\ 
			DPD-Lasso, $\alpha=$ 1.6 & 3.0 & 6.3 & 0.00 & 1.00 & 0.88 & 62.00 \\ 
			DPD-Lasso, $\alpha=$ 2 & 2.9 & 6.4 & 0.00 & 1.00 & 0.89 & 60.00 \\\hline 
			LDPD-Lasso, $\alpha=$ 0.2 & 1.9 & 5.0 & 0.06 & 1.00 & 0.98 & 15.14 \\ 
			LDPD-Lasso, $\alpha=$ 0.4 & 1.8 & 5.0 & 0.07 & 1.00 & 0.98 & 14.04 \\ 
			LDPD-Lasso, $\alpha=$ 0.6 & 1.8 & 5.1 & 0.07 & 1.00 & 0.98 & 14.03 \\ 
			LDPD-Lasso, $\alpha=$ 0.8 & 1.8 & 5.0 & 0.07 & 1.00 & 0.98 & 14.47 \\ 
			LDPD-Lasso, $\alpha=$ 1 & 1.8 & 5.0 & 0.07 & 1.00 & 0.98 & 13.90 \\ \hline 
			LASSO & 22.6 & 10.3 & 0.13 & 0.99 & 0.87 & 70.32 \\ 
			SCAD & 45.8 & 13.8 & 0.55 & 0.81 & 0.98 & 16.25 \\ 
			MCP & 45.2 & 12.8 & 0.49 & 0.81 & 0.97 & 16.45 \\  
			\hline
		\end{tabular}
		\caption{Table of outputs for $p=500$ and X-outliers}
		\label{table:outtable_p500e1x}
	\end{footnotesize}
\end{table}

\begin{table}[!h]
	\begin{footnotesize}
		\centering
		\begin{tabular}{r|rrrrrr}
			\hline
			\multicolumn{7}{c}{{\bf Setting A}}\\
			\hline
			Method          & $\text{MSEE} (\hat \bfbeta)$    & $\text{RMSPE}(\hat \bfbeta)$ & $\text{EE} (\widehat \sigma)$ & $\text{TP} (\hat \bfbeta)$   & $\text{TN} (\hat \bfbeta)$   & $\text{MS} (\hat \bfbeta)$\\ 
			~ & ($\times 10^{-4}$) & ($\times 10^{-2}$) &&&& \\\hline
			RLARS & 0.5 & 3.8 & 0.06 & 1.00 & 1.00 & 6.00 \\ 
			sLTS & 6.5 & 5.6 & 0.24 & 1.00 & 1.00 & 6.00 \\ 
			RANSAC & 6.5 & 4.6 & 0.09 & 1.00 & 0.99 & 10.50 \\ 
			LAD-Lasso & 4.7 & 3.9 & 0.44 & 1.00 & 1.00 & 6.70 \\ \hline 
			DPD-ncv, $\alpha=$ 0.2 & 0.5 & 3.8 & 0.06 & 1.00 & 1.00 & 6.00 \\ 
			DPD-ncv, $\alpha=$ 0.4 & 0.5 & 3.8 & 0.06 & 1.00 & 1.00 & 6.00 \\ 
			DPD-ncv, $\alpha=$ 0.6 & 0.5 & 3.8 & 0.06 & 1.00 & 1.00 & 6.00 \\ 
			DPD-ncv, $\alpha=$ 0.8 & 0.5 & 3.8 & 0.06 & 1.00 & 1.00 & 6.00 \\ 
			DPD-ncv, $\alpha=$ 1 & 0.5 & 3.8 & 0.06 & 1.00 & 1.00 & 6.00 \\\hline 
			DPD-Lasso, $\alpha=$ 0.4 & 1147.7 & 12.2 & 0.46 & 1.00 & 0.30 & 350.50 \\ 
			DPD-Lasso, $\alpha=$ 0.8 & 2.1 & 4.5 & 0.46 & 1.00 & 0.87 & 67.50 \\ 
			DPD-Lasso, $\alpha=$ 1.2 & 1.7 & 4.0 & 0.00 & 1.00 & 0.89 & 61.00 \\ 
			DPD-Lasso, $\alpha=$ 1.6 & 1.6 & 4.4 & 0.00 & 1.00 & 0.90 & 56.00 \\ 
			DPD-Lasso, $\alpha=$ 2 & 1.4 & 4.1 & 0.00 & 1.00 & 0.90 & 53.00 \\\hline 
			LDPD-Lasso, $\alpha=$ 0.2 & 1.7 & 3.4 & 0.06 & 1.00 & 0.99 & 9.90 \\ 
			LDPD-Lasso, $\alpha=$ 0.4 & 1.6 & 3.5 & 0.06 & 1.00 & 0.99 & 8.20 \\ 
			LDPD-Lasso, $\alpha=$ 0.6 & 1.6 & 3.4 & 0.06 & 1.00 & 0.99 & 7.80 \\ 
			LDPD-Lasso, $\alpha=$ 0.8 & 1.6 & 3.4 & 0.07 & 1.00 & 0.99 & 7.80 \\ 
			LDPD-Lasso, $\alpha=$ 1 & 1.6 & 3.3 & 0.07 & 1.00 & 0.99 & 7.60 \\ \hline 
			Lasso & 2.4 & 3.8 & 0.41 & 1.00 & 0.99 & 7.80 \\ 
			SCAD & 17.8 & 5.2 & 0.70 & 0.98 & 1.00 & 4.90 \\ 
			MCP & 12.6 & 4.7 & 0.57 & 0.98 & 1.00 & 4.90 \\ 
			\hline
			\multicolumn{7}{c}{{\bf Setting B}}\\
			\hline
			Method          & $\text{MSEE} (\hat \bfbeta)$    & $\text{RMSPE}(\hat \bfbeta)$ & $\text{EE} (\widehat \sigma)$ & $\text{TP} (\hat \bfbeta)$   & $\text{TN} (\hat \bfbeta)$   & $\text{MS} (\hat \bfbeta)$\\ 
			~ & ($\times 10^{-4}$) & ($\times 10^{-2}$) &&&& \\\hline
			RLARS & 1.4 & 4.7 & 0.12 & 1.00 & 0.99 & 10.00 \\ 
			sLTS & 7.9 & 5.7 & 0.24 & 1.00 & 0.93 & 42.00 \\ 
			RANSAC & 5.2 & 4.9 & 0.23 & 1.00 & 0.98 & 15.00 \\ 
			LAD-Lasso & 4.7 & 3.9 & 0.42 & 1.00 & 1.00 & 7.30 \\ \hline 
			DPD-ncv, $\alpha=$ 0.2 & 1.4 & 4.7 & 0.12 & 1.00 & 0.99 & 10.00 \\ 
			DPD-ncv, $\alpha=$ 0.4 & 1.4 & 4.7 & 0.12 & 1.00 & 0.99 & 10.00 \\ 
			DPD-ncv, $\alpha=$ 0.6 & 1.4 & 4.7 & 0.12 & 1.00 & 0.99 & 10.00 \\ 
			DPD-ncv, $\alpha=$ 0.8 & 1.4 & 4.7 & 0.12 & 1.00 & 0.99 & 10.00 \\ 
			DPD-ncv, $\alpha=$ 1 & 1.4 & 4.7 & 0.12 & 1.00 & 0.99 & 10.00 \\ \hline
			DPD-Lasso, $\alpha=$ 0.4 & 30.3 & 6.3 & 0.47 & 1.00 & 0.30 & 351.00 \\ 
			DPD-Lasso, $\alpha=$ 0.8 & 2.4 & 4.7 & 0.46 & 1.00 & 0.87 & 67.50 \\ 
			DPD-Lasso, $\alpha=$ 1.2 & 2.2 & 4.3 & 0.00 & 1.00 & 0.89 & 60.00 \\ 
			DPD-Lasso, $\alpha=$ 1.6 & 2.1 & 4.2 & 0.00 & 1.00 & 0.89 & 57.00 \\ 
			DPD-Lasso, $\alpha=$ 2 & 1.9 & 4.2 & 0.00 & 1.00 & 0.90 & 54.00 \\\hline 
			LDPD-Lasso, $\alpha=$ 0.2 & 1.9 & 3.9 & 0.06 & 1.00 & 0.97 & 17.50 \\ 
			LDPD-Lasso, $\alpha=$ 0.4 & 2.0 & 4.1 & 0.09 & 1.00 & 0.98 & 15.50 \\ 
			LDPD-Lasso, $\alpha=$ 0.6 & 2.0 & 4.2 & 0.09 & 1.00 & 0.98 & 16.90 \\ 
			LDPD-Lasso, $\alpha=$ 0.8 & 2.0 & 4.2 & 0.08 & 1.00 & 0.98 & 16.00 \\ 
			LDPD-Lasso, $\alpha=$ 1 & 2.0 & 4.1 & 0.08 & 1.00 & 0.98 & 16.40 \\ \hline 
			Lasso & 2.1 & 3.6 & 0.33 & 1.00 & 0.98 & 12.90 \\ 
			SCAD & 0.3 & 3.7 & 0.21 & 1.00 & 0.99 & 9.70 \\ 
			MCP & 0.3 & 3.7 & 0.20 & 1.00 & 1.00 & 6.80 \\ 
			\hline
		\end{tabular}
		\caption{Table of outputs for $p=500$ and no outliers}
		\label{table:outtable_p500e1}
	\end{footnotesize}
\end{table}

Comparing across different outlier settings, Y-outliers have a more severe negative effect on performances for all methods except DPD-Lasso and LDPD-Lasso, while for these two the trends are reversed. The severity of this increases as $p$ goes higher (comparing with tables B.1-B.6 in supplementary material). Comparing the large signal (Setting A) and small signal (Setting B) settings, the performance decrease of DPD-ncv is minimal compared to the competing methods.

It is important to note that, unlike classical MDPDEs, the effect of $\alpha$ in the performance of the DPD-ncv is not so prominent in cases of stronger signals and higher dimensions. For pure data, the DPD-ncv estimator generates almost the same values of the performance metrics 
at each distinct value of $\alpha$ used. The same is also true for variable selection metrics under contamination, 
where the smaller values of $\alpha$ appear to have competitive stability compared to its large values also in the errors in parameter estimates. 
The most probable reason could be that the changes in MDPDEs with varying $\alpha$  are relatively small for stronger signals 
and hence do not affect the variable selection performance (since actual values of the regression coefficients are reasonably away from zero)
leading to the same performance metrics. 
On the other hand, the higher variance of the MDPDE at a larger values of $\alpha$ might sometime affects  
the resulting variable selection even under contaminations. 
From the additional simulation results provided in the online supplement,  
the effects of $\alpha$ become more prominent as the signal gets weak (and dimension reduces) 
where an appropriate choices of $\alpha$ for DPD-ncv leads to the best performance.

\subsection{On the choice of robustness tuning parameter $\alpha$}

The tuning parameter $\alpha$ in the definition of DPD measure serves as a trade-off between robustness and efficiency of the resulting minimum DPD estimator under classical set-ups. The MDPDE at $\alpha=0$ is indeed the most efficient, but extremely non-robust MLE; the robustness of MDPDE under data contamination increases significantly with increasing values of $\alpha$ at a price of (slight) loss in efficiency \citep{GhoshBasu13,Basu/etc:2011}. The same continues to hold under the high-dimensional set-up, which is clear from our theoretical IFs and empirical analyses presented above. However, as noted earlier, the effect is observed to be lesser with SCAD penalty compared to the Lasso penalty 
for reasonably higher signal-to-noise ratio (at least in our simulations). In general, setting an appropriate value of $\alpha$ may be vital in the performance of the MNPDPDE. 

The ideal value of $\alpha$ should depend on the amount of contamination present in the sample data; 
we should take a larger $\alpha$ for higher degree or contamination and vice versa. 
However, in practice, the extent of contamination in the data is unknown and we need some appropriate guidelines 
to choose an optimal $\alpha$ value to apply the proposed  MNPDPDE. 
One straightforward way is to use an $\alpha$ value that provide the best result in an extensive simulation study with a similar set-up;
for example, in the simulations results presented in the paper (and several other not presented here for brevity), we can see that 
$\alpha\leq 0.5$ provides the best trade-off and hence, combining it with our experience, 
we can recommend the use of $\alpha\in[0.4, 0.5]$ in the MNPDPDE as an empirical suggestion.

However, since different datasets may have different degrees of contamination, it would be always helpful to use some data-driven choices of $\alpha$. Algorithms for choosing data-driven optimum $\alpha$ are available in MDPDE literature for classical low-dimensional models. One popular approach is to minimize and estimate of asymptotic mean squared error of the MDPDE, which was initially proposed by \cite{Warwick/Jones:2005} for IID data and later explored by \cite{GhoshBasu15} for {\it low-dimensional} LRM. A refinement of this approach is recently developed by \cite{Basu/etc:2020}.
This approach can be used for selection of optimal $\alpha$ for our MNPDPDE as well, as long as the final selected model is low-dimensional. But, naturally it does not take care of the effect of model selection in to account and would be ideal only if the final selected model has the same set of covariates for all $\alpha$.
 
An empirical way of choosing $\alpha$ taking care of the model selection performance would be 
to incorporate it as a second tuning parameter during the training stage, i.e. computing HBIC values, like in \eqref{eqn:hbic}, but for $(\lambda, \alpha)$ pairs chosen from a two-dimensional grid. 
However, since the conditions (A2)-(A4), (A2$^*$), and (A3$^*$) depend on $\alpha$, it is possible that there are more theoretically grounded ways of choosing $\alpha$. We plan to explore this in more detail in future research.

\section{Conclusion}
\label{sec:conclusion}

In this paper, we examined the penalized DPD loss function, proposed an estimator for the non-concave penalized loss, and established theoretical convergence and robustness properties in a high-dimensional context. As we demonstrate in the simulations, not all robust methods perform well in presence of data contamination of large magnitudes. To this end, an influence function-based approach, as taken by us in Section~\ref{SEC:IF}, gives provable guarantees for the robustness of the procedure being used. The simplicity of the DPD framework and the corresponding assumptions in Section~\ref{subsec:results} indicate a straightforward direction to extend all our theoretical results for any parametric regression models, including generalized linear models with non-identity link functions. 

Our work motivates two immediate theoretical directions to be pursued. Firstly, the relationship and proximity of local solutions obtained by Algorithm~\ref{algo:dpdncv-algo} may be explored in a rigorous manner. Using the tools provided by \cite{LohWainwright15} and \cite{LohWainwright17}, specifically by establishing Restricted Strong Convexity-type conditions for the DPD loss function and then generalizing the CCCP-based MM algorithm to a composite gradient descent algorithm, is one possible avenue that can be explored. Secondly, knowledge on the structure of the design matrix is essential in establishing non-asymptotic error bounds and hypothesis testing procedures in a high-dimensional regime, where parameter dimension, sparsity and sample size are allowed to diverge to infinity \citep{ZhangHuang08,BickelRitovTsybakov09}. To our knowledge, our results and conditions in Section~\ref{SEC:Theory} are the first attempt towards achieving these goals in high-dimensional robust analysis assuming a general penalty function. More work is needed to establish sufficient conditions on the data and contamination settings, such as error distributions, contamination proportion and magnitudes, for these results to hold.

It is highly important to extend our theory and computational algorithm for heavy tailed error distributions (e.g., Laplace or double-exponential) which are also useful in the context of robustness. Our assumptions (A1)-(A3) do not depend on the tail nature of the error distribution, but (A4) does. Therefore, one possible avenue to establish consistency of MNPDPDE under heavy-tailed error is by weakening Assumption (A4) appropriately, and then following in the lines of \cite{FanLiWang17} replacing their Huber loss by our DPD loss and $l_1$ penalization by the non-concave penalties.

There are at least two further extensions of our present work. Firstly, one can consider graphical models, where robust high-dimensional estimation is in its infancy. Little is known about the theoretical properties of such estimators and required conditions \citep{SunLi12,HiroseEtal17}, and our analysis in Section~\ref{subsec:results} provide a road-map towards obtaining parallel results in graphical models with generalized error distributions. Secondly, incorporating group penalties is of interest from a practical perspective, in order to perform robust analysis taking into account known structured sparsity patterns among predictors. We hope to pursue some of these extensions in future.

\section*{Acknowledgments}
Authors wish to thank the AE and two anonymous reviewers for their careful reading of the manuscript and several suggestions for improving the paper. Research of the first author (AG) is partially supported by the INSPIRE Faculty Research Grant from Department of Science and Technology, Government of India. SM was supported by Prof. George Michailidis during his time in University of Florida.

\bibliographystyle{apalike}
\bibliography{dpdncv-draft}

\newpage
\appendix
\section{Proofs of the Results of Section 4}\label{SEC:intro}

\subsection{Proof of Proposition 4.1}

From Equations (2.8), (3.4) and (4.9) of the main paper, we have 
\begin{eqnarray}
\nabla L_n^\alpha (\boldsymbol{\theta}) 
&=& \frac{1}{n} \sum_{i=1}^n \psi_\alpha\left((y_i, \boldsymbol{x}_i), \boldsymbol{\theta}\right),
\label{EQ:nabla_Ln}\\
\nabla^2 L_n^\alpha (\boldsymbol{\theta}) 
&=& \frac{1}{n} \sum_{i=1}^n \nabla\psi_\alpha\left((y_i, \boldsymbol{x}_i), \boldsymbol{\theta}\right)
= \frac{1}{n} \boldsymbol{X}^{\ast T}\boldsymbol{\Sigma}_{\alpha}(\boldsymbol{\theta})\boldsymbol{X}^{\ast},
\label{EQ:nabla2_Ln}
\end{eqnarray}
where $\nabla$ and $\nabla^2$ denote the first and second order partial derivatives with respect to $\boldsymbol{\theta}$,
respectively, $\boldsymbol{X}^\ast = \mbox{Block-diag}\left\{\boldsymbol{X}, \boldsymbol{1}_n \right\}$.
and $\boldsymbol{\Sigma}_{\alpha}(\boldsymbol{\theta})$ is as defined in Equation (4.9) of the main paper.

\bigskip
\noindent\textit{Only if part:}\\
Assume that $\widehat{\boldsymbol{\theta}} = (\widehat{\boldsymbol{\beta}}, \widehat{\sigma})$
with $\widehat{\boldsymbol{\beta} }=(\hat \beta_1, \ldots, \hat \beta_p)^T$ being a local minimizer of  $Q_{n,\lambda}^{(\alpha)}(\boldsymbol{\theta})$. 
Then, the necessary Karush-Kuhn-Tucker (KKT) conditions imply the existence of a $(p+1)$-vector 
$\boldsymbol{v}=(v_1, \ldots, v_p, v_{p+1})^T$ such that 
\begin{eqnarray}
\frac{1}{n} \sum_{i=1}^n \psi_\alpha\left((y_i, \boldsymbol{x}_i), \widehat{\boldsymbol{\theta}}\right)
+ \boldsymbol{v} = \boldsymbol{0}_{p+1}, 
\label{EQ:KKT}
\end{eqnarray}
where $v_{p+1}=0$ and for each $j=1, \ldots, p$, $v_j = p_\lambda'(|\widehat{\beta}_j|)$ if $\beta_j \neq 0$
and $v_j \in [-p_\lambda'(0+), p_\lambda'(0+)]$ if $\beta_j = 0$. 
Now, if we let $\widehat{\mathcal{S}}=\mbox{supp}(\widehat{\boldsymbol{\beta}})$ 
and $\widehat{\boldsymbol{\beta}}_1$, $\boldsymbol{x}_{i1}$ 
to be the corresponding partitions of $\widehat{\boldsymbol{\beta}}$ and $\boldsymbol{x}_i$, respectively,
formed only with components and columns with indices in $\widehat{\mathcal{S}}$,
then the above KKT conditions (\ref{EQ:KKT}) clearly leads to the required Equations (4.3)--(4.5)
of the main paper, with non-strict inequality in (4.4). 

Further, note that $\widehat{\boldsymbol{\theta}}$ is also a local minimizer of $Q_{n,\lambda}^{(\alpha)}(\boldsymbol{\theta})$ constrained on the $|S|$-dimensional subspace $\mathcal{B} = \left\{ (\boldsymbol{\beta}^T, \sigma)^T \in \mathbb{R}^p\times \mathbb{R}^{+} : \boldsymbol{\beta}_c = \boldsymbol{0} \right\}$,
 where $\boldsymbol{\beta}_c$ is the partition of $\widehat{\boldsymbol{\beta}}$  formed only with components with indices in $S^c$. Therefore, by the second order condition, we get that 
$$
\Lambda_{\min} \left(
\boldsymbol{X}_1^{\ast T} \boldsymbol{\Sigma}_{\alpha} (\widehat{\boldsymbol{\theta}})\boldsymbol{X}_1^\ast
\right)
\geq \zeta(P_{\lambda_n}, \widehat{\boldsymbol{\beta}}_1),
$$
which is the same as the desired Equation (4.6) of the main paper. Finally, both the non-strict inequalities will be strict inequalities when $\widehat{\boldsymbol{\theta}}$ is  a strict local minimizer of  $Q_{n,\lambda}^{(\alpha)}(\boldsymbol{\theta})$, completing the proof of the only if part. 

\bigskip
\noindent\textit{If part:}\\
To prove the if part, we assume that conditions (4.3)--(4.6) of the main paper hold and consider the objective function $Q_{n,\lambda}^{(\alpha)}(\boldsymbol{\theta})$ constrained on the subspace $\mathcal{B}$.
Condition (4.6) implies that $Q_{n,\lambda}^{(\alpha)}(\boldsymbol{\theta})$ is strictly convex in a neighborhood $\mathcal{N}_0\subseteq\mathcal{B}$ of $\widehat{\boldsymbol{\theta}}$. But, Conditions (4.3) and (4.5) imply that $\widehat{\boldsymbol{\theta}}$ is a critical point of $Q_{n,\lambda}^{(\alpha)}(\boldsymbol{\theta})$	in $\mathcal{B}$. Combining, we get that $\widehat{\boldsymbol{\theta}}$ is the unique minimizer of $Q_{n,\lambda}^{(\alpha)}(\boldsymbol{\theta})$	in $\mathcal{N}_0$.

Consider now a sufficiently small neighborhood $\mathcal{N}_1\subset \mathbb{R}^p\times\mathbb{R}^{+}$
centered at $\widehat{\boldsymbol{\theta}}$ such that $\mathcal{N}_1\cap \mathcal{B} \subset \mathcal{N}_0$. 
To prove that $\widehat{\boldsymbol{\theta}}$ is a strict local minimizer of $Q_{n,\lambda}^{(\alpha)}(\boldsymbol{\theta})$ over the full parameter space, it is enough to show that $Q_{n,\lambda}^{(\alpha)}(\widehat{\boldsymbol{\theta}}) < Q_{n,\lambda}^{(\alpha)}(\boldsymbol{\theta}_1)$ for any $\boldsymbol{\theta}_1 \in \mathcal{N}_1\backslash\mathcal{N}_0$. Let $\boldsymbol{\theta}_2$ denote the projection of $\boldsymbol{\theta}_1$ onto $\mathcal{B}$ so that  $\boldsymbol{\theta}_2\in \mathcal{N}_1$. Since $\widehat{\boldsymbol{\theta}}$ is the unique (strict) minimizer of $Q_{n,\lambda}^{(\alpha)}(\boldsymbol{\theta})$	in $\mathcal{N}_0$, we must have $Q_{n,\lambda}^{(\alpha)}(\widehat{\boldsymbol{\theta}})<Q_{n,\lambda}^{(\alpha)}(\boldsymbol{\theta}_2)$
when $\boldsymbol{\theta}_2 \neq \widehat{\boldsymbol{\theta}}$. Therefore, we just need to show that 
$Q_{n,\lambda}^{(\alpha)}(\boldsymbol{\theta}_2) < Q_{n,\lambda}^{(\alpha)}(\boldsymbol{\theta}_1)$.

Note that, an application of the mean-value theorem leads to 
\begin{eqnarray}
Q_{n,\lambda}^{(\alpha)}(\boldsymbol{\theta}_2) - Q_{n,\lambda}^{(\alpha)}(\boldsymbol{\theta}_1)
&=& \nabla Q_{n,\lambda}^{(\alpha)}(\boldsymbol{\theta}_3) (\boldsymbol{\theta}_2 - \boldsymbol{\theta}_1),
\label{EQLProp1_pf1}
\end{eqnarray}
for some $\boldsymbol{\theta}_3$ lying on the line segment joining $\boldsymbol{\theta}_1$ and $\boldsymbol{\theta}_2$. 
Let $\theta_{k,j}$ denote the $j$-th component of $\boldsymbol{\theta}_k$ for $k=1,2,3$. 
Then, we have 
\begin{eqnarray}
&&\theta_{1,j} = \theta_{2,j}, ~~~~~~~~~~~~~~~~~~~\mbox{ if } j\in \widehat{\mathcal{S}},~j=p+1, 
\nonumber\\
\mbox{and }&& sign(\theta_{1,j})=sign(\theta_{3,j}), ~~~\mbox{ if } j\notin \widehat{\mathcal{S}}.
\nonumber
\end{eqnarray}
Thus, we get from (\ref{EQLProp1_pf1}) that 
\begin{eqnarray}
Q_{n,\lambda}^{(\alpha)}(\boldsymbol{\theta}_2) - Q_{n,\lambda}^{(\alpha)}(\boldsymbol{\theta}_1)
&=& \frac{1+\alpha}{n\lambda_n\theta_{3,p+1}^{\alpha+1}}
\sum_{i=1}^n \psi_{1,\alpha}({r}_i(\boldsymbol{\theta}_3))\boldsymbol{x}_{2,i}^T\boldsymbol{\theta}_{1c}
- \sum_{j\notin \widehat{\mathcal{S}}}p_\lambda'(|\theta_{3,j}|)|\theta_{1,j}|,~~~~~~
\label{EQLProp1_pf2}
\end{eqnarray}
where $\boldsymbol{\theta}_{1c}$ is the partition of $\boldsymbol{\theta}_1$ having components 
only with indices in $\widehat{\mathcal{S}}^c$. \\
Note that $\boldsymbol{\theta}_{1c}\neq \boldsymbol{0}$ 
since $\boldsymbol{\theta}_1 \in \mathcal{N}_1\backslash\mathcal{N}_0$. 

But, by Assumption (P), $p_\lambda'(t)$ is continuous and decreasing (since $p_\lambda$ concave) on $(0,\infty)$.
So, Condition (4.4) implies the existence of a $\delta>0$ such that 
\begin{eqnarray}
\left|\left|\frac{1+\alpha}{n\lambda_n{\sigma}^{\alpha+1}}
\sum_{i=1}^n \psi_{1,\alpha}({r}_i({\boldsymbol{\theta}}))\boldsymbol{x}_{2,i}\right|\right|_{\infty}
< \frac{p_\lambda'(\delta)}{\lambda}, ~~~\mbox{ for all }~\boldsymbol{\theta}\in B^\ast_\delta,
\label{EQLProp1_pf3}
\end{eqnarray}
where $B^\ast_\delta = \left\{\boldsymbol{\theta}: ||\boldsymbol{\theta}-\widehat{\boldsymbol{\theta}}||<\delta \right\}$.
Now, with further shrinking if needed, we can assume that $\mathcal{N}_1\subset B^\ast_\delta$
so that $|\theta_{3,j}|\leq |\theta_{1,j}|<\delta$ for $j\notin \widehat{\mathcal{S}}$. 
Since $\boldsymbol{\theta}_3\in \mathcal{N}_1$ and $p_\lambda'(t)$ is monotone, 
using (\ref{EQLProp1_pf3}) in (\ref{EQLProp1_pf2}), we get 
 \begin{eqnarray}
 Q_{n,\lambda}^{(\alpha)}(\boldsymbol{\theta}_2) - Q_{n,\lambda}^{(\alpha)}(\boldsymbol{\theta}_1)
 &<& p_\lambda'(\delta)||\boldsymbol{\theta}_{1c}||_1 - p_\lambda'(\delta)||\boldsymbol{\theta}_{1c}||_1=0,
 \label{EQLProp1_pf4}
 \end{eqnarray}
which completes the proof.

\subsection{Proof of Theorem 4.1}
\label{SEC:Thm4.1}

Fix an $\alpha\geq 0$; for simplicity we will omit the subscript/superscript $\alpha$ in the relevant quantities, whenever clear. 
Let us denote 
\begin{eqnarray}
&&\boldsymbol{\xi}=(\xi_1, \ldots, \xi_p, \xi_{p+1})^T =\sum_{i=1}^n \psi_\alpha\left((y_i, \boldsymbol{x}_i), \boldsymbol{\theta}_0 \right)
= \boldsymbol{\Gamma}_\alpha(\boldsymbol{\beta}_{S0}, \sigma_0),
\label{EQ:Xi}\\
\mbox{and } && \boldsymbol{\eta}(\boldsymbol{\delta}) = n \left( p_{\lambda}'(|\delta_1|) \sign(\delta_1), \ldots, p_{\lambda}'(|\delta_s|)\sign (\delta_s)\right)^T, ~~~~
\nonumber
\end{eqnarray}
where  $\boldsymbol{\Gamma}_\alpha(\boldsymbol{\delta},\sigma)$ is as defined in (4.13) of the main paper
and $\boldsymbol{\delta}=(\delta_1, \ldots, \delta_s)^T\in\mathbb{R}^s$.
Also recall that ${\mathcal{S}}=\mbox{supp}(\boldsymbol{\beta}_0)=\left\{1, \ldots, s\right\}$. 
For ease of presentation of the proof, given any $(p+1)$-vector $\boldsymbol{a}=(a_1, \ldots, a_p, a_{p+1})^T$, 
we denote 
\begin{eqnarray}
\boldsymbol{a}_S =(a_1, \ldots, a_s)^T,
~~~~
\boldsymbol{a}_N =(a_{s+1}, \ldots, a_p)^T,
~~
\mbox{ and }~
\boldsymbol{a}_S^\ast =(a_1, \ldots, a_s, a_{p+1})^T.
\nonumber
\end{eqnarray} 
For example, ignoring the subscript $\alpha$, we have 
$\boldsymbol{\Gamma}_S(\boldsymbol{\delta},\sigma) = \left( {\Gamma}_{1, \alpha}(\boldsymbol{\delta},\sigma), \ldots, 
{\Gamma}_{s, \alpha}(\boldsymbol{\delta},\sigma)\right)^T$ and so on. 
Put $\boldsymbol{\eta}^\ast(\boldsymbol{\delta}) = (\boldsymbol{\eta}^T(\boldsymbol{\delta}), 0)^T$.

\bigskip
Now, in order to proof the theorem, we consider the events
\begin{eqnarray}
\mathcal{E}_1 &=& \left\{ ||\boldsymbol{\xi}_S||_{\infty} \leq \sqrt{\frac{n\log n}{c_1}} \right\},
\nonumber\\
\mathcal{E}_2 &=& \left\{ ||\boldsymbol{\xi}_N||_{\infty} \leq u_n \sqrt{n} \right\},\nonumber\\
\mathcal{E}_3 &=& \left\{ |{\xi}_{p+1}| \leq \sqrt{\frac{n\log n}{c_1}} \right\},\nonumber
\end{eqnarray}
where $u_n=c_1^{-1/2}n^{1/2-\tau^\ast}(\log n)^{1/2}$, and $\tau^\ast$ and $c_1$ are as defined in Assumptions (A3) and (A4), respectively. Note that, by Assumption (A4) along with Bonferroni's inequality, we see that
\begin{eqnarray}
P(\mathcal{E}_1\cap \mathcal{E}_2\cap \mathcal{E}_3) 
&\geq& 1- \sum_{j\in\mathcal{S}\cup\{p+1\}} P\left(|\xi_j|> \sqrt{\frac{n\log n}{c_1}}\right)
- \sum_{j\in\mathcal{S}^c} P\left(|\xi_j|> u_n \sqrt{n} \right)\nonumber\\
&\geq& 1 - 2\left[(s+1)n^{-1} + (p-s)e^{-c_1u_n^2}\right]
\nonumber\\
&=& 1 - 2\left[(s+1)n^{-1} + (p-s)e^{-n^{1-2\tau^\ast}\log n}\right]\nonumber
 \label{EQ:THM1_pf1}
\end{eqnarray}
We will now show the existence of $\widehat{\boldsymbol{\theta}}$, 
under the event $\mathcal{E}:=\mathcal{E}_1\cap \mathcal{E}_2\cap \mathcal{E}_3$, 
with the required properties as a minimizer of $Q_{n,\lambda}^{(\alpha)}(\boldsymbol{\theta})$, 
or equivalently as a solution to the Conditions (4.3)--(4.6) of Proposition 1 in the main paper. 

Let us first consider Equations (4.3) and (4.5)  and show that they posses 
a simultaneous solution $(\widehat{\boldsymbol{\beta}}_1, \widehat{\sigma})$ under the event $\mathcal{E}$.
For this purpose, note that, Equations (4.3) and (4.5) are equivalent to 
$\boldsymbol{\Psi}(\widehat{\boldsymbol{\beta}}_1, \widehat{\sigma})=\boldsymbol{0}_{s+1}$, where
\begin{eqnarray}
\boldsymbol{\Psi}(\boldsymbol{\delta}, \sigma) = \boldsymbol{\Gamma}_S^\ast(\boldsymbol{\delta}, \sigma) + \boldsymbol{\eta}^\ast(\boldsymbol{\delta})
= \boldsymbol{\Gamma}_S^\ast(\boldsymbol{\delta}, \sigma) - \boldsymbol{\Gamma}_S^\ast(\boldsymbol{\beta}_{S0}, \sigma_0) 
+ (\boldsymbol{\xi}_S^\ast +\boldsymbol{\eta}^\ast(\boldsymbol{\delta})).  
\label{EQ:Psi}
\end{eqnarray}
Now, consider the hypercube
$$
\mathcal{N} = \left\{(\boldsymbol{\delta}, \sigma)\in \mathbb{R}^s\times\mathbb{R}^{+}
~:~ ||\boldsymbol{\delta}-\boldsymbol{\beta}_{S0}||_\infty =n^{-\tau}\log n, |\sigma - \sigma_0| = n^{-\tau}\log n \right\},
$$
and take any arbitrary $\boldsymbol{\delta}=(\delta_1, \ldots, \delta_s)$ and $\sigma$ in $\mathcal{N}$.
Since $d_n \geq n^{-\tau}\log n$, we have 
\begin{eqnarray}
\min_{1\leq j\leq s}|\delta_j| \geq \min_{1\leq j\leq s} |\beta_{0,j}| -d_n = d_n,
~~\mbox{and}~sign(\boldsymbol{\delta})=sign(\boldsymbol{\beta}_{S0}).
\label{EQ:THM1_pf2}
\end{eqnarray}
This is because for any $j=1, \ldots, s$, we have 
$$
\left||\delta_j|-|\beta_{0,j}|\right| \leq |\delta_j - \beta_{0,j}| \leq
||\boldsymbol{\delta}-\boldsymbol{\beta}_{S0}||_\infty =n^{-\tau}\log n \leq d_n.
$$ 
Also,  by Assumption (P) and (\ref{EQ:THM1_pf2}), we get  $||\boldsymbol{\eta}(\boldsymbol{\delta})||_{\infty} \leq np_\lambda'(d_n)$.
Therefore, under $\mathcal{E}$, we get 
\begin{eqnarray}
||\boldsymbol{\xi}_{S}^\ast + \boldsymbol{\eta}^\ast(\boldsymbol{\delta})||_{\infty} \leq \sqrt{\frac{n\log n}{c_1}} + np_\lambda'(d_n).
\label{EQ:THM1_pf3}
\end{eqnarray}

Next, to handle the first two terms in $\boldsymbol{\Psi}(\boldsymbol{\delta}, \sigma)$, 
we consider a second Taylor series expansion of $\boldsymbol{\Gamma}_S^\ast(\boldsymbol{\delta}, \sigma)$
around $(\boldsymbol{\beta}_{S0}, \sigma_0)$ to get 
\begin{eqnarray}
 \boldsymbol{\Gamma}_S^\ast(\boldsymbol{\delta}, \sigma)=
\boldsymbol{\Gamma}_S^\ast(\boldsymbol{\beta}_{S0}, \sigma_0) 
+ \left(\mathbf{X}_S^{\ast T}\boldsymbol{\Sigma}_{\alpha}(\boldsymbol{\theta}_0)\mathbf{X}_S^\ast\right)
\left[(\boldsymbol{\delta}, \sigma)-(\boldsymbol{\beta}_{S0}, \sigma_0)\right]+\boldsymbol{r},  
\label{EQ:THM1_pf4}
\end{eqnarray}
where $\boldsymbol{r}=(r_1, \ldots, r_s, r_{s+1})^T$ with 
$$r_j = \frac{1}{2}\left[(\boldsymbol{\delta}, \sigma)-(\boldsymbol{\beta}_{S0}, \sigma_0)\right]^T
\nabla^2\Gamma_j(\widetilde{\boldsymbol{\delta}}, \widetilde{\sigma})
\left[(\boldsymbol{\delta}, \sigma)-(\boldsymbol{\beta}_{S0}, \sigma_0)\right],
$$
and $(\widetilde{\boldsymbol{\delta}}, \widetilde{\sigma})$ is some element in the line segment 
joining $(\boldsymbol{\delta}, \sigma)$ and $(\boldsymbol{\beta}_{S0}, \sigma_0)$.
But, Assumption (A3) implies that 
$$
||\boldsymbol{r}||_{\infty} \leq O\left(sn^{1-2\tau}(\log n)^2\right).
$$
But, by Assumption (A2) and (\ref{EQ:THM1_pf3}), for any $(\boldsymbol{\delta}, \sigma)\in \mathcal{N}$,  we get
\begin{eqnarray}
\left|\left|\left(\mathbf{X}_S^{\ast T}\boldsymbol{\Sigma}_{\alpha}(\boldsymbol{\theta}_0)\mathbf{X}_S^\ast\right)^{-1} 
\left[\boldsymbol{\xi}_S^\ast+\boldsymbol{\eta}^\ast(\boldsymbol{\delta})+\boldsymbol{r}\right]\right|\right|_{\infty} 
&\leq & \left|\left| \left(\mathbf{X}_S^{\ast T}\boldsymbol{\Sigma}_{\alpha}(\boldsymbol{\theta}_0)\mathbf{X}_S^\ast\right)^{-1}
\right|\right|_{\infty} \left[\left|\left|\boldsymbol{\xi}_S^\ast
+\boldsymbol{\eta}^\ast(\boldsymbol{\delta})\right|\right|_{\infty} + ||\boldsymbol{r}||_\infty\right]\nonumber\\
&=& O\left(b_s\sqrt{\frac{\log n}{n}} + b_s p_\lambda'(d_n) + b_s s n^{-2\tau}(\log n)^2\right)
=O\left(\frac{\log n}{n^{\tau}}\right),
\label{EQ:THM1_pf6}
\end{eqnarray}
where the last equality follows by Assumption (A3).
Combining (\ref{EQ:THM1_pf4}) and (\ref{EQ:THM1_pf6}), we get 
\begin{eqnarray}
\left(\mathbf{X}_S^{\ast T}\boldsymbol{\Sigma}_{\alpha}(\boldsymbol{\theta}_0)\mathbf{X}_S^\ast\right)^{-1}
\boldsymbol{\Psi}(\boldsymbol{\delta}, \sigma)
= \boldsymbol{k}+\boldsymbol{u},  
\label{EQ:THM1_pf5}
\end{eqnarray}
with $\boldsymbol{k}=(k_1, \ldots, k_s, k_{s+1})^T=\left[(\boldsymbol{\delta}, \sigma)-(\boldsymbol{\beta}_{S0}, \sigma_0)\right]$,  
and $||\boldsymbol{u}||_{\infty}=O\left({n^{-\tau}}{\log n}\right)$.

Therefore, denoting the left-hand side of (\ref{EQ:THM1_pf5}) as the vector 
$(\widetilde{{\Psi}}_1(\boldsymbol{\delta}, \sigma),\ldots, \widetilde{{\Psi}}_{s+1}(\boldsymbol{\delta}, \sigma))^T$, 
we have, for each $j=1, \ldots, s+1$,
\begin{eqnarray}
\widetilde{{\Psi}}_j(\boldsymbol{\delta}, \sigma)
&\geq& ~~n^{\tau}\sqrt{\log n} - ||\boldsymbol{u}||_{\infty} \geq 0, ~~~\mbox{ if }~{k}_j= n^{\tau}\sqrt{\log n},
\nonumber\\
\widetilde{{\Psi}}_j(\boldsymbol{\delta}, \sigma)
&\leq& -n^{\tau}\sqrt{\log n} + ||\boldsymbol{u}||_{\infty} \leq 0, ~~~\mbox{ if }~ {k}_j= -n^{\tau}\sqrt{\log n},
\nonumber
\end{eqnarray}
for sufficiently large $n$. Hence, by continuity, there is a solution $(\widehat{\boldsymbol{\beta}}_1, \widehat{\sigma})\in \mathcal{N}$
of the equation $\left(\mathbf{X}_S^{\ast T}\boldsymbol{\Sigma}_{\alpha}(\boldsymbol{\theta}_0)\mathbf{X}_S^\ast\right)^{-1}
\boldsymbol{\Psi}(\boldsymbol{\delta}, \sigma) = \boldsymbol{0}_{s+1}$, 
or equivalently of $\boldsymbol{\Psi}(\boldsymbol{\delta}, \sigma) = \boldsymbol{0}_{s+1}$.

Finally, using the above $(\widehat{\boldsymbol{\beta}}, \widehat{\sigma})\in \mathcal{N}$,
let us define the required solutions 
$\widehat{\boldsymbol{\theta}}=(\widehat{\boldsymbol{\beta}}^T, \widehat{\sigma})^T$
with $\widehat{\boldsymbol{\beta}}_S = \widehat{\boldsymbol{\beta}}_1$ 
and $\widehat{\boldsymbol{\beta}}_N = {\boldsymbol{0}}_{p-s}$. 
Note that, Assumption (A3) (in particular, Equation (4.15) of the main paper) 
ensures that this $\widehat{\boldsymbol{\theta}}$ satisfies Condition (4.6) of the main paper.
Thus, it remains only to show that it also satisfies Condition (4.4) and we are done.

To this end, we note that
\begin{eqnarray}
\boldsymbol{z} &:=& \frac{1+\alpha}{n\lambda_n\widehat{\sigma}^{\alpha+1}}
\sum_{i=1}^n \psi_{1,\alpha}({r}_i(\widehat{\boldsymbol{\theta}}))\boldsymbol{x}_{2,i}
=  \frac{1}{n\lambda_n}\left[ \boldsymbol{\xi}_{N}
+ \boldsymbol{\Gamma}_{N}(\widehat{\boldsymbol{\beta}}_S, \widehat{\sigma}) 
- \boldsymbol{\Gamma}_{N}(\boldsymbol{\beta}_{S0}, \sigma_0) \right].
\label{EQ:THM1_pf7}
\end{eqnarray}
But, by Assumption (A3), on the event $\mathcal{E}_2$, we have 
$$
\left|\left|\frac{1}{n\lambda_n}\boldsymbol{\xi}_{N}\right|\right|_{\infty} \leq 
O\left(u_n n^{-1/2}/\lambda_n\right) = o(1).
$$
Next,  a second  Taylor series expansion of $\boldsymbol{\Gamma}_N(\boldsymbol{\delta}, \sigma)$
around $(\boldsymbol{\beta}_{S0}, \sigma_0)$ yields
\begin{eqnarray}
\boldsymbol{\Gamma}_N(\widehat{\boldsymbol{\beta}}_S, \widehat{\sigma})=
\boldsymbol{\Gamma}_N(\boldsymbol{\beta}_{S0}, \sigma_0) 
+ \left(\mathbf{X}_N^{\ast T}\boldsymbol{\Sigma}_{\alpha}(\boldsymbol{\theta}_0)\mathbf{X}_S^\ast\right)
\left[(\widehat{\boldsymbol{\beta}}_S, \widehat{\sigma})-(\boldsymbol{\beta}_{S0}, \sigma_0)\right]
+\boldsymbol{w},  
\label{EQ:THM1_pf8}
\end{eqnarray}
where $\boldsymbol{w}=(w_{s+1}, \ldots, w_p)^T$ with 
$w_j = \frac{1}{2}\left[(\widehat{\boldsymbol{\beta}}_S, \widehat{\sigma})-(\boldsymbol{\beta}_{S0}, \sigma_0)\right]^T
\nabla^2\Gamma_j(\widetilde{\boldsymbol{\delta}}^\ast, \widetilde{\sigma}^\ast)
\left[(\widehat{\boldsymbol{\beta}}_S, \widehat{\sigma})-(\boldsymbol{\beta}_{S0}, \sigma_0)\right]$
and $(\widetilde{\boldsymbol{\delta}}^\ast, \widetilde{\sigma}^\ast)$ is some element in the line segment 
joining $(\widehat{\boldsymbol{\beta}}_S, \widehat{\sigma})$ and $(\boldsymbol{\beta}_{S0}, \sigma_0)$.
Noting that $\widehat{\boldsymbol{\beta}}_S\in \mathcal{N}$,
as before, Assumption (A3) leads to $||\boldsymbol{w}||_{\infty} \leq O\left(sn^{1-2\tau}(\log n)^2\right)$.
Further, since $(\widehat{\boldsymbol{\beta}}_S, \widehat{\sigma})$ satisfies 
 the equation  $\boldsymbol{\Psi}(\boldsymbol{\delta}, \sigma)=\boldsymbol{0}_{s+1}$,
 we have from (\ref{EQ:THM1_pf5}) that 
 \begin{eqnarray}
\left[(\widehat{\boldsymbol{\beta}}, \widehat{\sigma})-(\boldsymbol{\beta}_{S0}, \sigma_0)\right]
= -\left(\mathbf{X}_S^{\ast T}\boldsymbol{\Sigma}_{\alpha}(\boldsymbol{\theta}_0)\mathbf{X}_S^\ast\right)^{-1} 
\left[\boldsymbol{\xi}_S^\ast+\boldsymbol{\eta}^\ast(\boldsymbol{\delta})+\boldsymbol{r}\right].  
 \label{EQ:THM1_pf9}
 \end{eqnarray}
 Combining (\ref{EQ:THM1_pf7})--(\ref{EQ:THM1_pf9}) and using Assumption (A2), we finally get 
 \begin{eqnarray}
 ||\boldsymbol{z}||_{\infty} &\leq& o(1) 
 + \frac{1}{n\lambda_n}\left|\left| \boldsymbol{\Gamma}_{N}(\widehat{\boldsymbol{\beta}}_S, \widehat{\sigma}) 
 - \boldsymbol{\Gamma}_{N}(\boldsymbol{\beta}_{S0}, \sigma_0) \right|\right|_\infty
 \nonumber\\
 &\leq& o(1)  + \frac{1}{n\lambda_n}
 \left|\left| \left(\mathbf{X}_N^{\ast T}\boldsymbol{\Sigma}_{\alpha}(\boldsymbol{\theta}_0)\mathbf{X}_S^\ast\right)
 \left(\mathbf{X}_S^{\ast T}\boldsymbol{\Sigma}_{\alpha}(\boldsymbol{\theta}_0)\mathbf{X}_S^\ast\right)^{-1} \right|\right|_\infty
 \left(||\boldsymbol{\xi}_S^\ast+\boldsymbol{\eta}^\ast(\boldsymbol{\delta})||_\infty + ||\boldsymbol{r}||_{\infty} \right)
\nonumber\\
&&~~~~~~~~~~~~~~ +\frac{1}{n\lambda_n}||\boldsymbol{w}||_{\infty}
 \nonumber\\
 &\leq& o(1)  + \frac{1}{n\lambda_n}O\left( n^\tau_1\sqrt{n\log n}+ s n^{1-2\tau+\tau_1}(\log n)^2 + s n^{1-2\tau}(\log n)^2\right)
 \nonumber\\
 && ~~~~~~~~~~~~~~~~~~
+\left|\left| \left(\mathbf{X}_N^{\ast T}\boldsymbol{\Sigma}_{\alpha}(\boldsymbol{\theta}_0)\mathbf{X}_S^\ast\right)
\left(\mathbf{X}_S^{\ast T}\boldsymbol{\Sigma}_{\alpha}(\boldsymbol{\theta}_0)\mathbf{X}_S^\ast\right)^{-1} \right|\right|_\infty
\frac{p_{\lambda_n}'(d_n)}{\lambda_n}\nonumber\\
 &\leq& C\rho(p_\lambda) + o(1)\leq  \rho(p_\lambda),
\label{EQ:THM1_pf10}
\end{eqnarray}
for sufficiently large $n$.
Hence Condition (4.4) is satisfied, and this completes the proof.
 
\subsection{Proof of Theorem 4.2}
\label{SEC:Thm4.2}

We will proceed as in the proof of Theorem 4.1 with the same notation (described at the beginning of Section \ref{SEC:Thm4.1}). 
Let us first consider the $(s+1)$-dimensional subspace 
$\left\{(\boldsymbol{\beta}, \sigma)\in \mathbb{R}^p\times \mathbb{R}^{+} : \boldsymbol{\beta}_N = \boldsymbol{0} \right\}$
and the constrained objective function given by
\begin{eqnarray}
\overline{Q_{n,\lambda}^{\alpha}}(\boldsymbol{\delta}, \sigma) &=& 
 \overline{L_{n}^{\alpha}}(\boldsymbol{\delta}, \sigma) + \sum_{j=1}^{p} p_{\lambda}(|\delta_j|),
 \label{EQ:THM2_pf1}
\end{eqnarray}
where $\boldsymbol{\delta}=(\delta_1, \ldots, \delta_s)^T$, $\boldsymbol{x}_{i,S} = (x_{i1}, \ldots, x_{is})^T$ and
\begin{eqnarray}
\overline{L_n^\alpha} (\boldsymbol{\delta}, \sigma) 
= \frac{1}{\sigma^\alpha}M_f^{(\alpha)} 
- \frac{1+\alpha}{\alpha} \frac{1}{n\sigma^\alpha} 
\sum_{i=1}^n f^\alpha\left(\frac{y_i - \boldsymbol{x}_{i,S}^T\boldsymbol{\delta}}{\sigma}\right) + \frac{1}{\alpha}.
\label{EQ:THM2_pf2}
\end{eqnarray}
Let us define the closed set
$$
\mathcal{N}_r = \left\{ (\boldsymbol{\delta}, \sigma)\in \mathbb{R}^s\times \mathbb{R}^{+} : 
||\boldsymbol{\delta} - \beta_{S0}|| \leq r\sqrt{\frac{s}{n}}, |\sigma-\sigma_0|\leq \frac{r}{\sqrt{n}} \right\},
$$
for some $r \in (0,\infty)$ and the event 
$$
\mathcal{E}_n = \left\{ \overline{Q_{n,\lambda}^{\alpha}}(\boldsymbol{\beta}_{S0}, \sigma_0) 
< \min \limits_{(\boldsymbol{\delta}, \sigma)\in \partial\mathcal{N}_r} \overline{Q_{n,\lambda}^{\alpha}}(\boldsymbol{\delta}, \sigma) \right\},
$$
where $\partial\mathcal{N}_r$ denote the boundary $\mathcal{N}_r$.
Then, by definition, a local minimizer of $\overline{Q_{n,\lambda}^{\alpha}}(\boldsymbol{\delta}, \sigma)$ exists, 
say $(\widehat{\boldsymbol{\beta}}_1, \widehat{\sigma})$, in $\mathcal{N}_r$ on the event $\mathcal{E}_n$,
such that $||\widehat{\boldsymbol{\beta} }_1 - \boldsymbol{\beta}_{S0}|| = O(\sqrt{s/n})$ and 
$||\widehat{\sigma}-\sigma_0||=O(n^{-1/2})$.
So, we need to show that $P(\mathcal{E}_n)\rightarrow 1$ as $n\rightarrow\infty$
to prove the above consistency results with probability tending to one.

Now, let us take $n$ sufficiently large so that $r\sqrt{s/n}  \leq d_n$;
this is possible since $d_n \gg \sqrt{s/n}$ by Assumption (A3)$^\ast$.
Then, for any $\boldsymbol{\delta}=(\delta_1, \ldots, \delta_s)^T\in \mathcal{N}_r$, 
we have 
$$sign(\boldsymbol{\delta}) = sign(\boldsymbol{\beta}_{S0}),
~~~ 
||\boldsymbol{\delta}-\boldsymbol{\beta}_{S0}||_\infty \leq d_n,
~~~
|\sigma-\sigma_{0}| \leq d_n,
~~
\mbox{ and } \min_j |\delta_j| \geq d_n.
$$ 
Now, a Taylor theorem application yields, for any $(\boldsymbol{\delta}, \sigma)\in \mathcal{N}_r$,
\begin{eqnarray}
\overline{Q_{n,\lambda}^{\alpha}}(\boldsymbol{\delta}, \sigma)
- \overline{Q_{n,\lambda}^{\alpha}}(\boldsymbol{\beta}_{S0}, \sigma_0)
&=& \left[(\boldsymbol{\delta}, \sigma) - (\boldsymbol{\beta}_{S0}, \sigma_0)\right]\frac{1}{n}\boldsymbol{\Psi}(\boldsymbol{\delta}, \sigma)
\nonumber\\
&& +\frac{1}{2}\left[(\boldsymbol{\delta}, \sigma) - (\boldsymbol{\beta}_{S0}, \sigma_0)\right]^T\boldsymbol{D}
\left[(\boldsymbol{\delta}, \sigma) - (\boldsymbol{\beta}_{S0}, \sigma_0)\right],~~~~~~
\label{EQ:THM2_pf3}
\end{eqnarray}
where $\boldsymbol{\Psi}(\boldsymbol{\delta}, \sigma)$ is as defined in (\ref{EQ:Psi})
and 
and $\boldsymbol{D}=\frac{1}{n}
\left(\mathbf{X}_S^{\ast T}\boldsymbol{\Sigma}_{\alpha}(\widetilde{\boldsymbol{\delta}}, \widetilde{\sigma})\mathbf{X}_S^\ast\right)
+\boldsymbol{P}_\lambda^{\ast\ast}(\widetilde{\boldsymbol{\delta}}, 0)$,
with $(\widetilde{\boldsymbol{\delta}}, \widetilde{\sigma})$ being an element on the line segment joining 
$(\boldsymbol{\delta}, \sigma)$ and $(\boldsymbol{\beta}_{S0}, \sigma_0)$
and $\boldsymbol{P}_\lambda^{\ast\ast}$ being as defined in Section 3.2 of the main paper. 
But, by definition of $\mathcal{N}_0$, Assumption (A2)$^\ast$ implies that, for any  $(\boldsymbol{\delta}, \sigma)\in \mathcal{N}_r$,
$$
||\boldsymbol{\delta}-\boldsymbol{\beta}_{S0}||= r\sqrt{s/n},
~~~~|\sigma-\sigma_0|=n^{-1/2}r 
~~~~
\mbox{and }
(\widetilde{\boldsymbol{\delta}}, \widetilde{\sigma})\in \mathcal{N}_0.
$$
Then, by Condition (4.14) in Assumption (A2)$^\ast$ and Assumption (A3), we have
$$
\Lambda_{\min}(\boldsymbol{D}) \geq c - \max\limits_{(\boldsymbol{\delta},\sigma)\in \mathcal{N}_0} \zeta(P_{\lambda_n}; \boldsymbol{\delta})
\geq c/2,
$$
and hence, from (\ref{EQ:THM2_pf3}), we get
\begin{eqnarray}
\min\limits_{(\boldsymbol{\delta}, \sigma)\in \partial\mathcal{N}_r}
\overline{Q_{n,\lambda}^{\alpha}}(\boldsymbol{\delta}, \sigma)
- \overline{Q_{n,\lambda}^{\alpha}}(\boldsymbol{\beta}_{S0}, \sigma_0)
\geq   r\sqrt{s/n} \left(-||\boldsymbol{d}_1|| + r\frac{c}{4}\sqrt{s/n}\right)
+ r n^{-1/2} \left(-|{d}_2| + r\frac{c}{4}n^{-1/2}\right),\nonumber
\end{eqnarray}
where we denote the first $s$ elements of $\boldsymbol{\Psi}(\boldsymbol{\delta}, \sigma)$  as $\boldsymbol{d}_1$ and the last $(s+1)$-th element as  $d_2$.
Therefore, the probability of the event $\mathcal{E}_n$ is bounded below by
\begin{eqnarray}
P(\mathcal{E}_n) &\geq&  P\left(\left\{||\boldsymbol{d}_1||^2 < \frac{c^2sr^2}{16n}\right\}
\cap\left\{|{d}_2| < \frac{c^2r^2}{16n}~\and~\right\}|\right)\nonumber\\
&\geq&  1 - P\left(\left\{||\boldsymbol{d}_1||^2 \geq \frac{c^2sr^2}{16n}\right\}\right)
-P\left(\left\{|{d}_2| < \frac{c^2r^2}{16n}~\and~\right\}\right)\nonumber\\
&\geq&  1 -  \frac{16nE||\boldsymbol{d}_1||^2}{c^2sr^2} - \frac{16nE|{d}_2|^2}{c^2r^2},
\label{EQ:THM2_pf5}
\end{eqnarray}
where the last step follows by Markov's inequality.
But, by Assumption (A2)$^\ast$, we get 
\begin{eqnarray}
P(\mathcal{E}_n) &\geq&  1 -  O(r^{-2}) - O(r^{-2}) = 1- O(r^{-2}),
\nonumber
\end{eqnarray}
which indicates that 
$||\widehat{\boldsymbol{\beta} }_1 - \boldsymbol{\beta}_{S0}|| = O_P(\sqrt{s/n})$ 
and $||\widehat{\sigma}-\sigma_0||=O_P(n^{-1/2})$.

Now, with the above minimizers $(\widehat{\boldsymbol{\beta} }_1, \widehat{\sigma})$,
let us define the vector $\widehat{\boldsymbol{\beta}}=(\widehat{\boldsymbol{\beta}}_S^T, \widehat{\boldsymbol{\beta}}_N^T)^T$
with $\widehat{\boldsymbol{\beta} }_S=\widehat{\boldsymbol{\beta} }_1$ and $\widehat{\boldsymbol{\beta}}_N = \boldsymbol{0}_{p-s}$.
We claim that this $(\widehat{\boldsymbol{\beta}}, \widehat{\sigma})$ is indeed the required 
strict minimizer of $Q_{n,\lambda}^\alpha(\boldsymbol{\theta})$ over the whole parameter space. 
But, as in the proof of Proposition 4.1, it is enough only to show that Condition (4.4) is satisfied by this solution;
we can prove this in a similar fashion as we have done to prove Theorem 4.1.

Recalling the definition of $\boldsymbol{\xi}$ from Section \ref{SEC:Thm4.1}, we consider the events 
\begin{eqnarray}
\mathcal{E}_2 &=& \left\{ ||\boldsymbol{\xi}_N||_{\infty} \leq u_n \sqrt{n} \right\},\nonumber
\end{eqnarray}
with $u_n=c_1^{-1/2}n^{\tau^\ast/2}(\log n)^{1/2}$.
As in the proof of Theorem 4.1, using $\log p = O(n^{\tau^\ast})$, we can again show that 
$$
P(\mathcal{E}_2) \leq 1- 2pe^{-c_1u_n^2} \rightarrow 1, ~~~\mbox{as }n \rightarrow\infty.
$$
But, as in the derivation of (\ref{EQ:THM1_pf10}), on the event $\mathcal{E}_2$, we have
\begin{eqnarray}
&& \left|\left|\frac{1+\alpha}{n\lambda_n\widehat{\sigma}^{\alpha+1}}
\sum_{i=1}^n \psi_{1,\alpha}({r}_i(\widehat{\boldsymbol{\theta}}))\boldsymbol{x}_{2,i}\right|\right|_{\infty} \nonumber\\
&\leq& o(1)  + \frac{1}{n\lambda_n}
\left|\left| \left(\mathbf{X}_N^{\ast T}\boldsymbol{\Sigma}_{\alpha}(\boldsymbol{\theta}_0)\mathbf{X}_S^\ast\right)
\left[(\widehat{\boldsymbol{\beta}}_S, \widehat{\sigma})-(\boldsymbol{\beta}_{S0}, \sigma_0)\right] \right|\right|_\infty
+\frac{1}{n\lambda_n}||\boldsymbol{w}||_{\infty}
\nonumber\\
&=& o(1)  + \frac{O(n)}{n\lambda_n}
\left|\left|(\widehat{\boldsymbol{\beta}}_S, \widehat{\sigma})-(\boldsymbol{\beta}_{S0}, \sigma_0) \right|\right|_2
+\frac{O(n)}{n\lambda_n}
\left|\left|(\widehat{\boldsymbol{\beta}}_S, \widehat{\sigma})-(\boldsymbol{\beta}_{S0}, \sigma_0) \right|\right|_2
\nonumber\\
&\leq& o(1)  + \frac{r(\sqrt{s}+1)}{\lambda_n\sqrt{n}} = o(1),
~~~\mbox{ for sufficiently large }n,
\nonumber
\end{eqnarray}
which  completes the proof.

\subsection{Proof of Theorem 4.3}
\label{SEC:Thm4.3}

We continue to use the notation from sections \ref{SEC:Thm4.1} and \ref{SEC:Thm4.2}.
Consider the event $\mathcal{E}_n$ as defined in the proof of Theorem 4.2 in Section \ref{SEC:Thm4.2}
and the strict minimizers $\widehat{\boldsymbol{\beta}}=(\widehat{\boldsymbol{\beta}}_S^T, \widehat{\boldsymbol{\beta}}_N^T)^T$ 
and $\widehat{\sigma}$ of $Q_{n,\lambda}^\alpha(\boldsymbol{\beta}, \sigma)$ obtained there;
recall that $(\widehat{\boldsymbol{\beta} }_S, \widehat{\sigma})\in \mathcal{N}_r \subset \mathcal{N}_0$ 
is in fact a strict minimizer of $\overline{Q_{n,\lambda}^\alpha}(\boldsymbol{\delta}, \sigma)$
and $\widehat{\boldsymbol{\beta}}_N = \boldsymbol{0}_{p-s}$.
In view of Theorem 4.2, such a solution exists  on the event $\mathcal{E}_n$ having probability tending to one
and so we only need to prove the asymptotic normality of $(\widehat{\boldsymbol{\beta} }_S, \widehat{\sigma})$.

Note that, the first order condition of minimization along with definition of 
$\overline{Q_{n,\lambda}^\alpha}(\boldsymbol{\delta}, \sigma)$ yields 
$$
\nabla \overline{Q_{n,\lambda}^\alpha}(\widehat{\boldsymbol{\beta} }_S, \widehat{\sigma}) = 
 \nabla \overline{L_{n}^{\alpha}}(\widehat{\boldsymbol{\beta} }_S, \widehat{\sigma}) 
+ n^{-1}\boldsymbol{\eta}(\widehat{\boldsymbol{\beta} }_S) 
 = \boldsymbol{0},
$$
where $\boldsymbol{\eta}(\cdot)$ is as defined at the beginning of Section \ref{SEC:Thm4.1}. 
Now, using a Taylor series expansion of $\overline{L_{n}^\alpha}(\boldsymbol{\delta}, \sigma)$ around $\boldsymbol{\beta}_{S0}$
and Assumption (A2)$^\ast$, we get that
\begin{eqnarray}
\boldsymbol{0} &=& \nabla \overline{Q_{n,\lambda}^\alpha}(\widehat{\boldsymbol{\beta} }_S, \widehat{\sigma}) = 
\nabla \overline{L_{n}^{\alpha}}(\widehat{\boldsymbol{\beta} }_S, \widehat{\sigma}) + n^{-1}\boldsymbol{\eta}(\widehat{\boldsymbol{\beta} }_S) 
\nonumber\\
&=& \nabla \overline{L_{n}^{\alpha}}({\boldsymbol{\beta} }_{S0}, {\sigma}_0) + \frac{1}{n}
\left(\mathbf{X}_S^{\ast T}\boldsymbol{\Sigma}_{\alpha}(\boldsymbol{\theta}_0)\mathbf{X}_S^\ast\right)
\left[(\widehat{\boldsymbol{\beta}}_S, \widehat{\sigma})-(\boldsymbol{\beta}_{S0}, \sigma_0)\right]\nonumber\\
&&~~~~~~~~~~~~~~+O(1)\sqrt{s}\left|\left|(\widehat{\boldsymbol{\beta}}_S, \widehat{\sigma})-(\boldsymbol{\beta}_{S0}, \sigma_0) \right|\right|_2^2
+ \frac{1}{n}\boldsymbol{\eta}(\widehat{\boldsymbol{\beta} }_S) 
\nonumber\\
&=& \frac{1}{n}\sum_{i=1}^n\psi_\alpha((y_i, \boldsymbol{x}_i), \boldsymbol{\theta}_0) + \frac{1}{n}
\left(\mathbf{X}_S^{\ast T}\boldsymbol{\Sigma}_{\alpha}(\boldsymbol{\theta}_0)\mathbf{X}_S^\ast\right)
\left[(\widehat{\boldsymbol{\beta}}_S, \widehat{\sigma})-(\boldsymbol{\beta}_{S0}, \sigma_0)\right]
\nonumber\\
&&~~~~~~~~~~~~~~+O_p(s^{3/2}/n) + \frac{1}{n}\boldsymbol{\eta}(\widehat{\boldsymbol{\beta} }_S) 
\label{EQ:THM3_pf1}
\end{eqnarray}
But, since $\widehat{\boldsymbol{\beta} }_S\in \mathcal{N}_0$ and $p_{\lambda_n}'(d_n) = o((sn)^{-1/2})$ via Assumption (A5),
we get by the monotonicity of $p_{\lambda}(\cdot)$ that 
$$
\left|\left|n^{-1}\boldsymbol{\eta}(\widehat{\boldsymbol{\beta} }_S) \right|\right|_2
\leq \sqrt{s} p_{\lambda}'(d_n) = o_P(n^{-1/2}).
$$
Therefore, noting that $s=o(n^{1/3})$, we get from (\ref{EQ:THM3_pf1}) that 
\begin{eqnarray}
\left(\mathbf{X}_S^{\ast T}\boldsymbol{\Sigma}_{\alpha}(\boldsymbol{\theta}_0)\mathbf{X}_S^\ast\right)
\left[(\widehat{\boldsymbol{\beta}}_S, \widehat{\sigma})-(\boldsymbol{\beta}_{S0}, \sigma_0)\right]
 &=& - \sum_{i=1}^n\psi_\alpha((y_i, \boldsymbol{x}_i), \boldsymbol{\theta}_0) + 
+o_p(n^{1/2}),\nonumber
\end{eqnarray}
or equivalently, by Assumption (A5),  
\begin{eqnarray}
\boldsymbol{A}_n\boldsymbol{S}_{2,n}^{-1/2} \boldsymbol{S}_{1,n} 
\left[(\widehat{\boldsymbol{\beta}}_S, \widehat{\sigma})-(\boldsymbol{\beta}_{S0}, \sigma_0)\right]
&=& - \boldsymbol{A}_n\boldsymbol{S}_{2,n}^{-1/2}\sum_{i=1}^n\psi_\alpha((y_i, \boldsymbol{x}_i), \boldsymbol{\theta}_0) 
+o_p(1),\nonumber
\end{eqnarray}
where $\boldsymbol{S}_{1,n} = \left(\mathbf{X}_S^{\ast T}\boldsymbol{\Sigma}_{\alpha}(\boldsymbol{\theta}_0)\mathbf{X}_S^\ast\right)$, 
$\boldsymbol{S}_{2,n} = \left(\mathbf{X}_S^{\ast T}\boldsymbol{\Sigma}_{\alpha}^\ast(\boldsymbol{\theta}_0)\mathbf{X}_S^\ast\right)$
and $\boldsymbol{A}_n$ is as given in the statement of the theorem.
Therefore, we finally need to show that 
$$
\boldsymbol{u}_n := - \boldsymbol{A}_n\boldsymbol{S}_{2,n}^{-1/2}\sum_{i=1}^n\psi_\alpha((y_i, \boldsymbol{x}_i), \boldsymbol{\theta}_0)
\mathop{\rightarrow}^\mathcal{D} N_{q}\left(\boldsymbol{0}_{q}, \boldsymbol{G}\right).
$$

To this end, take a unit vector $\boldsymbol{e}\in \mathbb{R}^q$ and note that
$\boldsymbol{e}^T\boldsymbol{u}_n = \sum_{i=1}^n \zeta_i$,
where 
$$
\zeta_i = - \boldsymbol{e}^T\boldsymbol{A}_n\boldsymbol{S}_{2,n}^{-1/2}\psi_\alpha((y_i, \boldsymbol{x}_i), \boldsymbol{\theta}_0).
$$
But, by our assumptions each $\zeta_i$ are independent with mean $0$ and
$$
\sum_{i=1}^n \mbox{var}(\zeta_i)= \boldsymbol{e}^T\boldsymbol{A}_n\boldsymbol{S}_{2,n}^{-1/2}\boldsymbol{S}_{2,n}\boldsymbol{S}_{2,n}^{-1/2}\boldsymbol{A}_n^T\boldsymbol{e}
=\boldsymbol{e}^T\boldsymbol{A}_n\boldsymbol{A}_n^T\boldsymbol{e} \rightarrow \boldsymbol{e}^T\boldsymbol{G}\boldsymbol{e},
~~\mbox{as } ~ n\rightarrow\infty.
$$
Finally, by Assumption (A5), we get 
\begin{eqnarray}
\sum_{i=1}^n E|\zeta_i|^3 &\leq& \sum_{i=1}^n \left|\boldsymbol{e}^T\boldsymbol{A}_n\boldsymbol{S}_{2,n}^{-1/2}\boldsymbol{x}_{Si}^\ast\right|^3
\max_{k=1,2}\left|\psi_{k,\alpha}((y_i, \boldsymbol{x}_i), \boldsymbol{\theta}_0)\right|^3
\nonumber\\
&=& O(1)\sum_{i=1}^n \left|\boldsymbol{e}^T\boldsymbol{A}_n\boldsymbol{S}_{2,n}^{-1/2}\boldsymbol{x}_{Si}^\ast\right|^3
\nonumber\\
&\leq& O(1)\sum_{i=1}^n \left|\left|\boldsymbol{e}^T\boldsymbol{A}_n\right|\right|^3
\left|\left|\boldsymbol{S}_{2,n}^{-1/2}\boldsymbol{x}_{Si}^\ast\right|\right|^3
~~~~\mbox{[by Cauchy-Swartz inequality]}
\nonumber\\
&\leq& O(1)\sum_{i=1}^n \left[\boldsymbol{x}_{Si}^{\ast T}\boldsymbol{S}_{2,n}^{-1}\boldsymbol{x}_{Si}^\ast\right]^{3/2} = o(1).
\nonumber
\end{eqnarray}
Therefore, by Lyapunov's central limit theorem, 
we get $\boldsymbol{e}^T\boldsymbol{u}_n \displaystyle\mathop{\rightarrow}^\mathcal{D} N\left(0, \boldsymbol{e}^T\boldsymbol{G}\boldsymbol{e}\right)$
for any unit vector $\boldsymbol{e}\in \mathbb{R}^q$. Hence, 
$\boldsymbol{u}_n \displaystyle\mathop{\rightarrow}^\mathcal{D} N_{q}\left(\boldsymbol{0}_{q}, \boldsymbol{G}\right)$,
completing the proof.

\bigskip\bigskip

\section{Additional Numerical Results}\label{SEC:sim}
Tables \ref{table:outtable_p100e1y} -- \ref{table:outtable_p200} present the simulation results, under the set-up discussed in Section 5 of the main paper, for $p=100,200$ in presence of 10\% Y-outliers, X-outliers or no outliers, respectively. The findings are very similar to those reported in the main body of the paper. 

\begin{table}[t]
	\begin{footnotesize}
		\centering		
		\begin{tabular}{r|rrrrrr}
			\hline
			\multicolumn{7}{c}{{\bf Setting A}}\\\hline
			Method          & $\text{MSEE} (\hat \bfbeta)$    & $\text{RMSPE}(\hat \bfbeta)$ & $\text{EE} (\widehat \sigma)$ & $\text{TP} (\hat \bfbeta)$   & $\text{TN} (\hat \bfbeta)$   & $\text{MS} (\hat \bfbeta)$\\ 
			~ & ($\times 10^{-4}$) & ($\times 10^{-2}$) &&&& \\\hline
			RLARS           & 6.6    & 4.39  & 0.09      & 1.00 & 0.97 & 7.00  \\
			sLTS            & 19.5   & 5.26  & 0.32      & 1.00 & 0.99 & 6.00  \\
			RANSAC          & 22.1   & 5.08  & 0.08      & 1.00 & 0.95 & 10.00 \\
			LAD-Lasso       & 316.7  & 15.37 & 2.48      & 0.97 & 0.96 & 8.75  \\\hline
			DPD-ncv, $\alpha=$ 0.2 & 3.4    & 4.74  & 0.05      & 1.00 & 1.00 & 5.00  \\
			DPD-ncv, $\alpha=$ 0.4 & 3.4    & 4.41  & 0.05      & 1.00 & 1.00 & 5.00  \\
			DPD-ncv, $\alpha=$ 0.6 & 3.7    & 4.40  & 0.06      & 1.00 & 1.00 & 5.00  \\
			DPD-ncv, $\alpha=$ 0.8 & 3.6    & 4.66  & 0.07      & 1.00 & 1.00 & 5.00  \\
			DPD-ncv, $\alpha=$ 1   & 3.9    & 4.60  & 0.07      & 1.00 & 1.00 & 5.00  \\\hline
			DPD-Lasso, $\alpha=$ 0.4 & 10.1 & 4.22 & 0.37 & 1.00 & 0.67 & 36.00 \\ 
  			DPD-Lasso, $\alpha=$ 0.8 & 8.1 & 4.24 & 0.37 & 1.00 & 0.72 & 31.50 \\ 
  			DPD-Lasso, $\alpha=$ 1.2 & 7.5 & 4.29 & 0.46 & 1.00 & 0.75 & 29.00 \\ 
  			DPD-Lasso, $\alpha=$ 1.6 & 7.2 & 4.27 & 0.00 & 1.00 & 0.77 & 27.00 \\ 
  			DPD-Lasso, $\alpha=$ 2 & 6.9 & 4.26 & 0.00 & 1.00 & 0.78 & 25.50 \\\hline
			LDPD-Lasso, $\alpha=$ 0.2 & 7.8    & 4.56  & 0.04      & 1.00 & 0.96 & 8.68  \\
			LDPD-Lasso, $\alpha=$ 0.4 & 8.4    & 4.63  & 0.06      & 1.00 & 0.97 & 8.30  \\
			LDPD-Lasso, $\alpha=$ 0.6 & 8.9    & 4.72  & 0.07      & 1.00 & 0.97 & 7.49  \\
			LDPD-Lasso, $\alpha=$ 0.8 & 9.9    & 4.83  & 0.08      & 1.00 & 0.98 & 7.07  \\
			LDPD-Lasso, $\alpha=$ 1   & 10.9   & 4.91  & 0.10      & 1.00 & 0.98 & 6.83  \\\hline
			Lasso        & 1358.6 & 33.88 & 6.01      & 0.83 & 0.99 & 5.53  \\
			SCAD         & 478.4  & 15.63 & 2.65      & 0.88 & 0.94 & 10.36 \\
			MCP          & 539.4  & 16.36 & 2.83      & 0.81 & 0.97 & 6.52  \\ \hline
%
			\multicolumn{7}{c}{{\bf Setting B}}\\
			\hline
			Method          & $\text{MSEE} (\hat \bfbeta)$    & $\text{RMSPE}(\hat \bfbeta)$ & $\text{EE} (\widehat \sigma)$ & $\text{TP} (\hat \bfbeta)$   & $\text{TN} (\hat \bfbeta)$   & $\text{MS} (\hat \bfbeta)$\\ 
			~ & ($\times 10^{-4}$) & ($\times 10^{-2}$) &&&& \\\hline
			RLARS & 2.6 & 4.47 & 0.07 & 1.00 & 0.99 & 6.00 \\ 
			sLTS & 10.2 & 1.71 & 0.12 & 1.00 & 0.96 & 9.00 \\ 
			RANSAC & 24.9 & 4.59 & 0.07 & 1.00 & 0.94 & 11.00 \\ 
			LAD-Lasso & 274.4 & 14.59 & 2.40 & 0.77 & 0.97 & 6.78 \\ \hline
			DPD-ncv, $\alpha=$ 0.2 & 2.6 & 4.45 & 0.04 & 1.00 & 0.99 & 6.00 \\ 
			DPD-ncv, $\alpha=$ 0.4 & 2.3 & 4.60 & 0.03 & 1.00 & 0.99 & 6.00 \\ 
			DPD-ncv, $\alpha=$ 0.6 & 2.3 & 4.68 & 0.04 & 1.00 & 0.99 & 6.00 \\ 
			DPD-ncv, $\alpha=$ 0.8 & 2.5 & 4.72 & 0.05 & 1.00 & 0.99 & 6.00 \\ 
			DPD-ncv, $\alpha=$ 1 & 2.7 & 4.78 & 0.06 & 1.00 & 0.99 & 6.00 \\ \hline
			DPD-Lasso, $\alpha=$ 0.4 & 8.3 & 4.06 & 0.37 & 1.00 & 0.68 & 35.00 \\ 
  			DPD-Lasso, $\alpha=$ 0.8 & 7.1 & 3.90 & 0.37 & 1.00 & 0.73 & 31.00 \\ 
  			DPD-Lasso, $\alpha=$ 1.2 & 6.5 & 3.86 & 0.44 & 1.00 & 0.75 & 29.00 \\ 
  			DPD-Lasso, $\alpha=$ 1.6 & 6.2 & 3.66 & 0.00 & 1.00 & 0.77 & 27.00 \\ 
  			DPD-Lasso, $\alpha=$ 2 & 6.1 & 3.86 & 0.00 & 1.00 & 0.79 & 25.00 \\\hline
			LDPD-Lasso, $\alpha=$ 0.2 & 7.9 & 4.58 & 0.06 & 1.00 & 0.95 & 10.10 \\ 
			LDPD-Lasso, $\alpha=$ 0.4 & 7.9 & 4.62 & 0.06 & 1.00 & 0.97 & 7.74 \\ 
			LDPD-Lasso, $\alpha=$ 0.6 & 8.2 & 4.70 & 0.07 & 1.00 & 0.98 & 7.17 \\ 
			LDPD-Lasso, $\alpha=$ 0.8 & 8.5 & 4.71 & 0.09 & 1.00 & 0.98 & 6.90 \\ 
			LDPD-Lasso, $\alpha=$ 1 & 8.7 & 4.72 & 0.10 & 1.00 & 0.98 & 6.69 \\ \hline
			Lasso & 654.9 & 23.54 & 4.48 & 0.05 & 1.00 & 0.33 \\ 
			SCAD & 619.5 & 19.65 & 3.24 & 0.49 & 0.94 & 7.98 \\ 
			MCP & 653.8 & 19.47 & 3.50 & 0.34 & 0.97 & 4.24 \\ 
			\hline
		\end{tabular}
		\caption{Table of outputs for $p=100$ and Y-outliers}
		\label{table:outtable_p100e1y}
	\end{footnotesize}
\end{table}

\begin{table}[ht]
	\begin{footnotesize}
		\centering
		\begin{tabular}{r|rrrrrr}
			\hline
			\multicolumn{7}{c}{{\bf Setting A}}\\
			\hline
			Method          & $\text{MSEE} (\hat \bfbeta)$    & $\text{RMSPE}(\hat \bfbeta)$ & $\text{EE} (\widehat \sigma)$ & $\text{TP} (\hat \bfbeta)$   & $\text{TN} (\hat \bfbeta)$   & $\text{MS} (\hat \bfbeta)$\\ 
			~ & ($\times 10^{-4}$) & ($\times 10^{-2}$) &&&& \\\hline
			RLARS & 2.6 & 3.5 & 0.04 & 1.00 & 0.99 & 6.00 \\ 
			sLTS & 29.6 & 5.5 & 0.23 & 1.00 & 0.99 & 5.50 \\ 
			RANSAC & 18.9 & 5.1 & 0.10 & 1.00 & 0.95 & 9.50 \\ 
			LAD-Lasso & 2894.6 & 55.5 & 1.54 & 0.84 & 0.54 & 48.25 \\\hline 
			DPD-ncv, $\alpha=$ 0.2 & 2.0 & 3.7 & 0.03 & 1.00 & 1.00 & 5.00 \\ 
			DPD-ncv, $\alpha=$ 0.4 & 2.0 & 3.8 & 0.04 & 1.00 & 1.00 & 5.00 \\ 
			DPD-ncv, $\alpha=$ 0.6 & 2.0 & 3.7 & 0.05 & 1.00 & 1.00 & 5.00 \\ 
			DPD-ncv, $\alpha=$ 0.8 & 2.2 & 3.7 & 0.07 & 1.00 & 1.00 & 5.00 \\ 
			DPD-ncv, $\alpha=$ 1 & 2.3 & 3.7 & 0.08 & 1.00 & 1.00 & 5.00 \\\hline 
			DPD-Lasso, $\alpha=$ 0.4 & 11.6 & 4.36 & 0.37 & 1.00 & 0.64 & 39.50 \\ 
  			DPD-Lasso, $\alpha=$ 0.8 & 9.3 & 4.10 & 0.37 & 1.00 & 0.67 & 36.00 \\ 
  			DPD-Lasso, $\alpha=$ 1.2 & 9.7 & 4.22 & 0.46 & 1.00 & 0.69 & 34.50 \\ 
  			DPD-Lasso, $\alpha=$ 1.6 & 10.1 & 4.22 & 0.00 & 1.00 & 0.69 & 34.00 \\ 
  			DPD-Lasso, $\alpha=$ 2 & 10.2 & 4.02 & 0.00 & 1.00 & 0.72 & 32.00 \\\hline 
			LDPD-Lasso, $\alpha=$ 0.2 & 6.9 & 4.5 & 0.04 & 1.00 & 0.98 & 7.19 \\ 
			LDPD-Lasso, $\alpha=$ 0.4 & 7.0 & 4.5 & 0.04 & 1.00 & 0.98 & 6.79 \\ 
			LDPD-Lasso, $\alpha=$ 0.6 & 7.2 & 4.5 & 0.05 & 1.00 & 0.98 & 6.59 \\ 
			LDPD-Lasso, $\alpha=$ 0.8 & 7.3 & 4.5 & 0.05 & 1.00 & 0.98 & 6.56 \\ 
			LDPD-Lasso, $\alpha=$ 1 & 7.6 & 4.5 & 0.06 & 1.00 & 0.98 & 6.43 \\\hline 
			Lasso & 2784.6 & 54.6 & 4.81 & 0.82 & 0.61 & 41.45 \\ 
			SCAD & 4450.5 & 58.9 & 5.49 & 0.67 & 0.87 & 15.56 \\ 
			MCP & 4505.8 & 62.9 & 5.40 & 0.67 & 0.87 & 15.68 \\ 
			\hline
			\multicolumn{7}{c}{{\bf Setting B}}\\
			\hline
			Method          & $\text{MSEE} (\hat \bfbeta)$    & $\text{RMSPE}(\hat \bfbeta)$ & $\text{EE} (\widehat \sigma)$ & $\text{TP} (\hat \bfbeta)$   & $\text{TN} (\hat \bfbeta)$   & $\text{MS} (\hat \bfbeta)$\\ 
			~ & ($\times 10^{-4}$) & ($\times 10^{-2}$) &&&& \\\hline
			RLARS & 4.0 & 3.7 & 0.07 & 1.00 & 0.98 & 7.00 \\ 
  			sLTS & 17.2 & 4.2 & 0.14 & 1.00 & 0.81 & 23.00 \\ 
  			RANSAC & 16.0 & 4.0 & 0.20 & 1.00 & 0.92 & 13.00 \\ 
  			LAD-Lasso & 421.6 & 19.1 & 2.10 & 0.62 & 0.90 & 13.05 \\ \hline 
			DPD-ncv, $\alpha=$ 0.2 & 3.1 & 3.8 & 0.05 & 1.00 & 0.99 & 6.00 \\ 
			DPD-ncv, $\alpha=$ 0.4 & 2.8 & 3.8 & 0.06 & 1.00 & 0.99 & 6.00 \\ 
			DPD-ncv, $\alpha=$ 0.6 & 3.2 & 3.7 & 0.08 & 1.00 & 0.99 & 6.00 \\ 
			DPD-ncv, $\alpha=$ 0.8 & 3.3 & 3.6 & 0.10 & 1.00 & 0.99 & 6.00 \\ 
			DPD-ncv, $\alpha=$ 1 & 3.3 & 3.7 & 0.12 & 1.00 & 0.99 & 6.00 \\ \hline 
			DPD-Lasso, $\alpha=$ 0.4 & 9.5 & 4.20 & 0.37 & 1.00 & 0.67 & 36.00 \\ 
  			DPD-Lasso, $\alpha=$ 0.8 & 9.3 & 4.14 & 0.40 & 1.00 & 0.69 & 34.00 \\ 
  			DPD-Lasso, $\alpha=$ 1.2 & 9.0 & 4.13 & 0.44 & 1.00 & 0.72 & 32.00 \\ 
  			DPD-Lasso, $\alpha=$ 1.6 & 8.9 & 3.78 & 0.00 & 1.00 & 0.73 & 31.00 \\ 
  			DPD-Lasso, $\alpha=$ 2 & 9.3 & 3.80 & 0.00 & 1.00 & 0.74 & 30.00 \\\hline 
			LDPD-Lasso, $\alpha=$ 0.2 & 6.7 & 4.5 & 0.05 & 1.00 & 0.93 & 11.55 \\ 
			LDPD-Lasso, $\alpha=$ 0.4 & 6.4 & 4.4 & 0.05 & 1.00 & 0.95 & 9.99 \\ 
			LDPD-Lasso, $\alpha=$ 0.6 & 6.5 & 4.4 & 0.05 & 1.00 & 0.95 & 10.06 \\ 
			LDPD-Lasso, $\alpha=$ 0.8 & 6.6 & 4.4 & 0.05 & 1.00 & 0.95 & 9.93 \\ 
			LDPD-Lasso, $\alpha=$ 1 & 6.7 & 4.3 & 0.06 & 1.00 & 0.95 & 9.58 \\ \hline 
  			LASSO & 151.3 & 12.9 & 0.85 & 0.98 & 0.64 & 39.20 \\ 
  			SCAD & 239.2 & 13.0 & 0.99 & 0.83 & 0.88 & 15.31 \\ 
  			MCP & 232.1 & 12.6 & 0.94 & 0.84 & 0.88 & 16.01 \\ 
			\hline
		\end{tabular}
		\caption{Table of outputs for $p=100$ and X-outliers}
		\label{table:outtable_p100e1x}
	\end{footnotesize}
\end{table}

\begin{table}[ht]
	\begin{footnotesize}
		\centering
		\begin{tabular}{r|rrrrrr}
			\hline
			\multicolumn{7}{c}{{\bf Setting A}}\\
			\hline
			Method          & $\text{MSEE} (\hat \bfbeta)$    & $\text{RMSPE}(\hat \bfbeta)$ & $\text{EE} (\widehat \sigma)$ & $\text{TP} (\hat \bfbeta)$   & $\text{TN} (\hat \bfbeta)$   & $\text{MS} (\hat \bfbeta)$\\ 
			~ & ($\times 10^{-4}$) & ($\times 10^{-2}$) &&&& \\\hline
			RLARS & 2.6 & 3.48 & 0.04 & 1.00 & 0.99 & 6.00 \\ 
  			sLTS & 27.4 & 5.39 & 0.23 & 1.00 & 0.99 & 6.00 \\ 
  			RANSAC & 18.5 & 4.75 & 0.12 & 1.00 & 0.95 & 10.00 \\ 
  			LAD-Lasso & 11.1 & 4.95 & 0.38 & 1.00 & 0.99 & 5.89 \\ \hline 
			DPD-ncv, $\alpha=$ 0.2 & 1.9 & 3.64 & 0.03 & 1.00 & 1.00 & 5.00 \\ 
			DPD-ncv, $\alpha=$ 0.4 & 1.8 & 3.79 & 0.04 & 1.00 & 1.00 & 5.00 \\ 
			DPD-ncv, $\alpha=$ 0.6 & 1.8 & 3.77 & 0.05 & 1.00 & 1.00 & 5.00 \\ 
			DPD-ncv, $\alpha=$ 0.8 & 2.0 & 3.68 & 0.07 & 1.00 & 1.00 & 5.00 \\ 
			DPD-ncv, $\alpha=$ 1 & 2.0 & 3.79 & 0.08 & 1.00 & 1.00 & 5.00 \\ \hline 
			DPD-Lasso, $\alpha=$ 0.4 & 7.4 & 3.87 & 0.37 & 1.00 & 0.65 & 38.00 \\ 
  			DPD-Lasso, $\alpha=$ 0.8 & 5.9 & 3.60 & 0.43 & 1.00 & 0.71 & 33.00 \\ 
  			DPD-Lasso, $\alpha=$ 1.2 & 5.1 & 3.61 & 0.22 & 1.00 & 0.74 & 30.00 \\ 
  			DPD-Lasso, $\alpha=$ 1.6 & 4.9 & 3.61 & 0.00 & 1.00 & 0.75 & 29.00 \\ 
  			DPD-Lasso, $\alpha=$ 2 & 4.6 & 3.61 & 0.00 & 1.00 & 0.78 & 26.00 \\\hline 
			LDPD-Lasso, $\alpha=$ 0.2 & 5.4 & 5.44 & 0.05 & 1.00 & 0.98 & 6.56 \\ 
			LDPD-Lasso, $\alpha=$ 0.4 & 5.4 & 5.38 & 0.06 & 1.00 & 0.99 & 6.11 \\ 
			LDPD-Lasso, $\alpha=$ 0.6 & 5.4 & 5.33 & 0.07 & 1.00 & 0.99 & 6.11 \\ 
			LDPD-Lasso, $\alpha=$ 0.8 & 5.5 & 5.30 & 0.08 & 1.00 & 0.99 & 5.89 \\ 
			LDPD-Lasso, $\alpha=$ 1 & 5.5 & 5.29 & 0.09 & 1.00 & 0.99 & 5.67 \\ \hline 
  			LASSO & 8.8 & 5.72 & 0.41 & 1.00 & 0.99 & 6.33 \\ 
  			SCAD & 83.3 & 6.48 & 0.77 & 0.98 & 1.00 & 4.89 \\ 
  			MCP & 69.3 & 6.40 & 0.65 & 0.98 & 1.00 & 4.89 \\ 
			\hline
			\multicolumn{7}{c}{{\bf Setting B}}\\
			\hline
			Method          & $\text{MSEE} (\hat \bfbeta)$    & $\text{RMSPE}(\hat \bfbeta)$ & $\text{EE} (\widehat \sigma)$ & $\text{TP} (\hat \bfbeta)$   & $\text{TN} (\hat \bfbeta)$   & $\text{MS} (\hat \bfbeta)$\\ 
			~ & ($\times 10^{-4}$) & ($\times 10^{-2}$) &&&& \\\hline
			RLARS & 4.0 & 3.59 & 0.07 & 1.00 & 0.98 & 7.00 \\ 
  			sLTS & 17.3 & 4.13 & 0.15 & 1.00 & 0.81 & 23.00 \\ 
  			RANSAC & 17.2 & 4.39 & 0.18 & 1.00 & 0.93 & 12.00 \\ 
  			LAD-Lasso & 12.8 & 5.49 & 0.36 & 1.00 & 0.99 & 6.00 \\ \hline 
			DPD-ncv, $\alpha=$ 0.2 & 2.7 & 4.08 & 0.04 & 1.00 & 0.99 & 6.00 \\ 
			DPD-ncv, $\alpha=$ 0.4 & 2.8 & 3.74 & 0.06 & 1.00 & 0.99 & 6.00 \\ 
			DPD-ncv, $\alpha=$ 0.6 & 3.0 & 3.70 & 0.08 & 1.00 & 1.00 & 5.00 \\ 
			DPD-ncv, $\alpha=$ 0.8 & 3.2 & 3.60 & 0.09 & 1.00 & 0.99 & 5.50 \\ 
			DPD-ncv, $\alpha=$ 1 & 3.1 & 3.64 & 0.11 & 1.00 & 0.99 & 5.50 \\ \hline 
			DPD-Lasso, $\alpha=$ 0.4 & 7.9 & 3.88 & 0.37 & 1.00 & 0.64 & 39.00 \\ 
  			DPD-Lasso, $\alpha=$ 0.8 & 6.7 & 3.57 & 0.43 & 1.00 & 0.68 & 35.00 \\ 
  			DPD-Lasso, $\alpha=$ 1.2 & 6.3 & 3.77 & 0.00 & 1.00 & 0.73 & 31.00 \\ 
  			DPD-Lasso, $\alpha=$ 1.6 & 6.1 & 3.66 & 0.00 & 1.00 & 0.75 & 29.00 \\ 
  			DPD-Lasso, $\alpha=$ 2 & 5.9 & 3.72 & 0.00 & 1.00 & 0.77 & 27.00 \\\hline 
			LDPD-Lasso, $\alpha=$ 0.2 & 5.0 & 5.58 & 0.03 & 1.00 & 0.95 & 9.78 \\ 
			LDPD-Lasso, $\alpha=$ 0.4 & 5.1 & 5.58 & 0.03 & 1.00 & 0.96 & 8.67 \\ 
			LDPD-Lasso, $\alpha=$ 0.6 & 5.0 & 5.47 & 0.03 & 1.00 & 0.96 & 8.67 \\ 
			LDPD-Lasso, $\alpha=$ 0.8 & 5.0 & 5.42 & 0.04 & 1.00 & 0.96 & 8.78 \\ 
			LDPD-Lasso, $\alpha=$ 1 & 5.1 & 5.36 & 0.04 & 1.00 & 0.96 & 8.56 \\ \hline 
  			LASSO & 7.0 & 5.80 & 0.33 & 1.00 & 0.97 & 7.44 \\ 
  			SCAD & 1.9 & 4.90 & 0.20 & 1.00 & 0.97 & 8.00 \\ 
  			MCP & 2.0 & 4.83 & 0.21 & 1.00 & 0.99 & 6.33 \\
			\hline
		\end{tabular}
		\caption{Table of outputs for $p=100$ and no outliers}
		\label{table:outtable_p100}
	\end{footnotesize}
\end{table}

\begin{table}[ht]
	\begin{footnotesize}
		\centering
		\begin{tabular}{r|rrrrrr}
			\hline
			\multicolumn{7}{c}{{\bf Setting A}}\\
			\hline
			Method          & $\text{MSEE} (\hat \bfbeta)$    & $\text{RMSPE}(\hat \bfbeta)$ & $\text{EE} (\widehat \sigma)$ & $\text{TP} (\hat \bfbeta)$   & $\text{TN} (\hat \bfbeta)$   & $\text{MS} (\hat \bfbeta)$\\ 
			~ & ($\times 10^{-4}$) & ($\times 10^{-2}$) &&&& \\\hline
			RLARS & 5.7 & 5.67 & 0.14 & 1.00 & 0.99 & 6.00 \\ 
			sLTS & 10.7 & 4.74 & 0.32 & 1.00 & 0.99 & 6.00 \\ 
			RANSAC & 14.3 & 3.78 & 0.08 & 1.00 & 0.97 & 10.00 \\ 
			LAD-Lasso & 184.4 & 15.59 & 2.59 & 0.94 & 0.98 & 9.09 \\ \hline
			DPD-ncv, $\alpha=$ 0.2 & 4.5 & 4.85 & 0.10 & 1.00 & 1.00 & 5.00 \\ 
			DPD-ncv, $\alpha=$ 0.4 & 2.8 & 4.88 & 0.10 & 1.00 & 1.00 & 5.00 \\ 
			DPD-ncv, $\alpha=$ 0.6 & 3.2 & 4.98 & 0.12 & 1.00 & 1.00 & 5.00 \\ 
			DPD-ncv, $\alpha=$ 0.8 & 2.8 & 5.08 & 0.12 & 1.00 & 1.00 & 5.00 \\ 
			DPD-ncv, $\alpha=$ 1 & 3.4 & 5.15 & 0.13 & 1.00 & 1.00 & 5.00 \\ \hline
			DPD-Lasso, $\alpha=$ 0.4 & 7.4 & 5.85 & 0.37 & 1.00 & 0.76 & 51.00 \\ 
  			DPD-Lasso, $\alpha=$ 0.8 & 5.5 & 4.58 & 0.43 & 1.00 & 0.81 & 42.00 \\ 
  			DPD-Lasso, $\alpha=$ 1.2 & 4.8 & 4.28 & 0.46 & 1.00 & 0.85 & 34.00 \\ 
  			DPD-Lasso, $\alpha=$ 1.6 & 5.0 & 4.39 & 0.00 & 1.00 & 0.84 & 36.00 \\ 
  			DPD-Lasso, $\alpha=$ 2 & 5.7 & 4.44 & 0.00 & 1.00 & 0.86 & 32.00 \\\hline
			LDPD-Lasso, $\alpha=$ 0.2 & 4.6 & 4.53 & 0.05 & 1.00 & 0.97 & 11.13 \\ 
			LDPD-Lasso, $\alpha=$ 0.4 & 5.3 & 4.66 & 0.06 & 1.00 & 0.98 & 9.02 \\ 
			LDPD-Lasso, $\alpha=$ 0.6 & 6.4 & 4.82 & 0.07 & 1.00 & 0.98 & 7.99 \\ 
			LDPD-Lasso, $\alpha=$ 0.8 & 7.9 & 5.09 & 0.10 & 1.00 & 0.99 & 7.32 \\ 
			LDPD-Lasso, $\alpha=$ 1 & 9.2 & 5.29 & 0.14 & 1.00 & 0.99 & 6.84 \\ \hline
			Lasso & 757.0 & 32.81 & 6.20 & 0.80 & 0.99 & 5.70 \\ 
			SCAD & 289.5 & 19.60 & 3.05 & 0.84 & 0.96 & 11.26 \\ 
			MCP & 295.3 & 19.48 & 3.01 & 0.79 & 0.99 & 6.68 \\ 
			\hline
			\multicolumn{7}{c}{{\bf Setting B}}\\
			\hline
			Method          & $\text{MSEE} (\hat \bfbeta)$    & $\text{RMSPE}(\hat \bfbeta)$ & $\text{EE} (\widehat \sigma)$ & $\text{TP} (\hat \bfbeta)$   & $\text{TN} (\hat \bfbeta)$   & $\text{MS} (\hat \bfbeta)$\\ 
			~ & ($\times 10^{-4}$) & ($\times 10^{-2}$) &&&& \\\hline
			RLARS & 1.7 & 3.88 & 0.08 & 1.00 & 1.00 & 5.50 \\ 
			sLTS & 11.6 & 6.98 & 0.20 & 1.00 & 0.85 & 34.50 \\ 
			RANSAC & 15.7 & 4.03 & 0.12 & 1.00 & 0.96 & 12.00 \\ 
			LAD-Lasso & 156.2 & 15.74 & 2.55 & 0.72 & 0.99 & 6.35 \\ \hline
			DPD-ncv, $\alpha=$ 0.2 & 1.3 & 3.84 & 0.02 & 1.00 & 1.00 & 5.00 \\ 
			DPD-ncv, $\alpha=$ 0.4 & 1.1 & 4.27 & 0.02 & 1.00 & 1.00 & 5.00 \\ 
			DPD-ncv, $\alpha=$ 0.6 & 1.5 & 4.02 & 0.03 & 1.00 & 1.00 & 5.00 \\ 
			DPD-ncv, $\alpha=$ 0.8 & 1.6 & 4.03 & 0.04 & 1.00 & 1.00 & 5.00 \\ 
			DPD-ncv, $\alpha=$ 1 & 1.6 & 4.03 & 0.05 & 1.00 & 1.00 & 5.00 \\ \hline
			DPD-Lasso, $\alpha=$ 0.4 & 6.0 & 4.26 & 0.37 & 1.00 & 0.77 & 49.00 \\ 
  			DPD-Lasso, $\alpha=$ 0.8 & 4.4 & 4.51 & 0.43 & 1.00 & 0.80 & 44.00 \\ 
  			DPD-Lasso, $\alpha=$ 1.2 & 3.8 & 4.31 & 0.00 & 1.00 & 0.82 & 41.00 \\ 
  			DPD-Lasso, $\alpha=$ 1.6 & 3.4 & 3.88 & 0.00 & 1.00 & 0.84 & 37.00 \\ 
  			DPD-Lasso, $\alpha=$ 2 & 3.2 & 4.05 & 0.00 & 1.00 & 0.85 & 35.00 \\\hline
			LDPD-Lasso, $\alpha=$ 0.2 & 4.5 & 4.57 & 0.06 & 1.00 & 0.97 & 11.42 \\ 
			LDPD-Lasso, $\alpha=$ 0.4 & 4.6 & 4.61 & 0.07 & 1.00 & 0.98 & 8.30 \\ 
			LDPD-Lasso, $\alpha=$ 0.6 & 4.8 & 4.58 & 0.08 & 1.00 & 0.99 & 7.58 \\ 
			LDPD-Lasso, $\alpha=$ 0.8 & 4.8 & 4.56 & 0.09 & 1.00 & 0.99 & 7.31 \\ 
			LDPD-Lasso, $\alpha=$ 1 & 4.9 & 4.52 & 0.10 & 1.00 & 0.99 & 7.24 \\ \hline
			Lasso & 333.6 & 25.45 & 4.38 & 0.03 & 1.00 & 0.31 \\ 
			SCAD & 300.3 & 21.44 & 3.32 & 0.41 & 0.97 & 8.58 \\ 
			MCP & 351.0 & 23.84 & 3.54 & 0.28 & 0.98 & 4.63 \\ 
			\hline
		\end{tabular}
		\caption{Table of outputs for $p=200$ and Y-outliers}
		\label{table:outtable_p200e1y}
	\end{footnotesize}
\end{table}

%
\begin{table}[ht]
	\begin{footnotesize}
		\centering
		\begin{tabular}{r|rrrrrr}
			\hline
			\multicolumn{7}{c}{{\bf Setting A}}\\
			\hline
			Method          & $\text{MSEE} (\hat \bfbeta)$    & $\text{RMSPE}(\hat \bfbeta)$ & $\text{EE} (\widehat \sigma)$ & $\text{TP} (\hat \bfbeta)$   & $\text{TN} (\hat \bfbeta)$   & $\text{MS} (\hat \bfbeta)$\\ 
			~ & ($\times 10^{-4}$) & ($\times 10^{-2}$) &&&& \\\hline
			RLARS & 1.1 & 3.4 & 0.05 & 1.00 & 0.99 & 6.00 \\ 
  			sLTS & 14.9 & 5.2 & 0.24 & 1.00 & 0.99 & 6.00 \\ 
  			RANSAC & 10.5 & 4.0 & 0.10 & 1.00 & 0.98 & 9.00 \\ 
  			LAD-Lasso & 996.1 & 37.6 & 0.49 & 0.89 & 0.65 & 73.18 \\\hline 
			DPD-ncv, $\alpha=$ 0.2 & 0.9 & 3.3 & 0.04 & 1.00 & 1.00 & 5.00 \\ 
			DPD-ncv, $\alpha=$ 0.4 & 0.9 & 3.5 & 0.05 & 1.00 & 1.00 & 5.00 \\ 
			DPD-ncv, $\alpha=$ 0.6 & 1.0 & 3.5 & 0.06 & 1.00 & 1.00 & 5.00 \\ 
			DPD-ncv, $\alpha=$ 0.8 & 1.0 & 3.5 & 0.08 & 1.00 & 1.00 & 5.00 \\ 
			DPD-ncv, $\alpha=$ 1 & 1.1 & 3.4 & 0.10 & 1.00 & 1.00 & 5.00 \\  \hline 
			DPD-Lasso, $\alpha=$ 0.4 & 7.3 & 5.60 & 0.37 & 1.00 & 0.76 & 52.50 \\ 
  			DPD-Lasso, $\alpha=$ 0.8 & 6.0 & 4.44 & 0.43 & 1.00 & 0.78 & 47.00 \\ 
  			DPD-Lasso, $\alpha=$ 1.2 & 5.9 & 4.35 & 0.44 & 1.00 & 0.79 & 45.50 \\ 
  			DPD-Lasso, $\alpha=$ 1.6 & 5.7 & 4.35 & 0.00 & 1.00 & 0.79 & 45.00 \\ 
  			DPD-Lasso, $\alpha=$ 2 & 6.0 & 4.34 & 0.00 & 1.00 & 0.80 & 43.50 \\\hline 
			LDPD-Lasso, $\alpha=$ 0.2 & 3.8 & 4.4 & 0.04 & 1.00 & 0.98 & 7.97 \\ 
			LDPD-Lasso, $\alpha=$ 0.4 & 3.8 & 4.4 & 0.05 & 1.00 & 0.99 & 7.12 \\ 
			LDPD-Lasso, $\alpha=$ 0.6 & 3.7 & 4.3 & 0.05 & 1.00 & 0.99 & 6.97 \\ 
			LDPD-Lasso, $\alpha=$ 0.8 & 3.8 & 4.4 & 0.06 & 1.00 & 0.99 & 6.69 \\ 
			LDPD-Lasso, $\alpha=$ 1 & 3.9 & 4.4 & 0.07 & 1.00 & 0.99 & 6.54 \\ \hline 
  			LASSO & 1077.2 & 39.8 & 2.74 & 0.87 & 0.71 & 61.31 \\ 
  			SCAD & 2163.3 & 49.6 & 4.49 & 0.67 & 0.94 & 15.80 \\ 
  			MCP & 2150.3 & 51.8 & 4.47 & 0.65 & 0.94 & 15.88 \\ 
			\hline
			\multicolumn{7}{c}{{\bf Setting B}}\\
			\hline
			Method          & $\text{MSEE} (\hat \bfbeta)$    & $\text{RMSPE}(\hat \bfbeta)$ & $\text{EE} (\widehat \sigma)$ & $\text{TP} (\hat \bfbeta)$   & $\text{TN} (\hat \bfbeta)$   & $\text{MS} (\hat \bfbeta)$\\ 
			~ & ($\times 10^{-4}$) & ($\times 10^{-2}$) &&&& \\\hline
			RLARS & 2.4 & 3.6 & 0.09 & 1.00 & 0.98 & 8.00 \\ 
  			sLTS & 13.4 & 4.9 & 0.19 & 1.00 & 0.86 & 32.00 \\ 
  			RANSAC & 10.8 & 5.1 & 0.24 & 1.00 & 0.95 & 14.00 \\ 
  			LAD-Lasso & 255.8 & 20.6 & 2.52 & 0.49 & 0.97 & 9.01 \\ \hline 
			DPD-ncv, $\alpha=$ 0.2 & 2.0 & 3.6 & 0.06 & 1.00 & 0.99 & 6.00 \\ 
			DPD-ncv, $\alpha=$ 0.4 & 1.9 & 3.4 & 0.07 & 1.00 & 0.99 & 6.00 \\ 
			DPD-ncv, $\alpha=$ 0.6 & 2.1 & 3.6 & 0.09 & 1.00 & 0.99 & 6.50 \\ 
			DPD-ncv, $\alpha=$ 0.8 & 2.2 & 3.9 & 0.11 & 1.00 & 0.99 & 6.00 \\ 
			DPD-ncv, $\alpha=$ 1 & 2.1 & 3.7 & 0.12 & 1.00 & 0.99 & 6.00 \\ \hline 
			DPD-Lasso, $\alpha=$ 0.4 & 8.1 & 5.30 & 0.37 & 1.00 & 0.73 & 58.00 \\ 
  			DPD-Lasso, $\alpha=$ 0.8 & 6.5 & 4.13 & 0.43 & 1.00 & 0.77 & 50.00 \\ 
  			DPD-Lasso, $\alpha=$ 1.2 & 6.2 & 4.37 & 0.30 & 1.00 & 0.77 & 49.00 \\ 
  			DPD-Lasso, $\alpha=$ 1.6 & 6.1 & 4.47 & 0.00 & 1.00 & 0.79 & 46.00 \\ 
  			DPD-Lasso, $\alpha=$ 2 & 6.2 & 4.48 & 0.00 & 1.00 & 0.79 & 45.00 \\\hline 
			LDPD-Lasso, $\alpha=$ 0.2 & 3.8 & 4.2 & 0.05 & 1.00 & 0.96 & 12.07 \\ 
			LDPD-Lasso, $\alpha=$ 0.4 & 3.7 & 4.2 & 0.04 & 1.00 & 0.97 & 10.25 \\ 
			LDPD-Lasso, $\alpha=$ 0.6 & 3.7 & 4.2 & 0.05 & 1.00 & 0.97 & 10.20 \\ 
			LDPD-Lasso, $\alpha=$ 0.8 & 3.7 & 4.2 & 0.05 & 1.00 & 0.97 & 10.02 \\ 
			LDPD-Lasso, $\alpha=$ 1 & 3.8 & 4.1 & 0.05 & 1.00 & 0.97 & 9.94 \\ \hline 
  			LASSO & 59.4 & 10.4 & 0.36 & 0.99 & 0.74 & 56.49 \\ 
  			SCAD & 116.6 & 11.9 & 0.72 & 0.83 & 0.94 & 16.10 \\ 
  			MCP & 116.1 & 12.3 & 0.70 & 0.83 & 0.94 & 16.35 \\ 
			\hline
		\end{tabular}
		\caption{Table of outputs for $p=200$ and X-outliers}
		\label{table:outtable_p200e1x}
	\end{footnotesize}
\end{table}

\begin{table}[ht]
	\begin{footnotesize}
		\centering
		\begin{tabular}{r|rrrrrr}
			\hline
			\multicolumn{7}{c}{{\bf Setting A}}\\
			\hline
			Method          & $\text{MSEE} (\hat \bfbeta)$    & $\text{RMSPE}(\hat \bfbeta)$ & $\text{EE} (\widehat \sigma)$ & $\text{TP} (\hat \bfbeta)$   & $\text{TN} (\hat \bfbeta)$   & $\text{MS} (\hat \bfbeta)$\\ 
			~ & ($\times 10^{-4}$) & ($\times 10^{-2}$) &&&& \\\hline
			RLARS & 1.1 & 3.54 & 0.05 & 1.00 & 0.99 & 6.00 \\ 
  			sLTS & 14.9 & 5.55 & 0.24 & 1.00 & 0.99 & 6.00 \\ 
  			RANSAC & 11.1 & 4.43 & 0.09 & 1.00 & 0.97 & 10.00 \\ 
  			LAD-Lasso & 10.2 & 4.92 & 0.36 & 1.00 & 0.99 & 6.56 \\\hline 
			DPD-ncv, $\alpha=$ 0.2 & 0.9 & 3.42 & 0.04 & 1.00 & 1.00 & 5.00 \\ 
			DPD-ncv, $\alpha=$ 0.4 & 1.0 & 3.74 & 0.05 & 1.00 & 1.00 & 5.00 \\ 
			DPD-ncv, $\alpha=$ 0.6 & 1.1 & 3.57 & 0.07 & 1.00 & 1.00 & 5.00 \\ 
			DPD-ncv, $\alpha=$ 0.8 & 1.1 & 3.71 & 0.08 & 1.00 & 1.00 & 5.00 \\ 
			DPD-ncv, $\alpha=$ 1 & 1.1 & 3.52 & 0.10 & 1.00 & 1.00 & 5.00 \\ \hline 
			DPD-Lasso, $\alpha=$ 0.4 & 5.4 & 4.46 & 0.37 & 1.00 & 0.73 & 58.50 \\ 
  			DPD-Lasso, $\alpha=$ 0.8 & 3.5 & 3.50 & 0.43 & 1.00 & 0.78 & 48.50 \\ 
  			DPD-Lasso, $\alpha=$ 1.2 & 3.0 & 3.49 & 0.00 & 1.00 & 0.80 & 44.00 \\ 
  			DPD-Lasso, $\alpha=$ 1.6 & 2.7 & 3.44 & 0.00 & 1.00 & 0.82 & 40.00 \\ 
  			DPD-Lasso, $\alpha=$ 2 & 2.5 & 3.38 & 0.00 & 1.00 & 0.84 & 37.00 \\\hline 
			LDPD-Lasso, $\alpha=$ 0.2 & 4.1 & 4.81 & 0.04 & 1.00 & 0.98 & 8.44 \\ 
			LDPD-Lasso, $\alpha=$ 0.4 & 4.0 & 4.78 & 0.05 & 1.00 & 0.99 & 7.33 \\ 
			LDPD-Lasso, $\alpha=$ 0.6 & 3.9 & 4.82 & 0.05 & 1.00 & 0.99 & 7.33 \\ 
			LDPD-Lasso, $\alpha=$ 0.8 & 3.9 & 4.85 & 0.06 & 1.00 & 0.99 & 7.11 \\ 
			LDPD-Lasso, $\alpha=$ 1 & 3.9 & 4.91 & 0.07 & 1.00 & 0.99 & 7.00 \\ \hline 
  			LASSO & 6.7 & 5.10 & 0.39 & 1.00 & 0.99 & 6.44 \\ 
  			SCAD & 40.8 & 5.52 & 0.77 & 1.00 & 1.00 & 5.00 \\ 
  			MCP & 30.5 & 4.47 & 0.62 & 0.98 & 1.00 & 4.89 \\ 
			\hline
			\multicolumn{7}{c}{{\bf Setting B}}\\
			\hline
			Method          & $\text{MSEE} (\hat \bfbeta)$    & $\text{RMSPE}(\hat \bfbeta)$ & $\text{EE} (\widehat \sigma)$ & $\text{TP} (\hat \bfbeta)$   & $\text{TN} (\hat \bfbeta)$   & $\text{MS} (\hat \bfbeta)$\\ 
			~ & ($\times 10^{-4}$) & ($\times 10^{-2}$) &&&& \\\hline
			RLARS & 2.4 & 3.57 & 0.09 & 1.00 & 0.98 & 8.00 \\ 
  			sLTS & 14.0 & 5.03 & 0.19 & 1.00 & 0.86 & 32.00 \\ 
  			RANSAC & 9.8 & 4.42 & 0.20 & 1.00 & 0.96 & 13.00 \\ 
  			LAD-Lasso & 10.6 & 5.14 & 0.40 & 1.00 & 0.99 & 6.50 \\ \hline 
			DPD-ncv, $\alpha=$ 0.2 & 2.0 & 3.52 & 0.05 & 1.00 & 0.99 & 6.00 \\ 
			DPD-ncv, $\alpha=$ 0.4 & 2.0 & 3.27 & 0.07 & 1.00 & 0.99 & 6.00 \\ 
			DPD-ncv, $\alpha=$ 0.6 & 2.1 & 3.51 & 0.09 & 1.00 & 0.99 & 6.00 \\ 
			DPD-ncv, $\alpha=$ 0.8 & 2.2 & 3.86 & 0.11 & 1.00 & 0.99 & 6.00 \\ 
			DPD-ncv, $\alpha=$ 1 & 2.1 & 3.73 & 0.12 & 1.00 & 0.99 & 6.00 \\ \hline 
			DPD-Lasso, $\alpha=$ 0.4 & 6.1 & 4.01 & 0.37 & 1.00 & 0.72 & 60.00 \\ 
  			DPD-Lasso, $\alpha=$ 0.8 & 4.3 & 3.72 & 0.43 & 1.00 & 0.77 & 49.00 \\ 
  			DPD-Lasso, $\alpha=$ 1.2 & 3.7 & 3.85 & 0.00 & 1.00 & 0.80 & 44.00 \\ 
  			DPD-Lasso, $\alpha=$ 1.6 & 3.5 & 3.50 & 0.00 & 1.00 & 0.82 & 41.00 \\ 
  			DPD-Lasso, $\alpha=$ 2 & 3.3 & 3.48 & 0.00 & 1.00 & 0.83 & 38.00 \\\hline 
			LDPD-Lasso, $\alpha=$ 0.2 & 4.2 & 4.57 & 0.06 & 1.00 & 0.94 & 17.40 \\ 
			LDPD-Lasso, $\alpha=$ 0.4 & 4.0 & 4.78 & 0.05 & 1.00 & 0.96 & 12.20 \\ 
			LDPD-Lasso, $\alpha=$ 0.6 & 4.0 & 4.80 & 0.05 & 1.00 & 0.96 & 12.30 \\ 
			LDPD-Lasso, $\alpha=$ 0.8 & 3.9 & 4.88 & 0.06 & 1.00 & 0.97 & 11.50 \\ 
			LDPD-Lasso, $\alpha=$ 1 & 4.1 & 4.72 & 0.05 & 1.00 & 0.96 & 12.70 \\ \hline 
  			LASSO & 4.7 & 4.83 & 0.29 & 1.00 & 0.98 & 9.70 \\ 
  			SCAD & 1.6 & 4.21 & 0.19 & 1.00 & 0.98 & 8.00 \\ 
  			MCP & 1.6 & 4.39 & 0.18 & 1.00 & 0.99 & 7.20 \\ 
			\hline
		\end{tabular}
		\caption{Table of outputs for $p=200$ and no outliers}
		\label{table:outtable_p200}
	\end{footnotesize}
\end{table}

\end{document}